\documentclass[10pt,aps,showpacs,nofootinbib,prd,aps,epsf,floats,
               amsmath,amssymb,amsfonts,axodraw,floatfix,graphicx,twocolumn]{revtex4-1}
\usepackage{amsmath, amssymb}
\usepackage{paralist}
\newcommand{\mathsym}[1]{{}}

\usepackage{graphicx}
\usepackage{amsmath}
\usepackage{amssymb}
\usepackage{bm}
\newcommand{\be}{\begin{equation}}
\newcommand{\ee}{\end{equation}}
\newcommand{\bea}{\begin{eqnarray}}
\newcommand{\eea}{\end{eqnarray}}

\newsavebox{\PSLASH}
 \sbox{\PSLASH}{$p$\hspace{-1.8mm}/}
 
\renewcommand{\theequation}{\thesection.\arabic{equation}}
\newcounter{saveeqn}
\newcommand{\add}{\addtocounter{equation}{1}}
\newcommand{\alpheqn}{\setcounter{saveeqn}{\value{equation}}%
\setcounter{equation}{0}%
\renewcommand{\theequation}{\mbox{\thesection.\arabic{saveeqn}{\alph{equation}}}}}
\newcommand{\reseteqn}{\setcounter{equation}{\value{saveeqn}}%
\renewcommand{\theequation}{\thesection.\arabic{equation}}}

 \newsavebox{\notrightarrow}
 \sbox{\notrightarrow}{$\to$\hspace{-4mm}/}
 
 \newsavebox{\PARTIALSLASH}
 \sbox{\PARTIALSLASH}{$\partial$\hspace{-1.6mm}/}
 
 \newsavebox{\ASLASH}
 \sbox{\ASLASH}{$A$\hspace{-2.1mm}/}
 
 \newsavebox{\KSLASH}
 \sbox{\KSLASH}{$k$\hspace{-1.8mm}/}
 
 \newsavebox{\LSLASH}
 \sbox{\LSLASH}{$\ell$\hspace{-1.8mm}/}
 
 \newsavebox{\QSLASH}
 \sbox{\QSLASH}{$q$\hspace{-1.8mm}/}
 
 \newsavebox{\DSLASH}
 \sbox{\DSLASH}{$D$\hspace{-2.2mm}/}
 
 \newsavebox{\DbfSLASH}
 \sbox{\DbfSLASH}{${\mathbf D}$\hspace{-2.8mm}/}
 
 \newsavebox{\DELVECRIGHT}
 \sbox{\DELVECRIGHT}{$\stackrel{\rightarrow}{\partial}$}
 
 \newcommand{\blue}{\IfColor{\textCadetBlue}{}}
\newcommand{\black}{\IfColor{\textBlack}{}}
\newcommand{\red}{\IfColor{\textRed}{}}
\newcommand{\green}{\IfColor{\textOliveGreen}{}}
\newcommand{\lila}{\IfColor{\textRedViolet}{}}

\begin{document}
\title{Paramagnetic squeezing of a uniformly expanding quark-gluon plasma\\
in and out of equilibrium}
\author{N. Sadooghi}\email{sadooghi@physics.sharif.ir}
\author{S. M. A. Tabatabaee}\email{tabatabaeemehr\_sma@physics.sharif.ir}
\affiliation{Department of Physics, Sharif University of Technology,
P.O. Box 11155-9161, Tehran, Iran}
\begin{abstract}
The plasma of quarks and gluons created in ultrarelativistic heavy-ion collisions turns out  to be paramagnetic. In the presence of a background magnetic field, this paramagnetism thus leads to a pressure anisotropy, similar to anisotropies appearing in a viscous fluid. In the present paper, we use this analogy, and develop a framework similar to anisotropic hydrodynamics, to take the pressure anisotropy caused, in particular, by the nonvanishing magnetization of a plasma of quarks and gluons into account. We consider the first two moments of the classical Boltzmann equation in the presence of an electromagnetic source in the relaxation-time approximation, and derive a set of coupled differential equations for the anisotropy parameter $\xi_0$ and the effective temperature $\lambda_0$ of an ideal fluid with nonvanishing magnetization. We also extend this method to a dissipative fluid with finite magnetization in the presence of a strong and dynamical magnetic field. We present a systematic method leading to the one-particle distribution function of this magnetized dissipative medium in a first-order derivative expansion, and arrive at analytical expressions for the shear and bulk viscosities in terms of the anisotropy parameter $\xi$ and effective temperature $\lambda$. We then solve the corresponding differential equations for $(\xi_0,\lambda_0)$ and $(\xi,\lambda)$ numerically, and determine, in this way, the proper time and temperature dependence of the energy density, directional pressures, speed of sound, and the magnetic susceptibility of a longitudinally expanding magnetized quark-gluon plasma in and out of equilibrium.
\end{abstract}
\maketitle
\section{Introduction}\label{Introduction}\label{sec1}
\setcounter{equation}{0}
The past decade has witnessed enormous progress in the field of relativistic hydrodynamics, which finds important applications in the modern ultrarelativistic heavy-ion collision (HIC) experiments at the Relativistic Heavy Ion Collider (RHIC) and the Large Hadron Collider (LHC). The aim of these experiments is to produce a plasma of quarks and gluons, and to study its evolution from an early out of equilibrium stage, immediately after the collision, to a late hadronization stage, where the system is approximately thermalized. It is widely believed that in the early stage after the collision, the quark-gluon plasma (QGP) produced at the RHIC and LHC possesses a high degree of momentum-space anisotropy, that mainly arises from the initial state spatial anisotropies of the collision \cite{rajagopal2018,strickland2014,strickland2017}. These anisotropies are then converted into large pressure anisotropies in the transverse and longitudinal directions with respect to the beam direction.
The question how fast these anisotropies evolve during the hydrodynamical expansion of the QGP, in other words, how fast the isotropization process occurs, is extensively studied in the literature (see \cite{strickland2017} and references therein). In particular, in the framework of anisotropic hydrodynamics (aHydro) \cite{florkowski2010, strickland2010}, the small but nonvanishing ratio of the shear viscosity over entropy density of the QGP, $\eta/s$, is assumed to be the main source for the evolution of pressure anisotropies in this medium (for recent reviews of aHydro, see \cite{strickland2014,strickland2017}). In this framework, the momentum-space anisotropy is intrinsically implemented in an anisotropic one-particle distribution function $f(x,p;\xi,\lambda)$, including an anisotropy parameter $\xi$ and an effective temperature $\lambda$ \cite{strickland2010}. Taking the first two moments of the Boltzmann equation, satisfied by $f$, and using an appropriate relaxation time approximation (RTA), two coupled differential equations are derived for $\xi$ and $\lambda$. The numerical solution of these equations leads directly to the proper time dependence of $\xi$ and $\lambda$, and indirectly to the evolution of thermodynamic quantities, which are, in particular, expressed in terms of $f(x,p;\xi,\lambda)$ via kinetic theory relations. Choosing the relaxation time proportional to the shear viscosity of the medium, the effect of dissipation is considered, in particular, in the evolution of transverse and longitudinal pressures \cite{strickland2010}. It turns out that, in the local rest frame (LRF) of the fluid, the transverse pressure is larger than the longitudinal pressure, and that in the center of the fireball, the system needs many fm/c to become approximately isotropic \cite{strickland2017}. Subsequently, many efforts have been undertaken to study the effect of dissipation on the evolution of the ratio of longitudinal to transverse pressure in a more systematic manner (see e.g. \cite{strickland2013}).
\par
However, apart from finite dissipative corrections, the finite magnetization of the QGP created at the RHIC and LHC may be considered as another source of the aforementioned pressure anisotropies. It is the purpose of the present paper to focus on anisotropies caused, in particular, by the nonvanishing magnetization of a uniformly expanding QGP in and out of equilibrium. One of the main motivations for this study is the wide belief that the QGP produced in the early stages of noncentral HICs is the subject of an extremely large magnetic field \cite{warringa2007, skokov2009,gursoy2018}  (see also \cite{huang2015} and the references therein). Assuming the magnetic field to be aligned in a fixed direction, an anisotropy is naturally induced in any magnetized medium including charged fermions. This kind of anisotropy is previously studied, e.g., in \cite{fayazbakhsh2012,fayazbakhsh2013,fayazbakhsh2014, rojas2018,ferrer2010, martinez2014, endrodi2013, bali2013,delia2013}.
One of the consequences of this anisotropy is the difference between the longitudinal and transverse pressures with respect to the fixed direction of the magnetic field. This difference turns out to be proportional to the magnetization of the medium, and leads, in particular, to an anisotropic equation of state (EoS) for the magnetized QCD matter. The latter is studied, e.g., in \cite{ferrer2010, martinez2014, fayazbakhsh2014, rojas2018} in different contexts. In \cite{endrodi2013}, the EoS of the magnetized QCD is determined in the hadron resonance gas model. It is, in particular, shown that the magnetization of the QCD matter is positive. Several other results from lattice QCD \cite{delia2013,bali2013} agree qualitatively with this result, indicating that the QGP produced in HIC experiments is paramagnetic, and that the magnetic susceptibility of the medium increases with increasing temperature \cite{delia2013}. In \cite{bali2013}, it is argued that because of this paramagnetism, the ``QGP produced in noncentral HICs becomes elongated along the direction of the magnetic field'', and this paramagnetic squeezing may thus have a finite contribution to the elliptic flow $v_2$. The latter is one of the important observables in HIC experiments. Let us again emphasize that in all these computations, the magnitude and the direction of the background magnetic field are mainly assumed to be constant. Moreover, the quark matter produced in HICs is assumed to be static.
\par
In reality, however, the QGP created at the RHIC and LHC is expanding, and the relativistic hydrodynamics is one of the main tools to describe this specific expansion \cite{rajagopal2018}.
On the other hand, it is known that the magnetic field produced in noncentral HICs decays very fast \cite{warringa2007,skokov2009, huang2015, gursoy2018}. As concerns the aforementioned pressure anisotropies, it is thus necessary to consider the effect of the evolution of the QGP and the magnetic field on the paramagnetic squeezing of the QGP.
Recently, a number of attempts have been made to study the evolution of the magnetic field in ideal and nonideal fluids in the framework of magnetohydrodynamics (MHD) \cite{rischke2015, rischke2016, shokri2017, shokri2018, shokri2018-2}. The idea in all these papers is to combine hydrodynamic equations with the Maxwell equations, and to solve them simultaneously using a number of assumptions. In \cite{rischke2015}, it is assumed that (i) the external magnetic field is transverse to the fluid velocity (transverse MHD), (ii) the system is invariant under a longitudinal boost transformation, and (iii) the evolution of the system occurs longitudinally with respect to the beam direction. The last two assumptions are necessary for the $1+1$ dimensional Bjorken flow to be applicable \cite{bjorken}. Using Bjorken's velocity profile, it is found that in an ideal fluid with infinitely large conductivity, the magnitude of the magnetic field evolves as $B(\tau)\propto \tau^{-1}$, with $\tau$ being the proper time, and, moreover, the direction of the magnetic field is frozen, and thus unaffected by the expansion of the fluid \cite{rischke2015}. In \cite{rischke2016,shokri2017}, using the same assumptions as in \cite{rischke2015}, the deviation from the frozen flux theorem in a magnetized fluid with finite conductivity is studied within a $1+1$ dimensional ultrarelativistic nonideal and nondissipative MHD.\footnote{In a nonideal and nondissipative fluid, because of the finite electric conductivity of the medium, the electric field cannot be neglected. Moreover, the system is assumed to be nondissipative.}
\par
In the present paper, we use the analogy between the energy-momentum tensor of an ideal paramagnetic fluid in the presence of a magnetic field and the energy-momentum tensor of a longitudinally expanding fluid in the framework of aHydro \cite{strickland2010,strickland2017}, and study paramagnetic squeezing of a uniformly expanding QGP with and without dissipation. To do this, we make the same three assumptions as is used in the $1+1$ dimensional transverse MHD (see above). Moreover, we assume that the system includes massless particles (conformal symmetry). Using the above-mentioned analogy, we identify the unit vector in the direction of the magnetic field, $b^{\mu}$, with the anisotropy direction that appears in aHydro. Similar to aHydro, we introduce an anisotropic one-particle distribution function, $f_{b}$,\footnote{In the rest of this paper, we refer to $f_{b}$ as magneto-anisotropic one-particle distribution function.} which is expressed in terms of $b_{\mu}$, an anisotropy parameter and an effective temperature.
We then consider the Boltzmann equation in the presence of an electromagnetic source in the RTA. Taking the first two moments of this equation, we arrive, similar to aHydro, to two differential equations whose solutions lead to the proper time evolution of the anisotropy parameter and effective temperature in the nondissipative and dissipative cases. The only free parameter here is the relaxation time, which is chosen to be different in these two cases. Using the kinetic theory relations, it is then possible to determine numerically the ratio of transverse to longitudinal pressures with respect to $b^{\mu}$. This ratio can be regarded as a measure for the anisotropy caused by the magnetization of the fluid in the nondissipative case, and by the magnetization together with the dissipation in the dissipative case. In the latter case, we combine the method used in \cite{tuchin2013, roy2018} to determine the dissipative part of the one-particle distribution function in the first-order derivative expansion around $f_{b}$.
We hereby use a number of results from \cite{rischke2009,tuchin2013,roy2018, tabatabaee2016, kineticbook}.  Let us notice, that in both nondissipative and dissipative cases, apart from the aforementioned ratio of transverse to longitudinal pressures, the proper time evolution of the energy density and, in particular, the magnetic susceptibility $\chi_{m}$ of the paramagnetic QGP can be determined. To have a link to the previously found temperature dependence of $\chi_{m}$ from lattice QCD \cite{delia2013}, we combine our results for the $\tau$ dependence of the effective temperature and the magnetic susceptibility of the QGP, and determine the effective temperature dependence of $\chi_{m}$. We show that, as in \cite{delia2013}, $\chi_{m}$ increases up to a maximum value with increasing temperature. After reaching the maximum, it decreases with increasing temperature. This is in contrast to the lattice QCD results from \cite{delia2013}. This specific feature is, however, expected in our setup, bearing in mind that in \cite{delia2013}, in contrast to our case, the magnetic field is constant, and the quark matter is assumed to be static.
Apart from the above mentioned thermodynamic quantities, we determine, as by-products, the proper time evolution and the effective temperature dependence of the shear and bulk viscosities, and compare the results with the existing results in the literature \cite{arnold2000, buchel2005}.
\par
The organization of the paper is as follows: In Sec. \ref{sec2}, we present a brief review on ideal MHD and the $1+1$ dimensional Bjorkenian solution to the ideal transverse MHD (see also \cite{iqbal2016,hattori2017} for a new systematic formulation of relativistic MHD). In Sec. \ref{sec3}, we first introduce the magneto-anisotropic one-particle distribution function of an ideal magnetized fluid, and then, using the first two moments of the Boltzmann equation in the RTA, we derive the corresponding differential equations for the anisotropy parameter and effective temperature. In Sec. \ref{sec4}, we extend our method to a dissipative magnetized fluid. In Sec. \ref{sec4a}, the dissipative part of $f_{b}$ is determined, and analytical expressions for the shear and bulk viscosities of a dissipative and magnetized fluid are presented. Recently, nonresistive and resistive dissipative MHD are formulated via kinetic theory in the 14-moment approximation in \cite{denicol2018, denicol2019}. However, the effects of magnetization, which are of particular interest in the present paper, are not discussed in these papers.  In Sec. \ref{sec4b}, the corresponding differential equations for the anisotropy parameter and the effective temperature are derived. In Sec. \ref{sec5}, choosing appropriate initial values for the anisotropy parameter at the initial proper time, we numerically solve the two sets of differential equations arising in Secs. \ref{sec3} and \ref{sec4} for nondissipative and dissipative fluids. We then present numerical results for the proper time dependence of anisotropic pressures, the energy density, speed of sound, and magnetic susceptibility of a longitudinally expanding magnetized fluid. We discuss the effect of paramagnetic squeezing on these observables, and compare the results with the existing results in the literature. Section \ref{sec6} is devoted to concluding remarks.
\section{Review material}\label{sec2}
\setcounter{equation}{0}
\subsection{Ideal MHD}\label{sec2A}
Ideal relativistic hydrodynamics is a useful tool to describe the evolution of an ideal and locally equilibrated fluid, which is mainly characterized by its long-wavelength degrees of freedom, the four-velocity $u^{\mu}(x)$ and the temperature $T(x)$. Here, $u^{\mu}=\gamma\left(1,\boldsymbol{v}\right)$ is derived from $u^{\mu}=dx^{\mu}/d\tau$, with $x^{\mu}\equiv (t,\boldsymbol{x})$ being the four-coordinate of each fluid parcel in a flat Minkowski space and $\tau\equiv \sqrt{t^2-\boldsymbol{x}^{2}}$ is the proper time. It satisfies $u_{\mu}u^{\mu}=1$. In the absence of external electromagnetic fields, the ideal fluid is described by the local entropy density $s^{\mu}$ and fluid energy-momentum tensor ${\cal{T}}_{f,0}^{\mu\nu}$,\footnote{In this paper, quantities with subscripts $0$ are defined in nondissipative magnetized fluid, which is described by ideal MHD.}
\begin{eqnarray}\label{N1}
s^{\mu}\equiv su^{\mu},\quad\mbox{and}\quad
{\cal{T}}_{f,0}^{\mu\nu}\equiv \epsilon u^{\mu}u^{\nu}-p\Delta^{\mu\nu},
\end{eqnarray}
where the transverse projector $\Delta^{\mu\nu}\equiv g^{\mu\nu}-u^{\mu}u^{\nu}$, and the spacetime metric $g^{\mu\nu}=\mbox{diag}\left(1,-1,-1,-1\right)$.
In the ideal case, they satisfy
\begin{eqnarray}\label{N2}
\partial_{\mu}s^{\mu}=0,\quad\mbox{and}\quad \partial_{\mu}{\cal{T}}_{f,0}^{\mu\nu}=0.
\end{eqnarray}
In the presence of external magnetic fields, an ideal magnetized fluid is described by a total energy-momentum tensor
\begin{eqnarray}\label{N3}
T_{0}^{\mu\nu}\equiv T_{f,0}^{\mu\nu}+T_{em}^{\mu\nu},
\end{eqnarray}
including the fluid and electromagnetic energy-momentum tensors, $T_{f,0}^{\mu\nu}$ and $T_{em}^{\mu\nu}$. They are given by
\begin{eqnarray}\label{N4}
T_{f,0}^{\mu\nu}&=&\epsilon u^{\mu}u^{\nu}-p\Delta^{\mu\nu}-\frac{1}{2}\left(M^{\mu\lambda}F_{\lambda}^{~\nu}+M^{\nu\lambda}
F_{\lambda}^{~\mu}\right),\nonumber\\
T_{em}^{\mu\nu}&=&F^{\mu\lambda}F_{\lambda}^{~\nu}+\frac{1}{4}g^{\mu\nu}F^{\rho\sigma}F_{\rho\sigma}.
\end{eqnarray}
The total energy-momentum tensor \eqref{N3} satisfies the conservation relation
\begin{eqnarray}\label{N5}
\partial_{\mu}T^{\mu\nu}_{0}=0.
\end{eqnarray}
In \eqref{N4}, the field strength tensor $F^{\mu\nu}$ and a magnetization tensor $M^{\mu\nu}$ are expressed in terms of the magnetic field as\footnote{In this paper, we focus on fluids with infinitely large electric conductivity. We thus neglect the electric field (see \cite{rischke2009} for a similar treatment and \cite{tabatabaee2016, shokri2018,shokri2017} for more details).}
\begin{eqnarray}\label{N6}
F^{\mu\nu}\equiv-Bb^{\mu\nu},\quad\mbox{and}\quad
M^{\mu\nu}\equiv-Mb^{\mu\nu},
\end{eqnarray}
where $b^{\mu\nu}\equiv \epsilon^{\mu\nu\alpha\beta}b_{\alpha}u_{\beta}$,
and $b^{\mu}\equiv \frac{B^{\mu}}{B}$. Here, $B^{\mu}\equiv \frac{1}{2}\epsilon^{\mu\nu\alpha\beta}F_{\nu\alpha}u_{\beta}$. This leads to  $B_{\mu}B^{\mu}=-B^{2}$. In \eqref{N6}, $B$ is the strength of the magnetic field and $M$ is the magnetization of the fluid. In the LRF of the fluid with $u^{\mu}=(1,\boldsymbol{0})$, the magnetic field $B^{\mu}=(0,\boldsymbol{B})$. Using $B^{\mu}=Bb^{\mu}$ with  $B\equiv |\boldsymbol{B}|$, we thus obtain $b^{\mu}b_{\mu}=-1$. Similarly, the antisymmetric polarization tensor $M^{\mu\nu}$, which describes the response of the fluid to an applied electromagnetic field strength tensor $F^{\mu\nu}$, defines the magnetization four-vector $M^{\mu}\equiv \frac{1}{2}\epsilon^{\mu\nu\alpha\beta}M_{\nu\alpha}u_{\beta}$. In the LRF of the fluid, $M^{\mu}\equiv (0,\boldsymbol{M})$ with $M^{\mu}M_{\mu}=-M^2$ and $M\equiv |\boldsymbol{M}|$. The magnetic susceptibility of the fluid $\chi_{m}$ is then defined by $\boldsymbol{M}\equiv \chi_{m}\boldsymbol{B}$.
\par
Plugging $F^{\mu\nu}$ and $M^{\mu\nu}$ from \eqref{N6} into $T^{\mu\nu}_{f,0}$ and $T^{\mu\nu}_{em}$ from \eqref{N4}, we arrive after some work at \cite{rischke2009,tabatabaee2016}
\begin{eqnarray}\label{N7}
T^{\mu\nu}_{f,0}&=&\epsilon u^{\mu}u^{\nu}-p_{\perp}\Xi^{\mu\nu}_{B}+p_{\|}b^{\mu}b^{\nu},\nonumber\\
T^{\mu\nu}_{em}&=&\frac{1}{2}B^{2}\left(u^{\mu}u^{\nu}-\Xi_{B}^{\mu\nu}-b^{\mu}b^{\nu}\right).
\end{eqnarray}
Here, $p_{\perp}\equiv p-BM, p_{\|}\equiv p$ and $\Xi_{B}^{\mu\nu}\equiv \Delta^{\mu\nu}+b^{\mu}b^{\nu}$. Transverse and longitudinal directions, denoted by the subscripts $\perp$ and $\|$, are defined with respect to the direction of the external magnetic field. Contracting first \eqref{N5} together with $T_{f,0}^{\mu\nu}$ and $T_{em}^{\mu\nu}$ from \eqref{N7}, with $u_{\nu}$, we arrive at the energy equation
\begin{eqnarray}\label{N8}
&&\hspace{-1cm}D\epsilon+\theta\left(\epsilon+p_{\perp}\right)-B^{2}(1-\chi_{m})u_{\nu}b^{\mu}\partial_{\mu}b^{\nu}\nonumber\\
&&\hspace{-1cm}+B\left(DB+\theta B\right)=0,
\end{eqnarray}
where $D\equiv u_{\mu}\partial^{\mu}$ and $\theta\equiv \partial_{\mu}u^{\mu}$.
Contracting then \eqref{N5} with $\Delta_{\rho\nu}$, we arrive at the Euler equation
\begin{eqnarray}\label{N9}
&&\left(\epsilon+p_{\perp}+B^{2}\right)Du_{\rho}-\nabla_{\rho}\left(p_{\perp}+\frac{1}{2}B^{2}\right)\nonumber\\
&&+B^{2}(1-\chi_{m})u_{\rho}u_{\nu}b^{\mu}\partial_{\mu}b^{\nu}-\partial_{\mu}[(1-\chi_{m})B^{2}b^{\mu}b_{\rho}]=0,\nonumber\\
\end{eqnarray}
with $\nabla_{\rho}\equiv \Delta_{\rho\nu}\partial^{\nu}$.
\par
Apart from the energy and Euler equations, \eqref{N8} and \eqref{N9}, the magnetized fluid is described by homogeneous and inhomogeneous Maxwell equations,
\begin{eqnarray}\label{N10}
\partial_{\mu}\tilde{F}^{\mu\nu}=0,\quad\mbox{and}\quad \partial_{\mu}F^{\mu\nu}=J^{\nu},
\end{eqnarray}
where the dual field strength tensor and the electromagnetic current are given by
\begin{eqnarray}\label{N11}
\tilde{F}^{\mu\nu}=B^{\mu}u^{\nu}-B^{\nu}u^{\mu},
\end{eqnarray}
and
\begin{eqnarray}\label{N12}
J^{\mu}=\rho_{e}u^{\mu}+\partial_{\rho}M^{\rho\mu}.
\end{eqnarray}
Here, $\rho_{e}$ is the electric charge density, and $\partial_{\rho}M^{\rho\mu}$ is the magnetization current. It is given by contracting the inhomogeneous Maxwell equation $\partial_{\mu}F^{\mu\nu}=J^{\nu}$ from \eqref{N10} with $u_{\nu}$,\footnote{We use the notation $a\cdot b\equiv a_{\mu}b^{\mu}$. }
\begin{eqnarray}\label{N13}
\rho_{e}=2(1-\chi_{m})(B\cdot \omega),
\end{eqnarray}
where $\omega^{\mu}\equiv \frac{1}{2}\epsilon^{\mu\nu\alpha\beta}u_{\nu}\partial_{\alpha}u_{\beta}$ is the vorticity of the fluid. Contracting the homogeneous Maxwell equation $\partial_{\mu}\left(B^{\mu}u^{\nu}-B^{\nu}u^{\mu}\right)=0$ with $b^{\mu}$, we also obtain
\begin{eqnarray}\label{N14}
D\ln B+\theta-u^{\nu}b^{\mu}\partial_{\mu}b_{\nu}=0.
\end{eqnarray}
In what follows, we use these relations to determine the evolution of the magnetic field in transverse $1+1$ dimensional MHD.
\subsection{Bjorken flow and the ideal transverse MHD}\label{sec2B}
\begin{center}
\begin{figure}[t]
\includegraphics[width=8.5cm,height=5.8cm]{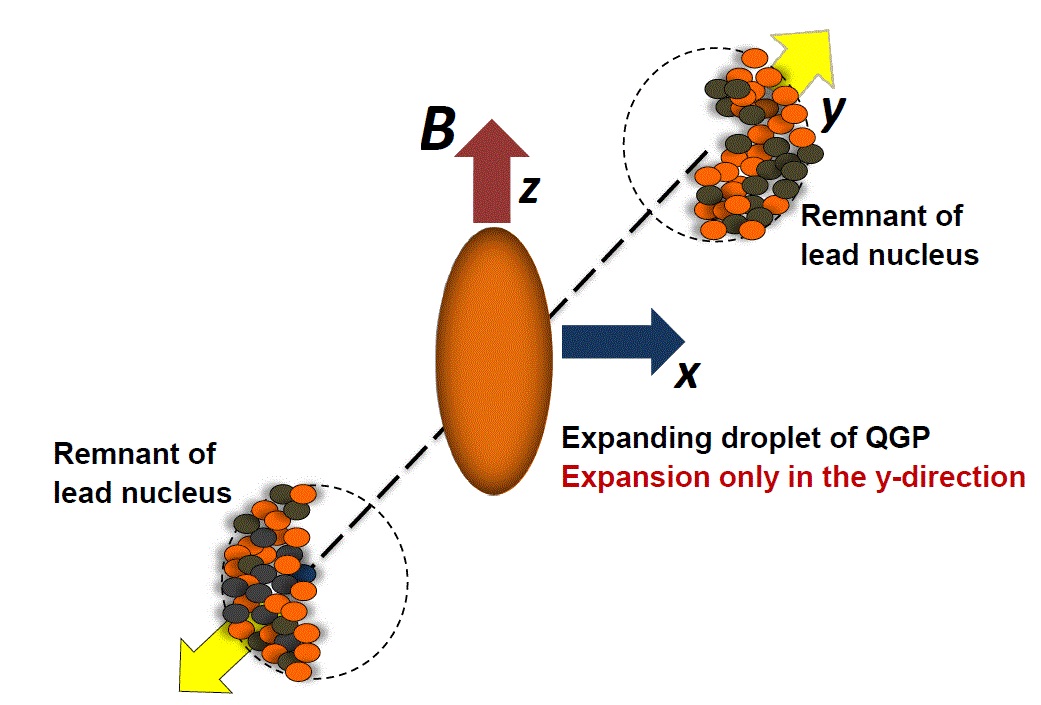}
\caption{(color online). Creation of magnetic fields in HIC experiments. The beam line is in the $y$ direction, and the magnetic field is aligned in the $z$ direction, perpendicular to the $x$-$y$ reaction plane. A uniform expansion of the QGP (the fluid droplet) occurs in the longitudinal $y$ direction. This system is described by a $1+1$ dimensional transverse MHD. }\label{fig1}
\end{figure}
\end{center}

In the present paper, we mainly focus on the effect of magnetic fields on the plasma of quarks and gluons created in the early stages of HICs. It is believed that they are created in a plane perpendicular to the reaction plane. For later convenience, let us assume the beam line to be in the longitudinal $y$ direction, and the magnetic field $\boldsymbol{B}$ in the LRF of the fluid being directed in the transverse $z$ direction perpendicular to the reaction plane in the $x$-$y$ (see the sketch in Fig. \ref{fig1}).\footnote{This choice is in contrast to the common practice where the beam line is assumed to be in the longitudinal $z$ direction and the magnetic field $\boldsymbol{B}$ aligned in the transverse $y$ direction. In the framework of $1+1$ dimensional approximation, it is assumed that the QGP expands uniformly in the longitudinal $z$ direction, and the system remains translational invariant in the transverse $x$-$y$ plane. Its expansion is then described by the Bjorken flow $u^{\mu}=\gamma(1,0,0,v_{z})$ with $v_z=z/t$. In the Milne parametrization $u^{\mu}$ is thus given by $u^{\mu}=(\cosh\eta,0,0,\sinh\eta)$ (see e.g. \cite{shokri2017} for more details).}  In this setup, the transverse MHD \cite{rischke2015,rischke2016,shokri2017,shokri2018} is characterized by
 \begin{itemize}
\item[(1)] translational invariance in the transverse $x$-$z$ plane,
\item[(2)] a uniform expansion of the fluid in the longitudinal beam direction, leading to a nonaccelerated flow,
\item[(3)] boost invariance along the beam line in the $y$ direction, and
\item[(4)] boost invariance of the pressure $p_{0}$.
\end{itemize}
Using the above assumption, and replacing $v_{y}$ in $u^{\mu}=\gamma(1,0,v_{y},0)$ with $v_{y}=y/t$, we arrive after an appropriate parametrization of the four-coordinate $x^{\mu}$ in terms of the Milne variables, the proper time $\tau\equiv \sqrt{t^{2}-y^{2}}$ and the boost variable $\eta\equiv \frac{1}{2}\ln\frac{t+y}{t-y}$, at the $1+1$ dimensional Bjorken flow
\begin{eqnarray}\label{N15}
u^{\mu}=\left(\cosh\eta,0,\sinh\eta,0\right).
\end{eqnarray}
Using this parametrization, we obtain
\begin{eqnarray}\label{N16}
\partial_{\mu}=(\partial_{t},0,\partial_y,0),
\end{eqnarray}
with
\begin{eqnarray}\label{N17}
\frac{\partial}{\partial t}&=&+\cosh\eta\frac{\partial}{\partial\tau}-\frac{1}{\tau}\sinh\eta\frac{\partial}{\partial\eta},\nonumber\\
\frac{\partial}{\partial y}&=&-\sinh\eta\frac{\partial}{\partial\tau}+\frac{1}{\tau}\cosh\eta\frac{\partial}{\partial\eta}.
\end{eqnarray}
These specific features of $u^{\mu}$ and $\partial_{\mu}$ leads, in particular, to vanishing vorticity $\omega^{\mu}$ in transverse MHD. Plugging $\omega^{\mu}=0$ into \eqref{N13}, the electric charge density $\rho_e$ in transverse MHD vanishes.
To determine the evolution of the magnetic field, we combine, at this stage, $u\cdot B=0$ with $\boldsymbol{v}\cdot\boldsymbol{B}=0$, which is valid in $1+1$ dimensional transverse MHD, and arrive at $B^{\mu}=(0,B_{x},0,B_{z})$.\footnote{In the specific setup, demonstrated in Fig. \ref{fig1}, the magnetic field is aligned in the third direction. We thus have $B_{x}=0$.}
Using then these relations together with $B\cdot \partial=0$ and $\partial\cdot B=0$, the homogeneous Maxwell equation $\partial_{\mu}\left(B^{\mu}u^{\nu}-B^{\nu}u^{\mu}\right)=0$ in transverse MHD reads
\begin{eqnarray}\label{N18}
\partial_{\mu}(Bu^{\mu})=0,\quad\mbox{or}\quad DB+\theta B=0.
\end{eqnarray}
Using $D=\frac{\partial}{\partial\tau}$ and $\theta=\frac{1}{\tau}$ arising from $\partial_{\mu}$ from \eqref{N17}, we then arrive at a simple differential equation for $B=|\boldsymbol{B}|$,
\begin{eqnarray}\label{N19}
\frac{\partial B}{\partial \tau}+\frac{B}{\tau}=0.
\end{eqnarray}
This leads immediately to the evolution of $B$ in the ideal transverse MHD
\begin{eqnarray}\label{N20}
B=\bar{B}\left(\frac{\bar{\tau}}{\tau}\right).
\end{eqnarray}
Here, $\bar{B}\equiv B(\bar{\tau})$ and $\bar{\tau}$ is initial time of the hydrodynamic expansion. Bearing in mind that in transverse MHD $B\cdot \partial=0$, the differential equation \eqref{N19} is consistent with \eqref{N14}. We emphasize at this stage, that the above $\tau$ dependence of $B(\tau)$ is also valid in dissipative MHD. This is mainly because \eqref{N20} arises from the homogeneous Maxwell equation, which is unaffected by dissipative terms in the energy-momentum tensor of the fluid.
\par
Using $DB+\theta B=0$ from \eqref{N18}, $B\cdot\partial=0$ and $\partial\cdot B=0$ in transverse MHD, it turns out that the electromagnetic part of the energy-momentum tensor $T^{\mu\nu}_{em}$ from \eqref{N7} is conserved. Plugging $u_{\nu}\partial_{\mu}T^{\mu\nu}_{em}=0$ in \eqref{N5}, we are therefore left with
\begin{eqnarray}\label{N21}
u_{\nu}\partial_{\mu}T_{f,0}^{\mu\nu}=0,
\end{eqnarray}
with $T_{f,0}^{\mu\nu}$ given in \eqref{N7}. Using $DB+\theta B=0$ from \eqref{N18} and $b\cdot \partial=0$ as well as $\partial\cdot b=0$ in transverse MHD, the energy equation \eqref{N8} is modified as
\begin{eqnarray}\label{N22}
D\epsilon+\theta\left(\epsilon+p-BM\right)=0.
\end{eqnarray}
As concerns the Euler equation, we use the fact that in the Bjorken setup the fluid is nonaccelerated, and obtain $Du_{\rho}=0$. Plugging this relation into \eqref{N9}, the Euler equation in ideal transverse MHD reads
\begin{eqnarray}\label{N23}
\frac{\partial}{\partial\eta}\left(p-BM\right)=0.
\end{eqnarray}
Here, $\nabla_{\mu}=-\frac{1}{\tau}\left(\sinh\eta,0,-\cosh\eta,0\right)\frac{\partial}{\partial\eta}$ is used. Using then the assumed boost invariance ($\eta$ independence) of $p$, the boost invariance of $B$ arising from \eqref{N20}, and $M=\chi_{m}B$, we arrive at the boost invariance of the magnetic susceptibility $\chi_{m}$.
\par
Using the same method, it is also possible to determine the evolution of the entropy. To do this, let us consider the conservation equation of the entropy current $\partial_{\mu}(s u^{\mu})=0$ from \eqref{N2} leading to $Ds+\theta s=0$. In the Milne coordinates, we thus arrive at
\begin{eqnarray}\label{N24}
\frac{\partial s}{\partial \tau}+\frac{s}{\tau}=0,
\end{eqnarray}
whose solution reads
\begin{eqnarray}\label{N25}
s=\bar{s}\left(\frac{\bar{\tau}}{\tau}\right).
\end{eqnarray}
Here, $\bar{s}\equiv s(\bar{\tau})$. As concerns the evolution of the energy density $\epsilon$, let us consider the energy equation \eqref{N22}. In the case of vanishing magnetic susceptibility, one usually uses the ideal gas EoS, $\epsilon=\kappa p$, with $\kappa=\mbox{const}$, to write \eqref{N22} with vanishing magnetic susceptibility as
\begin{eqnarray}\label{N26}
\frac{\partial p}{\partial\tau}+\left(1+\frac{1}{\kappa}\right)\frac{p}{\tau}=0,
\end{eqnarray}
whose solution is given by
\begin{eqnarray}\label{N27}
p=\bar{p} \left(\frac{\bar{\tau}}{\tau}\right)^{1+1/\kappa},\quad\mbox{for}\quad \chi_{m}=0.
\end{eqnarray}
Here, $\bar{p}=p(\bar{\tau})$. Using $\epsilon=\kappa p$, we have
\begin{eqnarray}\label{N28}
\epsilon=\bar{\epsilon} \left(\frac{\bar{\tau}}{\tau}\right)^{1+1/\kappa},\quad\mbox{for}\quad \chi_{m}=0,
\end{eqnarray}
with $\bar{\epsilon}=\kappa \bar{p}$. Combining the EoS $\epsilon=\kappa p$, \eqref{N24}, \eqref{N27}, and the thermodynamic relation $\epsilon+p=Ts$, the evolution of the temperature $T$ is given by
\begin{eqnarray}\label{N29}
T=\bar{T} \left(\frac{\bar{\tau}}{\tau}\right)^{1/\kappa},\quad\mbox{for}\quad \chi_{m}=0,
\end{eqnarray}
with $\bar{T}\equiv \left(1+\kappa\right)\frac{\bar{p}}{\bar{s}}$.
\par
In the next section, we determine the evolution of $\epsilon,p$ and $T$ in the case of nonvanishing and $\tau$ dependent $\chi_{m}$ in ideal MHD.\footnote{The evolution of thermodynamic functions for constant $\chi_m$ is studied in \cite{rischke2016}.} As a by-product, the evolution of the magnetic susceptibility is also found. To do this, we use the method used in \cite{strickland2010} in the framework of aHydro (see also \cite{strickland2017} and references therein).  
\section{Paramagnetic anisotropy in a nondissipative magnetized QGP}\label{sec3}
\setcounter{equation}{0}
\subsection{Boltzmann equation and ideal transverse MHD}\label{sec2C}
Let us start by considering the fluid part of the energy-momentum tensor $T_{f,0}^{\mu\nu}$ from \eqref{N7}, which can also be given as
\begin{eqnarray}\label{E1}
T_{f,0}^{\mu\nu}=(\epsilon+p_{\perp})u^{\mu}u^{\nu}-p_{\perp}g^{\mu\nu}+\left(p_{\|}-p_{\perp}\right)b^{\mu}b^{\nu}.
\nonumber\\
\end{eqnarray}
This relation is in many aspects comparable with the energy-momentum tensor \begin{eqnarray}\label{E2}
T_{f}^{\mu\nu}=(\epsilon+p_{T})u^{\mu}u^{\nu}-p_{T}g^{\mu\nu}+\left(p_{L}-p_{T}\right)Y^{\mu}Y^{\nu},\nonumber\\
\end{eqnarray}
appearing, e.g. in \cite{strickland2017} in the context of aHydro. Here, $Y^{\mu}$ is the beam direction.\footnote{Let us notice that in \cite{strickland2010, strickland2017}, the beam line is chosen to be in the $z$ direction. Hence, \eqref{E2} is formulated in terms of $Z^{\mu}$, the unit vector in this direction, instead of $Y^{\mu}$. In the present paper, however, we take the beam line in the $y$ direction, perpendicular to the magnetic field. The latter is chosen to be in the $z$ direction.} Let us emphasize at this stage that whereas subscripts $T$ and $L$ in \eqref{E2} correspond to transverse and longitudinal directions with respect to the beam direction, $\perp$ and $\|$ in \eqref{E1} correspond to transverse and longitudinal directions with respect to the direction of the magnetic field in the $z$ direction (see Fig. \ref{fig1}). Hence, $\{\|, \perp\}$ correspond to $\{T,L\}$ in \cite{strickland2010,strickland2017}, respectively.
\par
Using the analogy between \eqref{E1} and \eqref{E2}, we identify $Y^{\mu}$, appearing in \eqref{E2}, with $b^{\mu}=B^{\mu}/B$. Here, $b^{\mu}$ satisfies $b_{\mu}b^{\mu}=-1$, and, in the LRF of the fluid, we have $b^{\mu}=(0,0,0,1)$.
Physically, the main difference between $T^{\mu\nu}$ from \eqref{E2} and $T^{\mu\nu}_{f,0}$ from \eqref{E1} lies in the difference between the longitudinal
and the transverse pressures. Whereas $p_{L}-p_{T}$ in \eqref{E2} is brought in connection with the dissipative nature of the fluid, in particular, its shear viscosity \cite{strickland2017}, $p_{\|}-p_{\perp}$ in \eqref{E1} is related to the magnetization of the fluid through $p_{\|}-p_{\perp}=BM=\chi_{m}B^{2}$. It is therefore possible to follow the method presented in \cite{strickland2010,strickland2017}, and to determine the evolution of the energy density, the pressure and the effective temperature in the ideal nondissipative case, in order to focus on the effect of nonvanishing magnetization of the fluid on the evolution of anisotropies arising in the early stages of HICs. Similar to aHydro, the anisotropy induced by nonvanishing magnetization of the QGP is intrinsically implemented in the momentum distribution of the system, and can be considered as a new source for the pressure anisotropy appearing in the QGP created in HICs. In this section, we consider the anisotropy caused by the nonvanishing magnetization of a nondissipative fluid. We use the same method of moments of Boltzmann equation in the RTA as in \cite{strickland2010,strickland2017}, and derive, in this way, two differential equations whose solutions lead to the proper-time dependence of the anisotropy parameter and effective temperature.
\par
To do this, we introduce, as in aHydro, the one-particle distribution function
\begin{eqnarray}\label{E3}
f_{0}=\exp\left(-\sqrt{k^{\mu}\Xi_{\mu\nu}^{(0)}k^{\nu}}/\lambda_{0}\right),
\end{eqnarray}
with $\lambda_{0}$ being the effective temperature, and
\begin{eqnarray}\label{E4}
\Xi_{\mu\nu}^{(0)}\equiv u_{\mu}u_{\nu}+\xi_{0} b_{\mu}b_{\nu}.
\end{eqnarray}
Here, $\xi_{0}$ is the anisotropy parameter that is induced by the magnetization of the fluid. In transverse MHD, $\lambda_{0}$ and $\xi_{0}$ depend, in general, on $\tau$ and $\eta$. However, by the assumption of boost invariance, they depend only on $\tau$. In the presence of an external magnetic field described by the field strength tensor $F^{\mu\nu}$, $f_{0}$ satisfies the Boltzmann equation,
\begin{eqnarray}\label{E5}
k^{\mu}\partial_{\mu}f_{0}+q_{f}eF^{\mu\nu}k_{\nu}\frac{\partial f_{0}}{\partial k^{\mu}}=C[f_{0}],
\end{eqnarray}
with $q_{f}$ being the number of flavors. Using the RTA, we set, as in \cite{strickland2017},
\begin{eqnarray}\label{E6}
C[f_{0}]=-(k\cdot u)\left(\frac{f_{0}-f_{eq}}{\tau_{r,0}}\right),
\end{eqnarray}
with
\begin{eqnarray}\label{E7}
f_{eq}=\exp\left(-(k\cdot u)/T\right),
\end{eqnarray}
and $\tau_{r,0}$ the relaxation time. In \cite{strickland2017}, the relaxation time is brought in connection with the shear viscosity over the entropy density ratio. In ideal MHD, however, the fluid is dissipationless. The relaxation time $\tau_{r,0}$ is thus only related to the magnetization of the fluid, that, because of the induced anisotropy, affects $f_{eq}$.
\par
Contracting, at this stage, $T_{f,0}^{\mu\nu}$ from \eqref{E1} with  $u_{\mu}u_{\nu}$,  $b_{\mu}b_{\nu}$, and using $b\cdot b=-1$ and $u\cdot b=0$, we obtain
\begin{eqnarray}\label{E8}
\epsilon_{0}= u_{\mu}u_{\nu}T_{f,0}^{\mu\nu},\quad\mbox{and}\quad  p_{0}=b_{\mu}b_{\nu}T^{\mu\nu}_{f,0}.
\end{eqnarray}
Using, moreover, $T_{f,0}^{\mu\nu}g_{\mu\nu}=\epsilon_{0}-3p_{0}+2BM_{0}$, we arrive at\footnote{We replace $\epsilon$, $p$ and $M$ from \eqref{E1} with $\epsilon_{0}$, $p_{0}$ and $M_0$ to denote that these quantities are computed with anisotropic $f_{0}$ through standard definitions \eqref{E11}.}
\begin{eqnarray}\label{E9}
\Delta_{\mu\nu}T_{f,0}^{\mu\nu}=-3p_{0}+2BM_0.
\end{eqnarray}
Using the standard definition of the energy-momentum tensor in terms of the one-particle distribution function
\begin{eqnarray}\label{E10}
T_{f,0}^{\mu\nu}=\int d\tilde{k}~k^{\mu}k^{\nu} f_{0}(x,k),
\end{eqnarray}
with $d\tilde{k}\equiv \frac{d^{3}k}{(2\pi)^{3}|\boldsymbol{k}|}$, the energy density, the pressure and the magnetization of the fluid including massless particles, satisfying $k^{2}=0$,  read
\begin{eqnarray}\label{E11}
\hspace{-0.8cm}\epsilon_{0}&=&\int d\tilde{k}\left(k\cdot u\right)^{2}f_{0},\nonumber\\
\hspace{-0.8cm}p_{0}&=&\int d\tilde{k}\left(k\cdot b\right)^{2}f_{0},\nonumber\\
\hspace{-0.8cm}M_{0}&=&-\frac{1}{2B}\int d\tilde{k}~\big[\left(k\cdot u\right)^{2}-3\left(k\cdot b\right)^{2}\big]f_{0}.
\end{eqnarray}
Plugging $f_{0}$ from \eqref{E3} into these expressions, and performing the integrations, we arrive after some computation at
\begin{eqnarray}\label{E12}
\hspace{-0.5cm}\epsilon_{0}&=&\frac{3\lambda_{0}^{4}}{\pi^{2}}{\cal{R}}(\xi_{0}),\nonumber\\
\hspace{-0.5cm}p_{0}&=&\frac{3\lambda_{0}^{4}}{\pi^{2}\xi_{0}}\left({\cal{R}}(\xi_{0})-\frac{1}{1+\xi_{0}}\right),\nonumber\\
\hspace{-0.5cm}M_{0}&=&\frac{3\lambda_{0}^{4}}{2\pi^{2}\xi_{0}B}\bigg[\left(3-\xi_{0}\right){\cal{R}}(\xi_{0})-\frac{3}{1+\xi_{0}}\bigg],
\end{eqnarray}
where
\begin{eqnarray}\label{E13}
{\cal{R}}(\xi)&\equiv&\frac{1}{2(1+\xi)}\int_{0}^{\pi}d\theta\sin\theta\sqrt{1+\xi\sin^{2}\theta}\nonumber\\
&=&\frac{1}{2}\left(\frac{1}{1+\xi}+\frac{\arctan\sqrt{\xi}}{\sqrt{\xi}}\right).
\end{eqnarray}
In \eqref{E12}, $\lambda_{0}$ and $\xi_{0}$ satisfy differential equations that can be determined by making use of the Boltzmann equation \eqref{E5}.
\subsection{Differential equations leading to $\boldsymbol{\xi_{0}}$ and $\boldsymbol{\lambda_{0}}$}\label{sec3b}
We start by considering the zeroth moment of the Boltzmann equation \eqref{E5},
\begin{eqnarray}\label{E14}
\int d\tilde{k}\left(k^{\mu}\partial_{\mu}f_{0}+q_{f}eF^{\mu\nu}k_{\nu}\frac{\partial f_{0}}{\partial k^{\mu}}\right)=\int d\tilde{k}~C[f_{0}], \nonumber\\
\end{eqnarray}
with $C[f_{0}]$ given in \eqref{E6}. Using
\begin{eqnarray}\label{E15}
\hspace{-0.5cm}n^{\mu}_{0}\equiv n_{0}u^{\mu} =\int d\tilde{k}~k^{\mu}f_{0}=\frac{\lambda_{0}^{3}}{\pi^{2}\sqrt{1+\xi_{0}}}u^{\mu},
\end{eqnarray}
and
\begin{eqnarray}\label{E16}
\hspace{-0.5cm}\int d\tilde{k}~F^{\mu\nu}k_{\nu}\frac{\partial f_{0}}{\partial k^{\mu}}=\int d\tilde{k}~F^{\mu\nu}g_{\mu\nu}f_{0}=0,
\end{eqnarray}
as well as
\begin{eqnarray}\label{E17}
\int d\tilde{k}~C[f_{0}]=-\frac{1}{\tau_{r,0}}\left(n_{0}-n_{eq}\right),
\end{eqnarray}
with
\begin{eqnarray}\label{E18}
n_{eq}^{\mu}=n_{eq}u^{\mu}=\int d\tilde{k}~k^{\mu}f_{eq}=\frac{T^{3}}{\pi^{2}} u^{\mu},
\end{eqnarray}
we arrive at
\begin{eqnarray}\label{E19}
Dn_{0}+\theta n_{0}=-\frac{1}{\tau_{r,0}}\left(n_{0}-n_{eq}\right).
\end{eqnarray}
Using $D=\partial_{\tau}\equiv \frac{\partial}{\partial\tau}$, $\theta=\frac{1}{\tau}$, and plugging $n_{0}$ and $n_{eq}$ into \eqref{E19}, we arrive first at
\begin{eqnarray}\label{E20}
\frac{\partial_{\tau}\xi_{0}}{1+\xi_{0}}-\frac{6\partial_{\tau}\lambda_{0}}{\lambda_{0}}-\frac{2}{\tau}=\frac{2}{\tau_{r,0}}\left(1-\left(\frac{T}{\lambda_0}\right)^{3}\sqrt{1+\xi_{0}}\right).\nonumber\\
\end{eqnarray}
The relation between $T$ and the effective temperature $\lambda_{0}$ arises from the first moment of the Boltzmann equation \eqref{E5},
\begin{eqnarray}\label{E21}
\int d\tilde{k}~k^{\rho}\left(k^{\mu}\partial_{\mu}f_{0}+q_{f}eF^{\mu\nu}k_{\nu}
\frac{\partial f_{0}}{\partial k^{\mu}}\right)=\int d\tilde{k}~k^{\rho}C[f_{0}], \nonumber\\
\end{eqnarray}
with $C[f_{0}]$ given in \eqref{E6}. Using \eqref{E10}, the first term on the left-hand side (lhs) of \eqref{E21} is given by $\partial_{\mu}T^{\mu\rho}_{f,0}$. Using $n_{0}u^{\mu}=\int d\tilde{k}k^{\mu}f_{0}$ and $F^{\mu\nu}u_{\nu}=E^{\mu}$, the second term on the lhs of \eqref{E21} reads
$$
q_{f}eF^{\mu\nu}\int d\tilde{k}~k^{\rho}k_{\nu}\frac{\partial f_{0}}{\partial k^{\mu}}=-q_{f}eF^{\rho}_{~\nu}n_{0}^{\nu}=-q_{f}en_{0}E^{\rho}=0.$$
We thus arrive at
\begin{eqnarray}\label{E22}
\partial_{\mu}T^{\rho\mu}_{f,0}=\int d\tilde{k}~k^{\rho}C[f_{0}]=-\frac{1}{\tau_{r,0}}u_{\mu}\left(T^{\rho\mu}_{f,0}-T^{\rho\mu}_{f,eq}\right),\nonumber\\
\end{eqnarray}
where $T^{\mu\nu}_{f,eq}$ is defined by \eqref{E10} with $f_{0}$ replaced with $f_{eq}$ from \eqref{E7}. Using, at this stage, $\partial_{\mu}T^{\mu\nu}_{0}=\partial_{\mu}T^{\mu\nu}_{f,0}=0$ from \eqref{N21}, we arrive at
\begin{eqnarray}\label{E23}
u_{\mu}T^{\rho\mu}_{f,0}=u_{\mu}T_{f,eq}^{\rho\mu},
\end{eqnarray}
which leads to
\begin{eqnarray}\label{E24}
\epsilon_{0}=\epsilon_{eq},
\end{eqnarray}
upon multiplying \eqref{E23} by $u_{\rho}$. Here, $\epsilon_{0}$ is defined in \eqref{E11} and
\begin{eqnarray}\label{E25}
\epsilon_{eq}=\int d\tilde{k}~(k\cdot u)^{2}f_{eq}=\frac{3T^{4}}{\pi^{2}}.
\end{eqnarray}
Plugging $\epsilon_{0}$ and $\epsilon_{eq}$ from \eqref{E12} and \eqref{E25} into \eqref{E24}, we arrive at
\begin{eqnarray}\label{E26}
T=\lambda_{0} {\cal{R}}^{1/4}(\xi_{0}),
\end{eqnarray}
with ${\cal{R}}(\xi)$ given in \eqref{E13}. Plugging this expression into \eqref{E20}, we obtain
\begin{eqnarray}\label{E27}
\frac{\partial_{\tau}\xi_{0}}{1+\xi_{0}}-\frac{6\partial_{\tau}\lambda_{0}}{\lambda_{0}}-\frac{2}{\tau}=\frac{2}{\tau_{r,0}}\left(1-{\cal{R}}^{3/4}(\xi_{0})\sqrt{1+\xi_{0}}\right).\nonumber\\
\end{eqnarray}
Let us now consider the energy equation \eqref{N22}. Plugging $\epsilon_{0}, p_{0}$ and $BM_{0}$ from \eqref{E12} into \eqref{N22}, we arrive at the second differential equation leading to the evolution of $\lambda_0$ and $\xi_0$,
\begin{eqnarray}\label{E28}
\lefteqn{\hspace{-0.5cm}\frac{\partial {\cal{R}}(\xi_{0})}{\partial\xi_0}\frac{\partial_{\tau}\xi_0}{{\cal{R}}(\xi_{0})}+\frac{4\partial_{\tau}\lambda_0}{\lambda_0}}\nonumber\\
&&=-\frac{1}{2\tau\xi_0}\left(3\xi_0-1+\frac{1}{(1+\xi_0){\cal{R}}(\xi_{0})}\right).
\end{eqnarray}
In Sec. \ref{sec5}, we solve the above differential equations \eqref{E27} and \eqref{E28} numerically for a given relaxation time $\tau_{r,0}$, and determine the evolution of $\lambda_{0}$ and $\xi_{0}$ in terms of the proper time $\tau$. The resulting solutions of $\lambda_{0}$ and $\xi_{0}$ are then used to determine the evolution of thermodynamic quantities $\epsilon_{0}, p_{0}$ and $M_{0}$. Using then the relation $M_0=\chi_{m,0}B$ and the evolution \eqref{N20} of $B$ in terms of $\tau$, the evolution of the magnetic susceptibility $\chi_{m,0}$ in a nondissipative QGP is determined.
\section{Paramagnetic anisotropy in a dissipative magnetized QGP}\label{sec4}
\setcounter{equation}{0}
In this section, we extend the method described in the previous section to a dissipative QGP in the presence of a strong but fast decaying magnetic field. To do this, we first use the method described in \cite {roy2018}, and determine the dissipative part of the one-particle distribution function in a magnetized fluid in the first-order derivative expansion. Then, using the same method as in the previous section, we determine the differential equations corresponding to the anisotropy parameter $\xi$ and the effective temperature $\lambda$ in the RTA. Choosing appropriate initial values, these equations are then solved numerically for various relaxation times. The $\tau$ dependence of $\xi$ and $\lambda$, arising from this procedure, leads eventually to the $\tau$ dependence of thermodynamic quantities $\epsilon,p$ and $M$ in a dissipative and magnetized QGP.
\subsection{The dissipative part of the magneto-anisotropic one-particle distribution function, shear and bulk viscosities}\label{sec4a}
To determine the dissipative part of the one-particle distribution function, we start by plugging $f=f_b+\delta f_d$ including the nondissipative and dissipative one-particle distribution functions, $f_b$ and $\delta f_{d}$, into the Boltzmann equation
\begin{eqnarray}\label{D1}
k^{\mu}\partial_{\mu}f+q_{f}eF^{\mu\nu}k_{\nu}\frac{\partial f}{\partial k^{\mu}}=C[f].
\end{eqnarray}
Here, as in the previous section, $f_{b}$ is given by
\begin{eqnarray}\label{D2}
f_{b}\equiv \exp\left(-\sqrt{k^{\mu}\Xi_{\mu\nu}k^{\nu}}/\lambda\right),
\end{eqnarray}
with
\begin{eqnarray}\label{D3}
\Xi_{\mu\nu}=u_{\mu}u_{\nu}+\xi b_{\mu}b_{\nu}.
\end{eqnarray}
In \eqref{D2} and \eqref{D3}, $\lambda$ and $\xi$ are the effective temperature and anisotropy parameter, receiving, in contrast to $\lambda_{0}$ and $\xi_{0}$ from the previous section, dissipative contributions. Keeping the dissipative part up to the first-order derivative expansion, and using  $F^{\mu\nu}=-Bb^{\mu\nu}$ and the RTA ansatz
\begin{eqnarray}\label{D4}
C[f]=-(k\cdot u)\frac{(f-f_{eq})}{\tau_{r}},
\end{eqnarray}
we arrive at
\begin{eqnarray}\label{D5}
&&\hspace{-0.5cm}k^{\mu}\partial_{\mu}f_{b}-q_{f}eB b^{\mu\nu}k_{\nu}\frac{\partial f_{b}}{\partial k^{\mu}}-q_{f}eB b^{\mu\nu}k_{\nu}\frac{\partial \delta f_{d}}{\partial k^{\mu}}
\nonumber\\
&&=-(k\cdot u)\frac{(f-f_{eq})}{\tau_{r}},
\end{eqnarray}
where the relaxation time $\tau_{r}$ and the equilibrium one-particle distribution function $f_{eq}$ are defined in \eqref{D4} and \eqref{E7}. Let us notice, at this stage, that in the dissipative case $\tau_{r}$ is different from $\tau_{r,0}$, which appeared in \eqref{E6}. In what follows, we consider $\tau_{r}$ as a free parameter, and study qualitatively the effect of different choices of $\tau_{r}>\tau_{r,0}$ and $\tau_{r}\leq\tau_{r,0}$ on the evolution of thermodynamical quantities, which are separately affected by the anisotropy induced by the magnetization and the first-order dissipation.\footnote{The exact determination of $\tau_{r,0}$ and $\tau_{r}$ in the presence of external magnetic fields and in terms of magnetization and dissipative coefficients is beyond the scope of this paper.}
\par
Plugging $f_{b}$ from \eqref{D2} into \eqref{D5}, and using $u_{\mu}b^{\mu\nu}=0$,  $b_{\mu}b^{\mu\nu}=0$, as well as $\partial_{\mu}=\nabla_{\mu}+u_{\mu}D$, we arrive after a straightforward computation at\footnote{In a nonaccelerating system  $Du_{\mu}=0$ is assumed. For a boost invariant system $\nabla_{\mu}\xi\sim\frac{\partial\xi}{\partial\eta}=0$ and $\nabla_{\mu}\lambda\sim\frac{\partial\lambda}{\partial\eta}=0$ are also assumed. }
\begin{eqnarray}\label{D6}
&&-\frac{f_{b}}{\lambda H_{b}}\left(k^{\mu}k^{\nu}w_{\mu\nu}+(k\cdot b)^{2}D\xi-(k\cdot u)^{2}H_{b}^{2}\frac{D\lambda}{\lambda}\right)\nonumber\\
&&-q_{f}eB b^{\mu\nu}k_{\nu}\frac{\partial \delta f_{d}}{\partial k^{\mu}}=-(k\cdot u)\frac{(f-f_{eq})}{\tau_{r}}.
\end{eqnarray}
Here, $w_{\mu\nu}\equiv \frac{1}{2}\left(\nabla_{\mu}u_{\nu}+\nabla_{\nu}u_{\mu}\right)$, and
\begin{eqnarray}\label{D7}
 H_{b}(k)\equiv\sqrt{1+\xi\frac{(k\cdot b)^{2}}{(k\cdot u)^{2}}}.
\end{eqnarray}
Let us consider, at this stage, $f-f_{eq}$ with $f=f_{b}+\delta f_{d}$ on the right-hand side (rhs) of \eqref{D6}.
Bearing in mind that $\delta f_{d}$ may include terms consisting of derivatives with respect to $\lambda,\xi$ and $u^{\mu}$, we introduce
\begin{eqnarray}\label{D8}
\delta f_{d}=\delta f_{d}^{(a)}+\delta f_{d}^{(b)}-(f_{b}-f_{eq}),
\end{eqnarray}
with $\delta f_{d}^{(a)}$ and $\delta f_{d}^{(b)}$ including derivatives with respect to  $\lambda,\xi$ and $u^{\mu}$. Plugging \eqref{D8} into \eqref{D6}, and comparing the terms including $\lambda,\xi$ and $u^{\mu}$, we arrive at an algebraic equation, satisfied by $\delta f_{d}^{(a)}$
\begin{eqnarray}\label{D9}
\delta f_{d}^{(a)}=\nu_{H} \left((k\cdot b)^{2}D\xi-(k\cdot u)^{2}H_{b}^{2}\frac{D\lambda}{\lambda}\right),
\end{eqnarray}
with $\nu_{H}\equiv \frac{\tau_{r}f_{b}}{\lambda(k\cdot u)H_{b}}$,
and a differential equation satisfied by $\delta f_{d}^{(b)}$
\begin{eqnarray}\label{D10}
\frac{f_{b}}{\lambda H_{b}}\left(k^{\mu}k^{\nu}w_{\mu\nu}\right)=(k\cdot u)\frac{\delta f_{d}^{(b)}}{\tau_{r}}-q_{f}eBb^{\mu\nu}k_{\nu}\frac{\partial\delta f_{d}^{(b)}}{\partial k^{\mu}}.\nonumber\\
\end{eqnarray}
Let us first consider \eqref{D9}. Defining three second rank tensors
\begin{eqnarray}\label{D11}
\hspace{-0.2cm}U_{\mu\nu}^{(0)}\equiv \Delta_{\mu\nu}, \quad U_{\mu\nu}^{(1)}\equiv b_{\mu}b_{\nu}, \quad U_{\mu\nu}^{(2)}\equiv b_{\mu\nu},
\end{eqnarray}
expanding $\delta f_{d}^{(a)}$ in terms of $U_{\mu\nu}^{(n)},~n=0,1,2$ as
\begin{eqnarray}\label{D12}
\delta f_{d}^{(a)}=\sum_{n=0}^{2}\ell_{n} k^{\mu}k^{\nu}U_{\mu\nu}^{(n)},
\end{eqnarray}
and plugging \eqref{D12} into \eqref{D9}, we obtain
\begin{eqnarray}\label{D13}
\ell_{0}&=&\nu_{H} \frac{D \lambda}{\lambda},\nonumber\\
\ell_{1}&=&\nu_{H}\xi\left(\frac{D \xi}{\xi} -\frac{D \lambda}{\lambda} \right),\nonumber\\
\ell_{2}&=&0.
\end{eqnarray}
This determines the final form of $\delta f_{d}^{(a)}$. As concerns $\delta f_{d}^{(b)}$, that satisfies \eqref{D10}, we follow the method presented recently in \cite{roy2018}. We start with the ansatz
\begin{eqnarray}\label{D14}
\delta f_{d}^{(b)}= \nu_{H}k^{\mu}k^{\nu}w^{\rho\sigma}C_{\mu\nu\rho\sigma}.
\end{eqnarray}
Plugging \eqref{D14} into \eqref{D10}, and using $u_{\mu}b^{\mu\nu}=0$, we arrive first at
\begin{eqnarray}\label{D15}
k^{\mu}k^{\nu}w_{\mu\nu}=\left(k^{\rho}k^{\sigma}C_{\rho\sigma\alpha\beta}
-2\chi_{H}b^{\mu\nu}k_{\nu}k^{\rho}C_{\mu\rho\alpha\beta}\right)w^{\alpha\beta},\nonumber\\
\end{eqnarray}
with $\chi_{H}\equiv \frac{q_{f}eB\tau_{r}}{(k\cdot u)}$.
Defining then the basis tensor of rank four (see also \cite{rischke2009,tuchin2013,tabatabaee2016} for similar bases)
\begin{eqnarray}\label{D16}
\xi_{\mu\nu\rho\sigma}^{(1)} &=& \Delta_{\mu \nu} \Delta_{\rho \sigma},   \nonumber \\
\xi_{\mu\nu\rho\sigma}^{(2)} &=&  \Delta_{\mu \rho} \Delta_{\nu \sigma} + \Delta_{\mu \sigma} \Delta_{\nu \rho},   \nonumber \\
\xi_{\mu\nu\rho\sigma}^{(3)} &=& \Delta_{\mu \nu} b_{\rho} b_{\sigma} + \Delta_{\rho \sigma} b_{\mu} b_{\nu}  ,   \nonumber \\
\xi_{\mu\nu\rho\sigma}^{(4)} &=& b_{\mu} b_{\nu} b_{\rho} b_{\sigma} ,  \nonumber \\
\xi_{\mu\nu\rho\sigma}^{(5)} &=& \Delta_{\mu \rho} b_{\nu} b_{\sigma} + \Delta_{\nu \rho} b_{\mu} b_{\sigma} + \Delta_{\mu \sigma} b_{\rho} b_{\nu} + \Delta_{\nu \sigma} b_{\rho} b_{\mu},   \nonumber \\
\xi_{\mu\nu\rho\sigma}^{(6)} &=& \Delta_{\mu \rho} b_{\nu \sigma}  + \Delta_{\nu \rho} b_{\mu \sigma}  + \Delta_{\mu \sigma}  b_{\nu \rho} + \Delta_{\nu \sigma} b_{\mu \rho},   \nonumber \\
\xi_{\mu\nu\rho\sigma}^{(7)} &=& b_{\mu \rho} b_{\nu} b_{\sigma} + b_{\nu \rho} b_{\mu} b_{\sigma} + b_{\mu \sigma} b_{\rho} b_{\nu} + b_{\nu \sigma} b_{\rho} b_{\mu}, \nonumber\\
\end{eqnarray}
where $\Delta_{\mu\nu}$ and $b_{\mu\nu}$ are defined in previous sections, it is straightforward to check that \eqref{D15} is given by
\begin{eqnarray}\label{D17}
\xi_{\mu\nu\rho\sigma}^{(2)}=\left(\xi_{\mu\nu\alpha\beta}^{(2)}+\chi_{H}
\xi_{\mu\nu\alpha\beta}^{(6)}\right)C^{\alpha\beta}_{~~~\rho\sigma}.
\end{eqnarray}
Expanding, at this stage, $C_{\mu\nu\rho\sigma}$ in terms of $\xi_{\mu\nu\rho\sigma}^{(n)}$  from \eqref{D16} as
\begin{eqnarray}\label{D18}
C_{\mu\nu\rho\sigma}=\sum\limits_{n=1}^{7}c_{n}\xi_{\mu\nu\rho\sigma}^{(n)},
\end{eqnarray}
and plugging \eqref{D18} into \eqref{D17}, we arrive after some work at\footnote{In \eqref{D19}, the indices $\mu\nu\rho\sigma$ of $\xi^{(n)}$ are skipped. }
\begin{eqnarray}\label{D19}
\xi^{(2)}&=&2\xi^{(1)}\left(c_{1}+4\chi_{H}c_{6}\right)+
2\xi^{(2)}\left(c_{2}-4\chi_{H}c_{6}\right)\nonumber\\
&&+2\xi^{(3)}\left(c_{3}+4\chi_{H}c_{6}\right)+2\xi^{(4)}\left(c_{4}-4\chi_{H}c_{7}\right)\nonumber\\
&&+2\xi^{(5)}\left(c_{5}-3\chi_{H}c_{6}-\chi_{H}c_{7}\right)+
2\xi^{(6)}\left(c_{6}+\chi_{H}c_{2}\right)\nonumber\\
&&+2\xi^{(7)}\left(c_{7}+\chi_{H}c_{5}\right).
\end{eqnarray}
Comparing both sides of this relation, and solving the set of algebraic equations that arises from this comparison, the coefficients $c_{n}$ from \eqref{D18} are determined as
\begin{eqnarray}\label{D20}
c_{1}&=&  \frac{2\chi_{H}^{2} }{1+ 4 \chi_{H}^{2} },  \hspace{1.5cm} c_{4}=  \frac{6 \chi_{H}^{4 } }{ \left( 1+  \chi_{H}^{2}\right) \left(1+ 4 \chi_{H}^{2} \right) }, \nonumber \\
c_{2}&=&  \frac{1 }{2 \left( 1+ 4 \chi_{H}^{2} \right) },   \hspace{1cm} c_{5}= - \frac{ 3\chi_{H}^{2 } }{ 2\left( 1+  \chi_{H}^{2}\right) \left(1+ 4 \chi_{H}^{2} \right) }, \nonumber \\
c_{3}&=&  \frac{2\chi_{H}^{2} }{ 1+ 4 \chi_{H}^{2} },   \hspace{1.6cm}  c_{6}= - \frac{ \chi_{H} }{2 \left(1+ 4 \chi_{H}^{2} \right) },\nonumber\\
c_{7}&=&  \frac{3 \chi_{H}^{3 } }{ 2\left( 1+  \chi_{H}^{2}\right) \left(1+ 4 \chi_{H}^{2} \right)}.
\end{eqnarray}
Plugging then $C_{\mu\nu\rho\sigma}$ from \eqref{D18} into \eqref{D14}, $\delta f_{d}^{(b)}$ is determined. In the next step, we express $\delta f_{d}^{(b)}$ in terms of traceless and traceful bases $V_{\mu\nu\rho\sigma}^{(n)}, n=0,\cdots, 4$ and $W_{\mu\nu\rho\sigma}^{(n)}, n=0,1$, defined by\footnote{In \eqref{D21} and \eqref{D22}, the indices $\mu\nu\rho\sigma$ of $\xi^{(n)}$ are skipped.}
\begin{eqnarray}\label{D21}
V^{(0)} &=& \xi^{(2)}-\frac{2}{3}\xi^{(1)}, \nonumber \\
V^{(1)} &=&\xi^{(2)}-\xi^{(1)} - \xi^{(3)} + \xi^{(4)} + \xi^{(5)}, \nonumber \\
V^{(2)} &=& - \left( \xi^{(5)} + 4 \xi^{(4)} \right), \nonumber \\
V^{(3)} &=&\xi^{(6)}+\xi^{(7)}, \nonumber \\
V^{(4)} &=& \xi^{(7)},
\end{eqnarray}
and
\begin{eqnarray}\label{D22}
W^{(0)}= \xi^{(1)}, \quad \mbox{and}\quad
W^{(1)}= \xi^{(3)},
\end{eqnarray}
with $\xi_{\mu\nu\rho\sigma}^{(n)}, n=1,\cdots,7$ given in \eqref{D16}.
The aim is to determine the shear and bulk viscosities of the magnetized fluid. To do this, we first introduce
\begin{eqnarray}\label{D23}
\hspace{-1.2cm}V_{\mu\nu}^{(n)}\equiv V_{\mu\nu\rho\sigma}^{(n)}w^{\rho\sigma}, \quad\mbox{and}\quad
W_{\mu\nu}^{(m)}\equiv W_{\mu\nu\rho\sigma}^{(m)}w^{\rho\sigma},
\end{eqnarray}
with $V_{\mu\nu\rho\sigma}^{(n)}, n=0,\cdots , 4$
and
$W_{\mu\nu\rho\sigma}^{(m)}, m=0,1$ from \eqref{D21} and \eqref{D22}. We then use
\begin{eqnarray}\label{D24}
\hspace{-1cm}\delta f_{d}^{(b)}=\sum_{n=0}^{4}g_{n}p^{\mu}p^{\nu}V_{\mu\nu}^{(n)}+
\sum_{n=0}^{1}h_{n}p^{\mu}p^{\nu}W_{\mu\nu}^{(n)}.
\end{eqnarray}
The coefficients $g_{n}$ and $h_{n}$, appearing in \eqref{D24} are determined by comparing $\delta f_{d}^{(b)}$ from \eqref{D24} with $\delta f_{d}^{(b)}$ that arises by plugging \eqref{D18} with $c_{n}$  from \eqref{D20} into \eqref{D14}. They are given by
\begin{eqnarray}\label{D25}
g_{0}&=&\nu_{H}\left(c_{2}+\frac{1}{3}c_{4}-\frac{4}{3}c_{5}\right) =  \frac{\nu_{H}}{2} , \nonumber \\
g_{1}&=&\nu_{H}\left( \frac{4 c_5 - c_4}{3} \right) =-  \frac{2\nu_{H}\chi_{H}^{2}}{(1+4\chi_{H}^{2})}, \nonumber \\
g_{2}&=&\nu_{H}\left(\frac{c_5 - c_4}{3}\right) =-  \frac{\nu_{H}\chi_{H}^{2}}{2(1+\chi_{H}^{2})} , \nonumber \\
g_{3}&=&\nu_{H} c_6 = - \frac{\nu_{H}\chi_{H}}{2(1+4\chi_{H}^{2})} , \nonumber \\
g_{4}&=&\nu_{H}\left( c_7 - c_6\right) =\frac{\nu_{H}\chi_{H}}{2(1+\chi_{H}^{2})},
\end{eqnarray}
and
\begin{eqnarray}\label{D26}
h_{0}&=&\nu_{H}\left(c_{1}+\frac{2}{3}c_{2}-\frac{1}{9}c_{4} + \frac{4}{9} c_{5}\right) =\frac{\nu_{H}}{3}, \nonumber \\
h_{1}&=&\nu_{H}\left(c_{3}-\frac{1}{3}c_{4}+\frac{4}{3}c_{5}\right)=0.
\end{eqnarray}
This determines $\delta f_{d}^{(b)}$ in terms of traceless and traceful bases $V_{\mu\nu\rho\sigma}^{(n)}, n=0,\cdots , 4$
and
$W_{\mu\nu\rho\sigma}^{(m)}, m=0,1$ from \eqref{D21} and \eqref{D22}. Plugging finally $\delta f_{d}^{(a)}$ from \eqref{D12} and $\delta f_{d}^{(b)}$ from \eqref{D24} into \eqref{D8}, the dissipative part of the magneto-anisotropic one-particle distribution function is given by
\begin{eqnarray}\label{D27}
\delta f_{d}&=&\sum_{n=0}^{2} \ell_{n}k^{\mu}k^{\nu}U_{\mu\nu}^{(n)}+\sum_{n=0}^{4}g_{n}
k^{\mu}k^{\nu}V_{\mu\nu}^{(n)}\nonumber\\
&&+\sum_{n=0}^{1}h_{n}k^{\mu}k^{\nu}W_{\mu\nu}^{(n)}-(f_{b}-f_{eq}),
\end{eqnarray}
with $\ell_n, g_n$ and $h_n$ from \eqref{D13}, \eqref{D25} and \eqref{D26}.
Combining finally $\delta f_{d}$ with $f_{b}$ from \eqref{D2}, we arrive at the one-particle distribution function of the magnetized QGP up to first-order derivative expansion.
\par
Let us notice, at this stage,  that $\delta f_{d}$ from \eqref{D27} can be used to determine the dissipative part $\tau^{\mu\nu}$ of the energy-momentum tensor $T^{\mu\nu}=T^{\mu\nu}_{b}+\tau^{\mu\nu}$, defined by
\begin{eqnarray}\label{D28}
T^{\mu\nu}\equiv \int d\tilde{k}~k^{\mu}k^{\nu}f.
\end{eqnarray}
Using $f=f_{b}+\delta f_{d}$, plugging $\delta f_{d}$ from \eqref{D27} into
\begin{eqnarray}\label{D29}
\tau^{\mu\nu}=\int d\tilde{k}~k^{\mu}k^{\nu}\delta f_{d},
\end{eqnarray}
and comparing the resulting expression with
\begin{eqnarray}\label{D30}
\tau_{\mu\nu}&=&\sum\limits_{n=0}^{1}\alpha_{n}U^{(n)}_{\mu\nu}+
\sum_{n=0}^{4}\eta_{n}V^{(n)}_{\mu\nu}+
\sum_{n=0}^{1}\tilde{\zeta}_{n}W^{(n)}_{\mu\nu}\nonumber\\
&& -(T^{b}_{\mu\nu}-T^{eq}_{\mu\nu}),
\end{eqnarray}
 we arrive after some work at
\begin{eqnarray}\label{D31}
\alpha_{n}&=&\frac{1}{3}\int d\tilde{k}~\ell_{n}|\boldsymbol{k}|^{4},\nonumber\\
\eta_{n}&=&\frac{2}{15}\int d\tilde{k}~g_{n}|\boldsymbol{k}|^{4},\nonumber\\
\tilde{\zeta}_{n}&=&\frac{1}{3}\int d\tilde{k}~h_{n}|\boldsymbol{k}|^{4},
\end{eqnarray}
where $\eta_{n}$ is the shear viscosity of the medium. An appropriate combination of $\tilde{\zeta}_{n}$ and $\alpha_{n}$ is then identified with the bulk viscosity of the medium (see below). In \eqref{D30}, $T^{\mu\nu}_{b}$ and $T^{\mu\nu}_{eq}$ are defined by \eqref{D28} with $f$ replaced with $f_{b}$ and $f_{eq}$, respectively.
\subsection{Differential equations leading to $\boldsymbol{\xi}$ and $\boldsymbol{\lambda}$}\label{sec4b}
We start, as in Sec. \ref{sec3b}, with the zeroth moment of the Boltzmann equation \eqref{D1},
\begin{eqnarray}\label{D32}
\int d\tilde{k}\left(k^{\mu}\partial_{\mu}f+q_{f}eF^{\mu\nu}k_{\nu}\frac{\partial f}{\partial k^{\mu}}\right)=\int d\tilde{k}~C[f],\nonumber\\
\end{eqnarray}
with $f=f_{b}+\delta f_{d}$ and $C[f]$ given in \eqref{D4}. Here, $f_{b}$ and $\delta f_{d}$ are given in \eqref{D2} and \eqref{D27}. Whereas the second term on the lhs of \eqref{D32} vanishes because of the same argument as in \eqref{E16}, the first term on the lhs of \eqref{D32} leads to
\begin{eqnarray}\label{D33}
\int d\tilde{k} k^{\mu}\partial_{\mu} f=\partial_{\mu}n^{\mu},
\end{eqnarray}
with $n_{\mu}\equiv n_{b}^{\mu}+\delta n_{d}^{\mu}$ consisting of two terms defined by
\begin{eqnarray}
n_{b}^{\mu}&\equiv&\int d\tilde{k}~k^{\mu}f_{b}=\frac{\lambda^{3}}{\pi^{2}\sqrt{1+\xi}}u^{\mu},\label{D34}\\
\delta n_{d}^{\mu}&\equiv& \int d\tilde{k}~\Delta^{\mu\nu}k_{\nu}\delta f_{d}.\label{D35}
\end{eqnarray}
The expression arising in \eqref{D34} is similar to \eqref{E15} with $(\xi_{0},\lambda_{0})$ replaced by $(\xi,\lambda)$. Plugging, on the other hand,  $\delta f_{d}$ from \eqref{D27} into \eqref{D35}, and comparing the resulting expression with \cite{kineticbook}
\begin{eqnarray}\label{D36}
\delta n_{d}^{\mu}&=&\sum_{n=0}^{2}\rho_{n} U^{(n)}_{\mu\nu}\partial^{\nu}\xi+\sum_{n=0}^{2}\rho^{\prime}_{n} U^{(n)}_{\mu\nu}\partial^{\nu}\lambda\nonumber\\
&&+\sum_{n=0}^{3}
\sigma_{n}T^{(n)}_{\mu\alpha\beta}w^{\alpha\beta},
\end{eqnarray}
where $U_{\mu\nu}^{(n)}, n=0,1,2$ are given in \eqref{D11} and $T_{\mu\nu}^{(n)}, n=0,\cdots,3$ are defined by \cite{kineticbook}
\begin{eqnarray}\label{D37}
\begin{array}{rclcrcl}
C^{(0)}_{\gamma\alpha\beta} &\equiv& b_{\gamma} b_{\alpha} b_{\beta} , &\qquad& C^{(2)}_{\gamma\alpha\beta} &\equiv& b_{\alpha} \Delta_{\beta \gamma} + b_{\beta} \Delta_{\alpha \gamma},\\
C^{(1)}_{\gamma\alpha\beta} &\equiv& b_{\gamma} \Delta_{\alpha \beta}, &\qquad& C^{(3)}_{\gamma\alpha\beta} &\equiv& b_{\alpha} b_{\beta \gamma} + b_{\beta} b_{\alpha \gamma},\nonumber\\
\end{array}\\
\end{eqnarray}
it turns out that $\delta n_{d}^{\mu}$ vanishes, as expected \cite{strickland2013}. We are therefore left with $\partial_{\mu}n^{\mu}=\partial_{\mu}n_{b}^{\mu}$ with $n_{b}^{\mu}$ given in \eqref{D34}. Plugging, at this stage, $C[f]$ from \eqref{D4} into the rhs of \eqref{D32}, the equation arising from the zeroth moment of the Boltzmann equation reads
\begin{eqnarray}\label{D38}
Dn_{b}+\theta n_{b}=-\frac{1}{\tau_{r}}\left(n_{b}-n_{eq}\right).
\end{eqnarray}
Here, we used $n_{b}^{\mu}=n_{b}u^{\mu}$. Using, as in Sec. \ref{sec3b}, $D=\partial_{\tau}$, $\theta=1/\tau$, and plugging $n_{b}$ and $n_{eq}$ from \eqref{D34} and \eqref{E18}, we arrive at
\begin{eqnarray}\label{D39}
\frac{\partial_{\tau} \xi}{1+ \xi} - \frac{6 \partial_{\tau} \lambda}{\lambda} - \frac{2}{\tau}= \frac{2}{\tau_{r}} \left(1-\left(\frac{T}{\lambda}\right)^{3}\sqrt{1+\xi}\right).
\nonumber\\
\end{eqnarray}
Similar to the nondissipative case in Sec. \ref{sec3b}, the relation between $T$ and $\lambda$ arises from the first moment of the Boltzmann equation,
\begin{eqnarray}\label{D40}
\int d\tilde{k}~k^{\rho}\left(k^{\mu}\partial_{\mu}f+q_{f}eF^{\mu\nu}k_{\nu}\frac{\partial f}{\partial k^{\mu}}\right)=\int d\tilde{k}~k^{\rho}C[f].\nonumber\\
\end{eqnarray}
Using \eqref{D28}, we arrive first at
\begin{eqnarray}\label{D41}
\hspace{-1cm}\partial_{\mu}T^{\rho\mu}=\int d\tilde{k}~k^{\rho}C[f]=-\frac{1}{\tau_{r}}u_{\mu}\left(T^{\rho\mu}-T^{\rho\mu}_{eq}\right),
\end{eqnarray}
where $T_{eq}^{\rho\mu}$ is defined in \eqref{D28} with $f$ replaced with $f_{eq}$. Using then the energy-momentum conservation $\partial_{\mu}T^{\mu\nu}=0$, we obtain $u_{\mu}T^{\rho\mu}=u_{\mu}T^{\mu\nu}_{eq}$. Moreover, using the definition $T^{\mu\nu}=T^{\mu\nu}_{b}+\tau^{\mu\nu}$ and the fact that $u_{\mu}\tau^{\mu\nu}=0$, we obtain
\begin{eqnarray}\label{D42}
\epsilon_{b}=\epsilon_{eq},
\end{eqnarray}
as in the nondissipative case [see \eqref{E24}]. Here,
\begin{eqnarray}\label{D43}
\epsilon_{b}\equiv \int d\tilde{k}~\left(k\cdot u\right)^{2}f_{b}=\frac{3\lambda^{4}}{\pi^{2}}{\cal{R}}(\xi),
\end{eqnarray}
and $\epsilon_{eq}=\frac{3T^{4}}{\pi^{2}}$ is given in \eqref{E25}. In \eqref{D43},  ${\cal{R}}(\xi)$ is defined in \eqref{E13}. Using \eqref{D42} thus leads to the Landau matching condition
\begin{eqnarray}\label{D44}
T=\lambda {\cal{R}}^{1/4}(\xi).
\end{eqnarray}
Plugging, at this stage, \eqref{D44} into \eqref{D39}, we obtain the first differential equation leading to the $\tau$ dependence of $\xi$ and $\lambda$,
\begin{eqnarray}\label{D45}
\frac{\partial_{\tau}\xi}{1+\xi}-\frac{6\partial_{\tau}\lambda}{\lambda}-\frac{2}{\tau}=\frac{2}{\tau_{r}}\left(1-{\cal{R}}^{3/4}(\xi)\sqrt{1+\xi}\right). \nonumber\\
\end{eqnarray}
The second differential equation arises from  $\partial_{\mu}T^{\mu\nu}=0$, leading to the energy equation
\begin{eqnarray}\label{D46}
D\epsilon_{b}+(\epsilon_{b}+p_{b}-BM_{b})\theta=w_{\mu\nu}\tau^{\mu\nu}.
\end{eqnarray}
Here, $\epsilon_{b}$ is given in \eqref{D43}, $p_{b}$ and $M_{b}$ are given by
\begin{eqnarray}\label{D47}
p_{b}&=&\int d\tilde{k}~(k\cdot b)^{2}f_{b}=\frac{3\lambda^{4}}{\pi^{2}\xi}\left({\cal{R}}(\xi)-\frac{1}{1+\xi}\right),\nonumber\\
M_{b}&=&-\frac{1}{2B}\int d\tilde{k}~\big[\left(k\cdot u\right)^{2}-3\left(k\cdot b\right)^{2}\big]f_{b}\nonumber\\
&=&\frac{3\lambda^{4}}{2\pi^{2}\xi B}\bigg[\left(3-\xi\right){\cal{R}}(\xi)-\frac{3}{1+\xi}\bigg],
\end{eqnarray}
[see also \eqref{E11} and \eqref{E12} for similar expressions]. The dissipative part of the energy-momentum tensor is given by \eqref{D30}. Plugging $U_{\mu\nu}^{(n)}, n=0,1,2$, $V_{\mu\nu}^{(n)}, n=0,\cdots,4$ and $W_{\mu\nu}^{(n)}, n=0,1,2$ from \eqref{D11}, \eqref{D21} and \eqref{D22} into \eqref{D30}, and using
\begin{eqnarray}\label{D48}
T_{eq}^{\mu\nu}&=&\left(\epsilon_{eq}+p_{eq}\right)u^{\mu}u^{\nu}-p_{eq}g^{\mu\nu},\nonumber\\
T_{b}^{\mu\nu}&=&\left(\epsilon_{b}+p_{b}\right)u^{\mu}u^{\nu}-\left(p_{b}-BM_{b}\right)g^{\mu\nu}+BM_{b}b^{\mu}b^{\nu}, \nonumber\\
\end{eqnarray}
we arrive first at
\begin{eqnarray}\label{D49}
\tau_{\mu\nu}&=&2\eta_{0}\left(w_{\mu\nu}-\frac{1}{3}\theta\Delta_{\mu\nu}\right)
+\eta_{1}\left(2w_{\mu\nu}-\theta\Xi_{\mu\nu}^{B}\right)\nonumber\\
&&+2\eta_{3}\left(\Delta_{\mu\rho}b_{\nu\sigma}+\Delta_{\nu\rho}b_{\mu\sigma}
\right)w^{\rho\sigma}+\tilde{\zeta}_{0}\theta\Delta_{\mu\nu}+\tilde{\zeta}_1\theta b_{\mu}b_{\nu}\nonumber\\
&&
+\alpha_{0}\Delta_{\mu\nu}+\alpha_{1}b_{\mu}b_{\nu}+\alpha_{2}b_{\mu\nu}-
BM_{b}\Xi^{B}_{\mu\nu}\nonumber\\
&&+(p_b-p_{eq})\Delta_{\mu\nu},
\end{eqnarray}
where  $\Xi_{\mu\nu}^{B}\equiv \Delta_{\mu\nu}+b_{\mu\nu}$, $p_{b}$ is given in \eqref{D47} and $p_{eq}=\frac{\epsilon_{eq}}{3}=\frac{T^{4}}{\pi^{2}}$ with $\epsilon_{eq}$ defined in \eqref{D25}. To arrive at \eqref{D49}, we mainly used $b^{\mu}w_{\mu\nu}=0$, which is valid in transverse MHD. Multiplying $\tau_{\mu\nu}$ from \eqref{D49} with $w^{\mu\nu}$, and using a number of algebraic relations like $w^{\mu\nu}\Delta_{\rho\nu}=w^{\mu}_{~\rho}$, we arrive after some work at
\begin{eqnarray}\label{D50}
\tau^{\mu\nu}w_{\mu\nu}&=&2(\eta_{0}+\eta_{1})w^{\mu\nu}w_{\mu\nu}-\left(\frac{2}{3}\eta_{0}+\eta_{1}-\tilde{\zeta}_{0}\right)\theta^{2}\nonumber\\
&&+ \alpha_{0} \theta + \left(p_{b}- p_{eq}- BM_{b}\right) \theta.
\end{eqnarray}
The relevant transport coefficients  $\alpha_{0}, \eta_{0},\eta_{1}$ and $\tilde{\zeta}_{0}$ can be determined using \eqref{D31} and the assumption of large magnetic field, leading to $\chi_{H}=\frac{q_{f}eB\tau_{r}}{(k\cdot u)}\gg 1$. Defining the relaxation frequency $\omega_{r}\equiv 1/\tau_{r}$, the limit $\chi_{H}\gg 1$ can be interpreted as $\omega_{r}\ll \omega_{L}$ with the Larmor frequency $\omega_{L}\equiv \frac{q_{f}eB}{(k\cdot u)}$. Plugging $\ell_{0}$ from \eqref{D13} into $\alpha_{0}$ from \eqref{D31}, and performing the corresponding integration, we obtain
\begin{eqnarray}\label{D51}
\alpha_{0}=\frac{\lambda^{4}\tau_{r}}{\pi^{2}}\left(3{\cal{R}}(\xi)+\frac{1}{(1+\xi)^{2}}\right)\frac{D\lambda}{\lambda}.
\end{eqnarray}
Similarly, plugging $g_{0}=\frac{\nu_{H}}{2}$ and $g_{1}\approx -\frac{\nu_{H}}{2}$ for $\chi_{H}\gg 1$ from \eqref{D25} into $\eta_{i}, i=0,1$ from \eqref{D31}, and performing the integration over $k$, we arrive at
\begin{eqnarray}\label{D52}
\eta_{0}=-\eta_{1}=\frac{\lambda^{4}\tau_{r}}{5\pi^{2}}\left(3{\cal{R}}(\xi)+\frac{1}{(1+\xi)^{2}}\right).
\end{eqnarray}
Finally, plugging $h_{0}=\frac{\nu_{H}}{3}$ from \eqref{D26} into $\tilde{\zeta}_{0}$ and performing the integration, we obtain
\begin{eqnarray}\label{D53}
\tilde{\zeta}_{0}=\frac{\lambda^{4}\tau_{r}}{3\pi^{2}}\left(3{\cal{R}}(\xi)+\frac{1}{(1+\xi)^{2}}\right).
\end{eqnarray}
Plugging all these results into \eqref{D46}, the second differential equation leading to $\xi$ and $\lambda$ reads
\begin{widetext}
\begin{eqnarray}\label{D54}
\frac{4\partial_{\tau}\lambda}{\lambda}\bigg[1-\frac{\tau_{r}}{12\tau}\left(3+\frac{1}{{\cal{R}}(\xi)(1+\xi)^{2}}\right)\bigg]+\frac{\partial_{\tau}\xi}{{\cal{R}}(\xi)}\frac{\partial {\cal{R}}(\xi)}{\partial\xi}+\frac{4}{3\tau}-\frac{2\tau_{r}}{15}\left(3+\frac{1}{{\cal{R}}(\xi)(1+\xi)^{2}}\right)\frac{1}{\tau^2}=0.
\end{eqnarray}
\end{widetext}
Equations \eqref{D45} and \eqref{D54} build a set of coupled differential equations, whose solution leads to $\lambda$ and $\xi$ for the case of dissipative QGP. In Sec. \ref{sec5}, we solve these differential equations numerically, and compare the corresponding results with those arising from \eqref{E27} and\eqref{E28} in the case of nondissipative QGP.
\section{Numerical results}\label{sec5}
\setcounter{equation}{0}
\begin{figure*}[hbt]
\includegraphics[width=8cm,height=6.5cm]{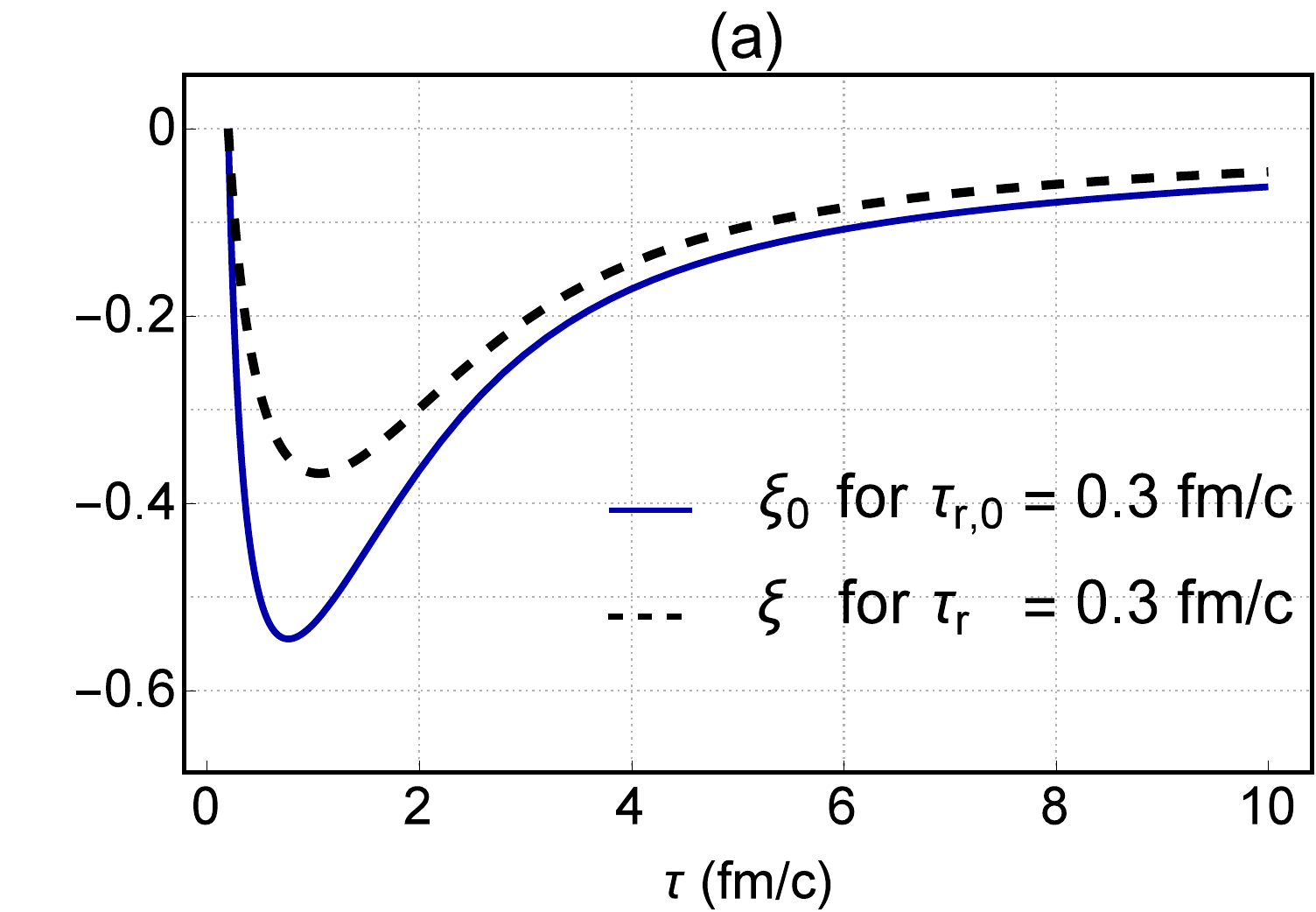}\hspace{0.8cm}
\includegraphics[width=8cm,height=6.5cm]{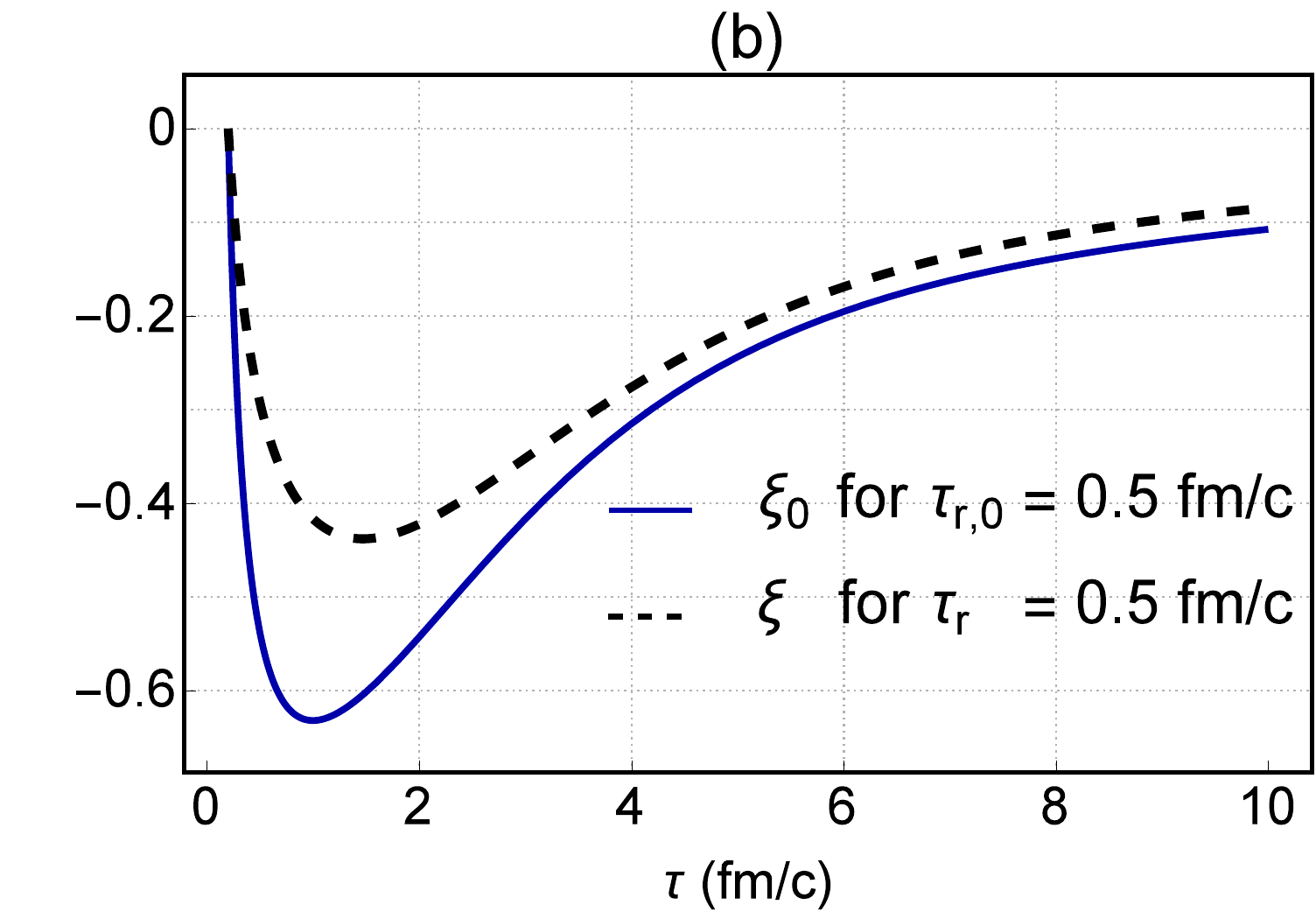}
\caption{(color online).  The $\tau$ dependence of the anisotropy function $\xi_{0}$ (nondissipative case) and $\xi$ (dissipative case) is plotted for relaxation times $\tau_{r,0}$ (blue solid curves) and $\tau_{r}$ (black dashed curves) equal to $0.3$ fm/c (panel a) and $0.5$ fm/c (panel b). For a comparison see the main text. }\label{fig-2}
\end{figure*}
\begin{figure*}[hbt]
\includegraphics[width=8cm,height=6.5cm]{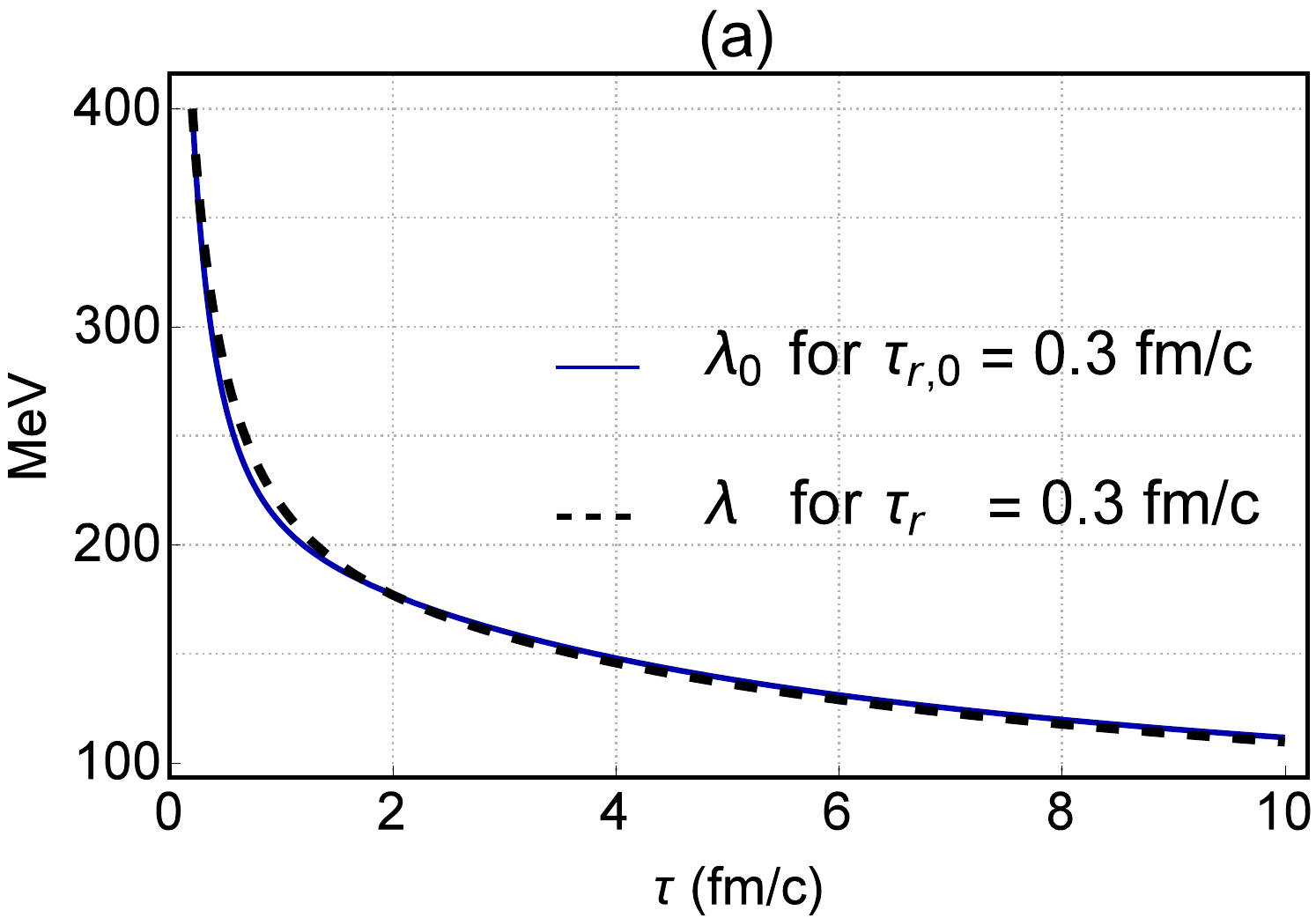}\hspace{0.8cm}
\includegraphics[width=8cm,height=6.5cm]{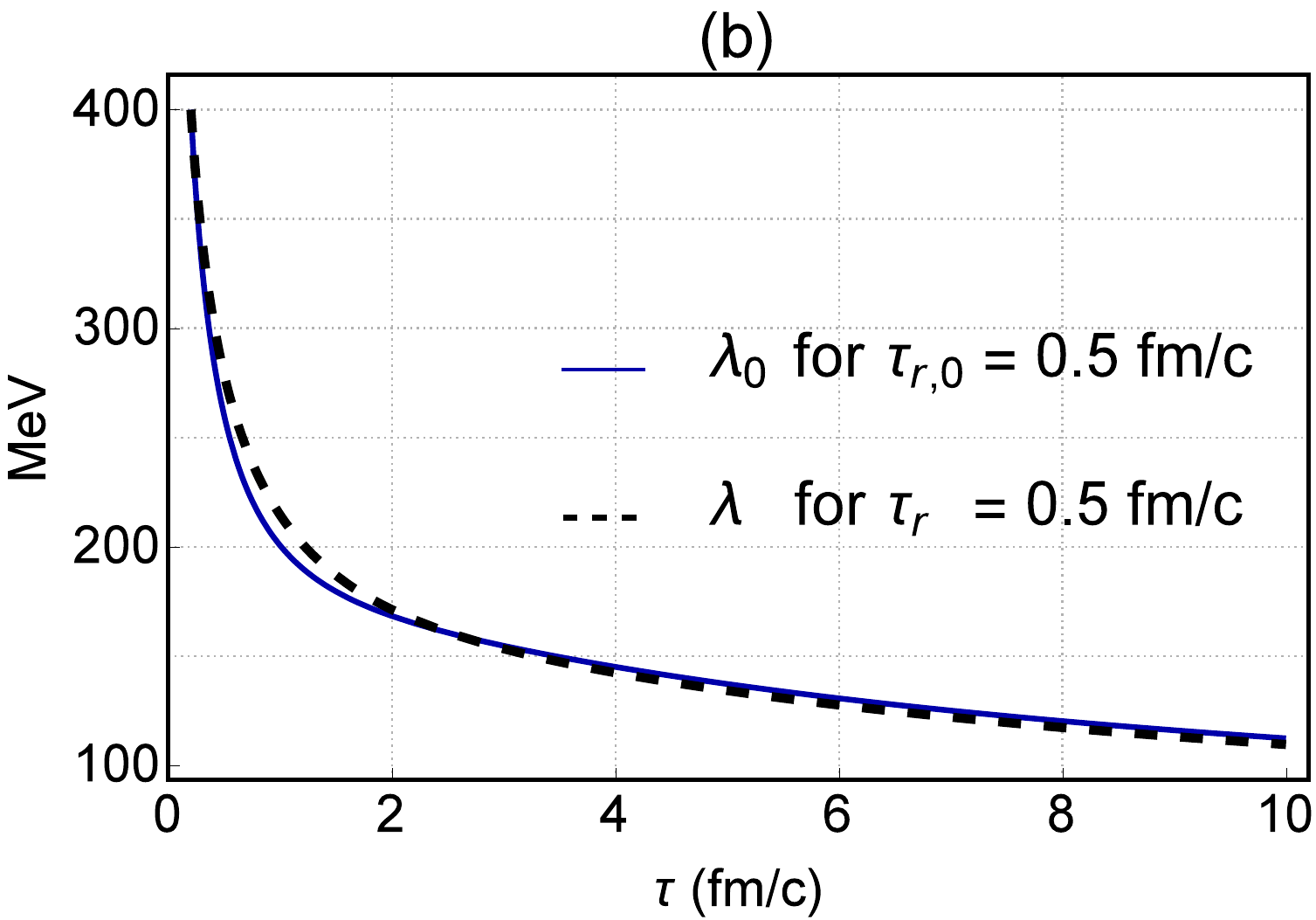}
\caption{(color online).  The $\tau$ dependence of the effective temperature in a magnetized fluid $\lambda_{0}$ (nondissipative case) and $\lambda$ (dissipative case) is plotted for relaxation times $\tau_{r,0}$ (blue solid curves) and $\tau_{r}$ (black dashed curves) equal to $0.3$ fm/c (panel a) and $0.5$ fm/c (panel b). It turns out that, independent of the value of the relaxation time, the effect of dissipation on the effective temperature of a magnetized fluid is negligible.}\label{fig-3}
\end{figure*}
\begin{figure}[hbt]
\includegraphics[width=8cm,height=6.5cm]{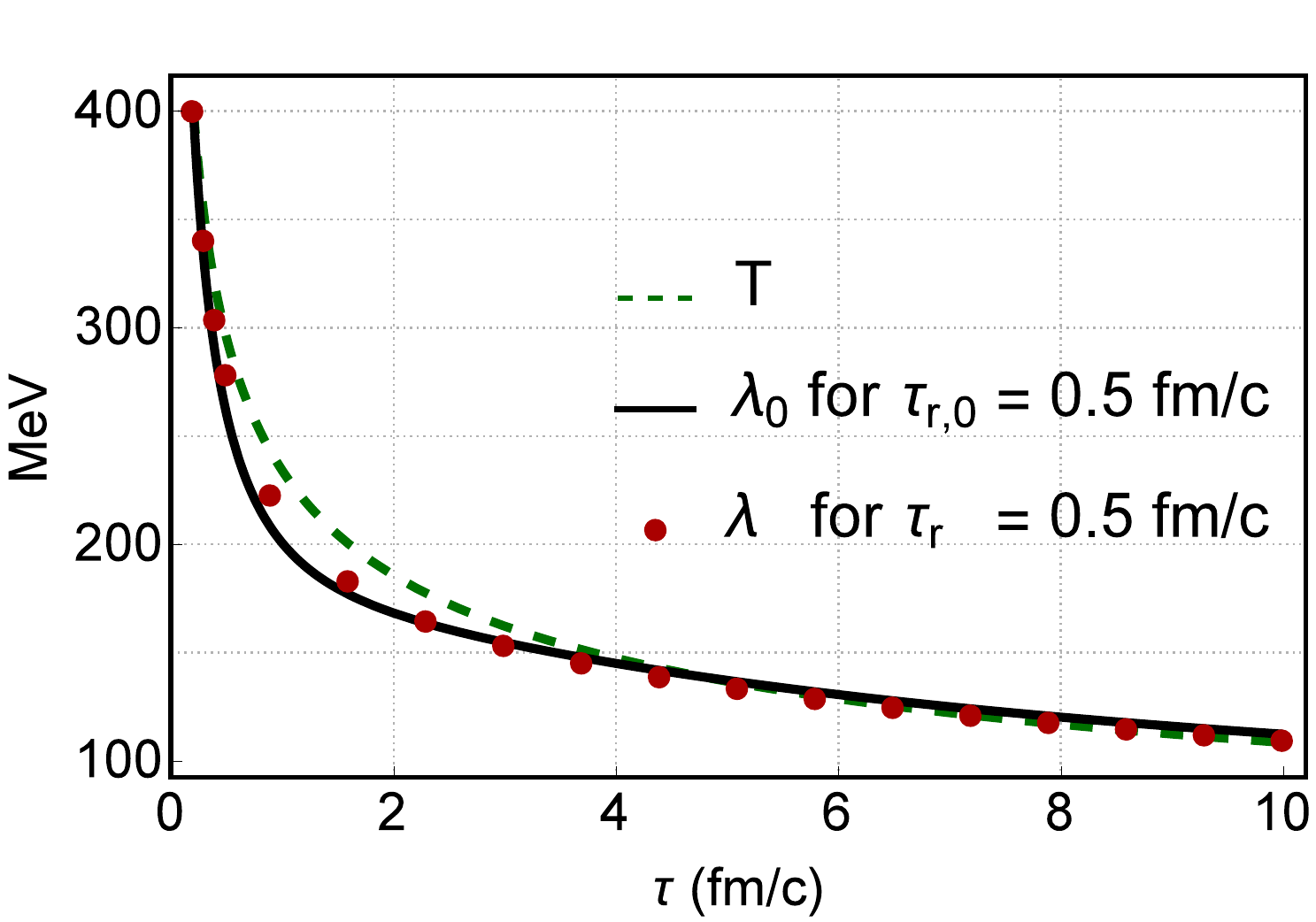}
\caption{(color online). The $\tau$ dependence of $T$ from \eqref{N29} for $\chi_m=0$ (green dashed curve), $\lambda_0$, the effective temperature for a nondissipative magnetized fluid (black solid curve), and $\lambda$, the effective temperature for a dissipative magnetized fluid (red dots) for $\tau_{r}=0.5$ fm/c is plotted. It turns out that, except in the interval $\tau\in [\sim 0.5,\sim 4]$ fm/c, the dynamics of $T,\lambda_{0}$, and $\lambda$ coincides. Hence, the effect of magnetization and dissipation on the proper time evolution of the temperature is negligible.}\label{fig-4}
\end{figure}
\begin{figure*}[hbt]
\includegraphics[width=8cm,height=6.5cm]{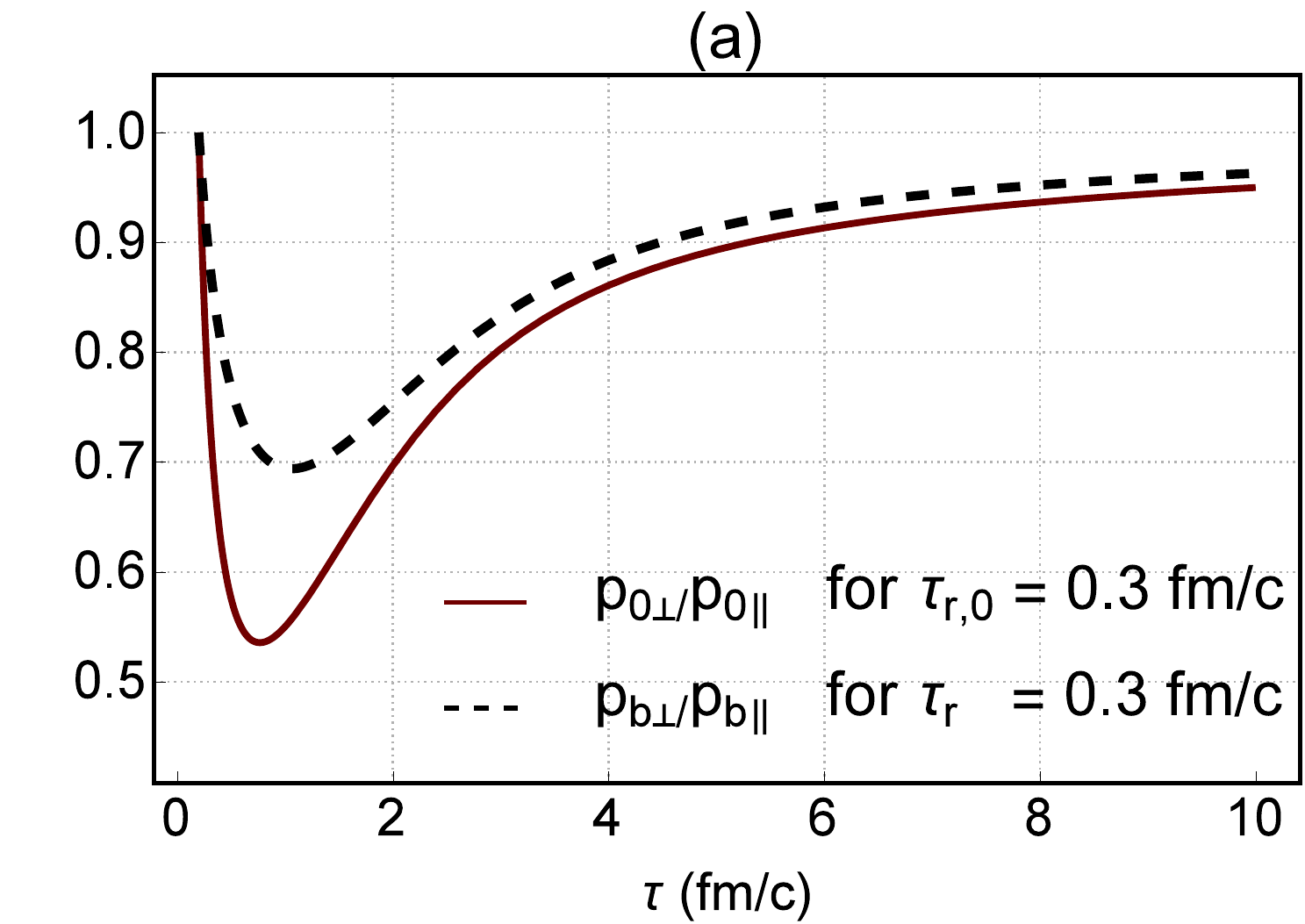}\hspace{0.8cm}
\includegraphics[width=8cm,height=6.5cm]{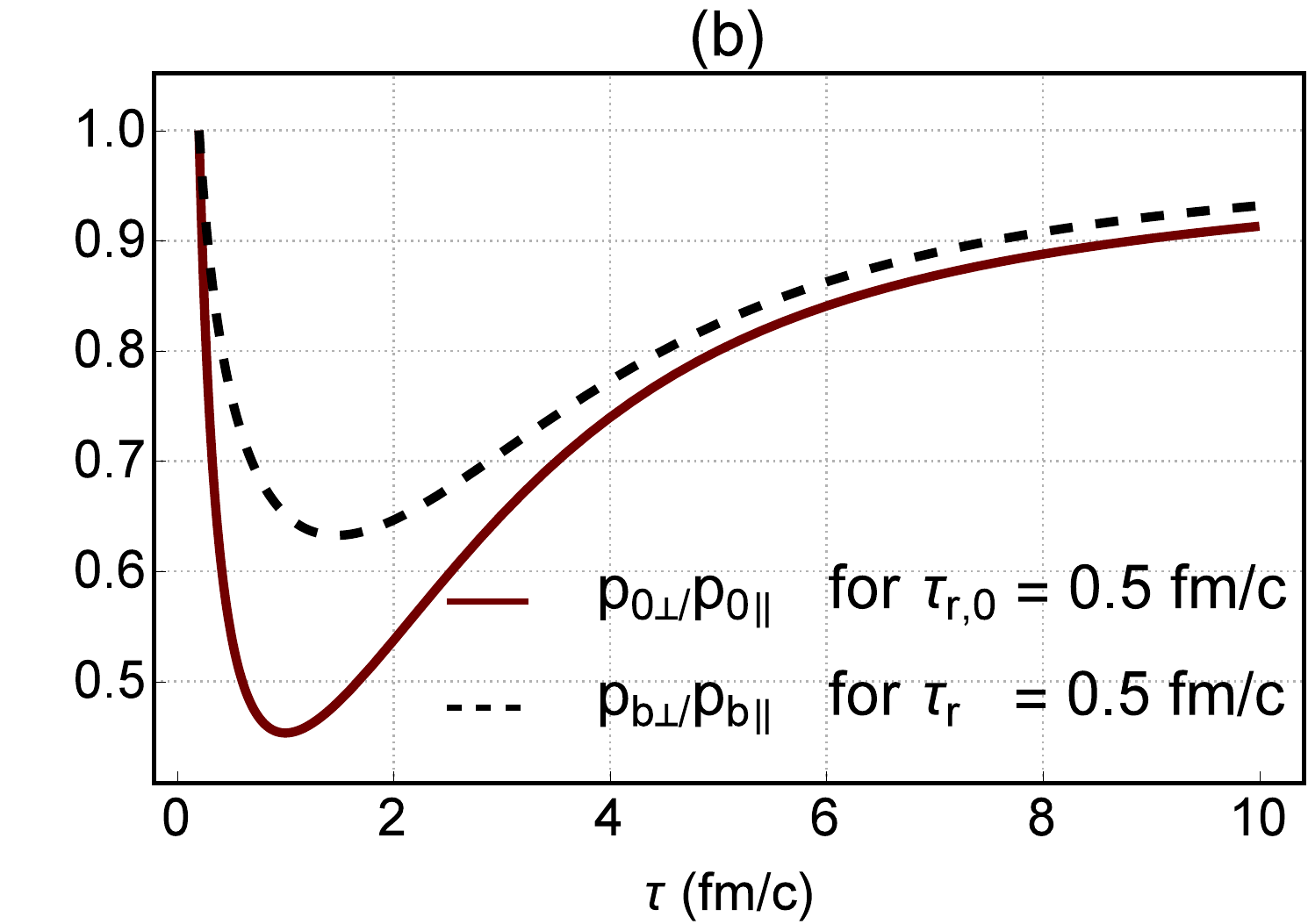}
\caption{(color online).  The $\tau$ dependence of the ratio $p_{0,\perp}/p_{0,\|}$ and  $p_{b,\perp}/p_{b,\|}$ for a nondissipative and a dissipative magnetized fluid is plotted for relaxation times  $\tau_{r,0}$ (red solid curves) and $\tau_{r}$ (black dashed curves) equal to  $0.3$ fm/c (panel a) and $0.5$ fm/c (panel b). For a comparison see the main text. }\label{fig-5}
\end{figure*}
\begin{figure}[hbt]
\includegraphics[width=8cm,height=6.5cm]{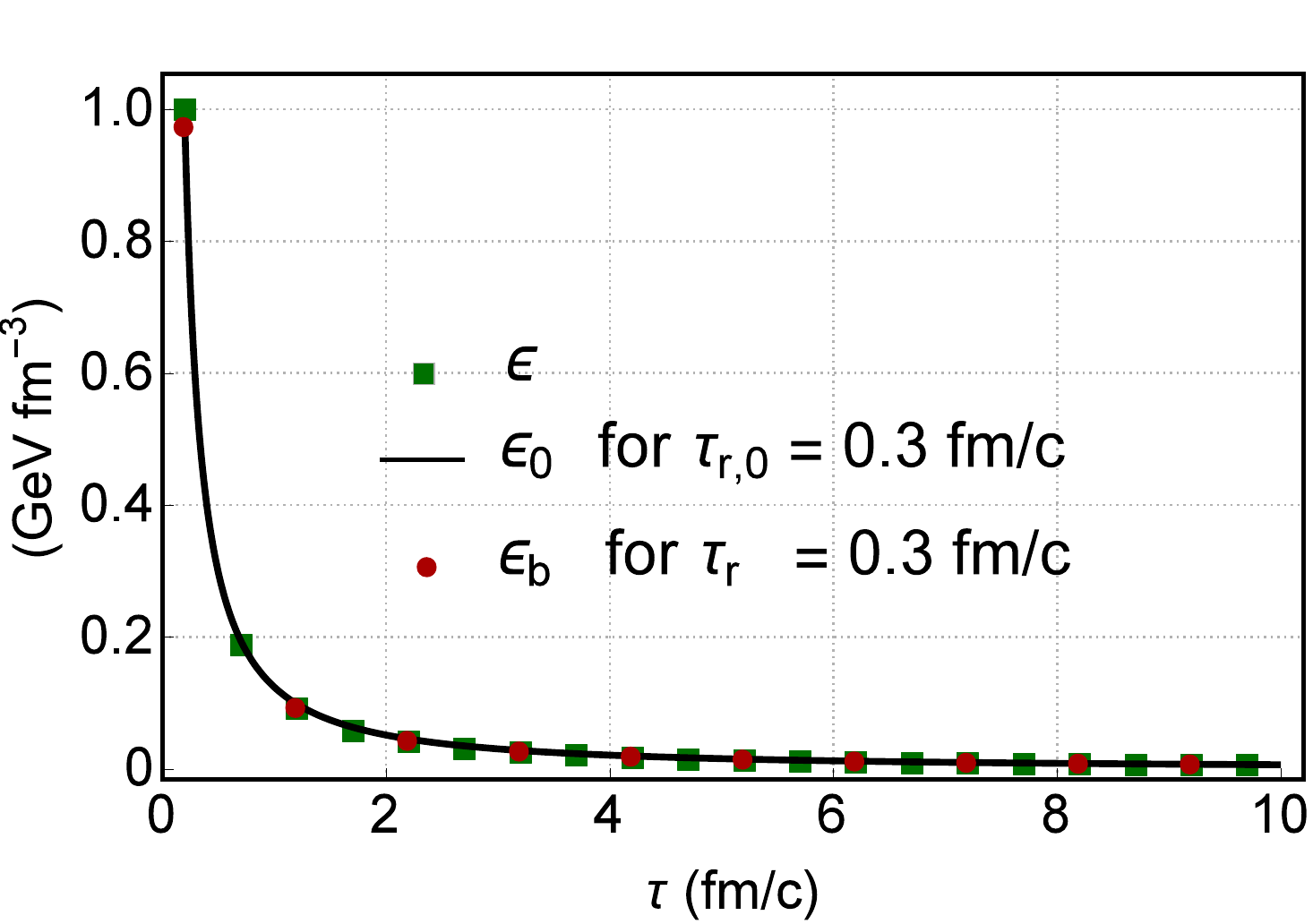}
\caption{(color online). The $\tau$ dependence of $\epsilon$ (green squares), $\epsilon_{0}$ (black solid curve), and $\epsilon_{b}$ (red circles) is plotted for relaxation times $\tau_{r,0}$ and $\tau_{r}$ equal to $0.3$ fm/c. As expected, $\epsilon=\epsilon_0=\epsilon_b$. }\label{fig-6} 
\end{figure}
\begin{figure*}[hbt]
\includegraphics[width=8cm,height=6.5cm]{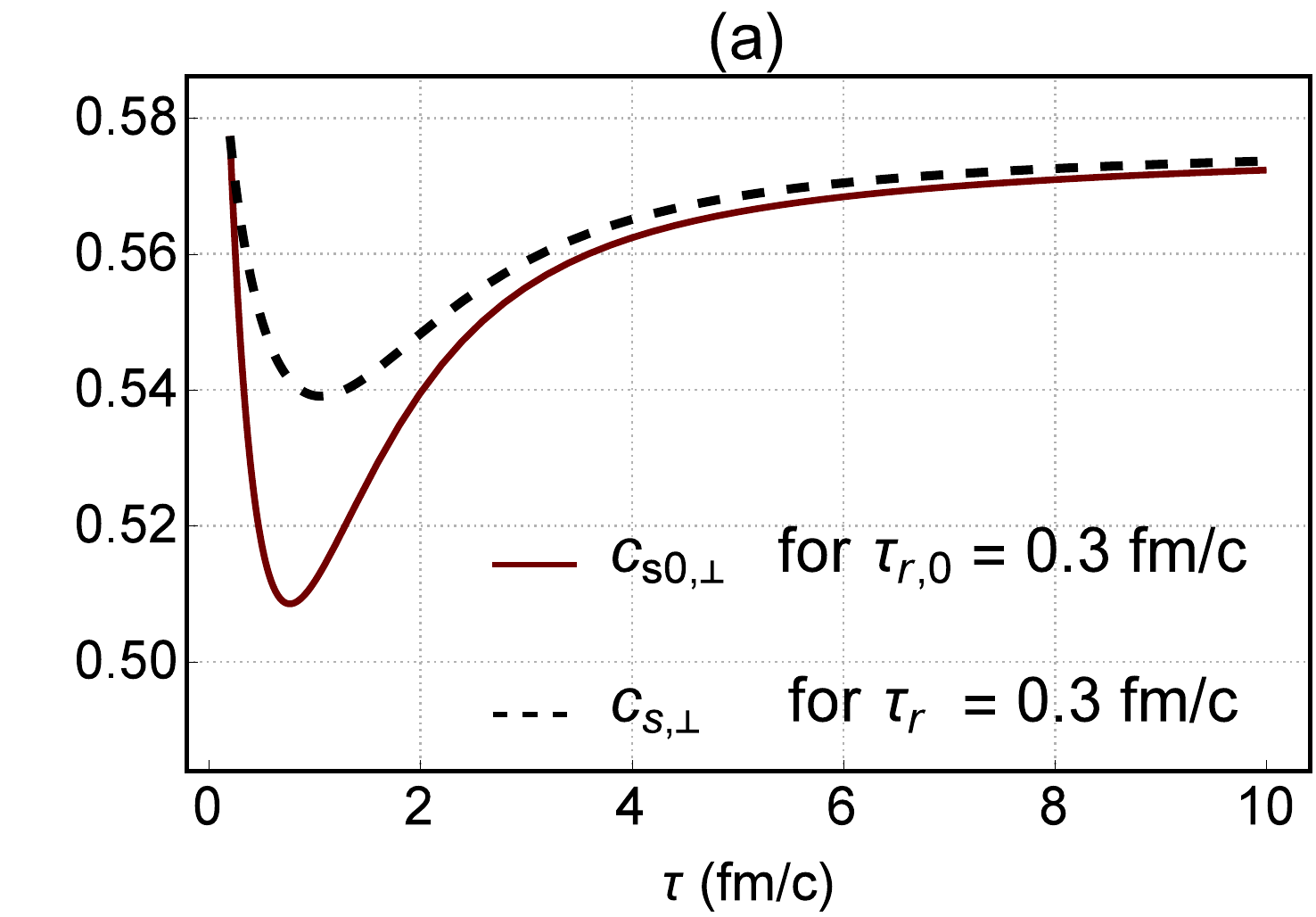}\hspace{0.8cm}
\includegraphics[width=8cm,height=6.5cm]{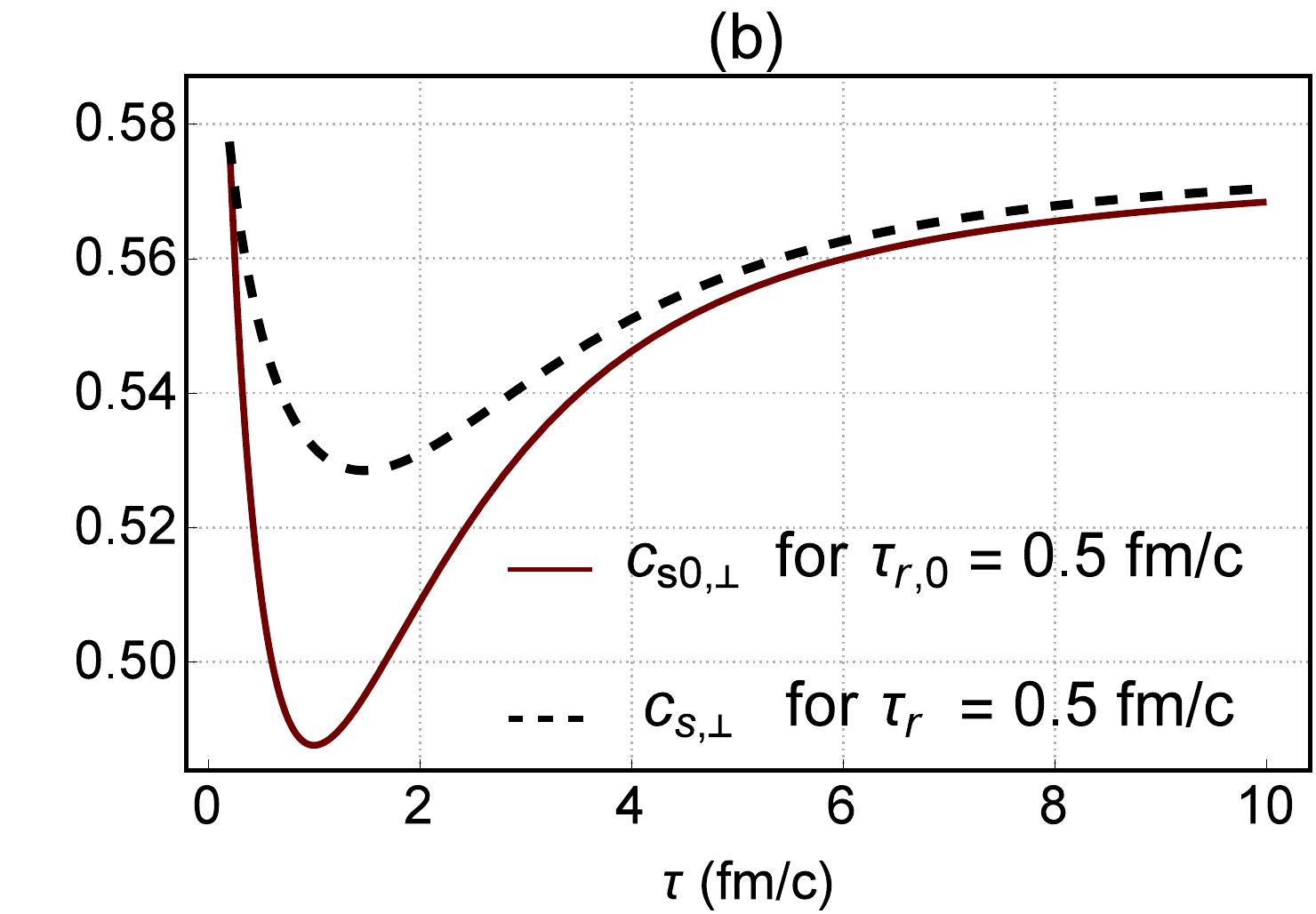}
\caption{(color online).  The $\tau$ dependence of the transverse speed of sound $c_{s0,\perp}$ and $c_{s,\perp}$ in a nondissipative (red solid curves) and dissipative (black dashed curves) fluid is plotted for the relaxation times $\tau_{r,0}$ and $\tau_{r}$ equal to $0.3$ fm/c (panel a), and $\tau_{r,0}$ as well as $\tau_{r}$ equal to $0.5$ fm/c (panel b). The evolution of $c_{s0,\perp}$ and $c_{s,\perp}$ reflects the dynamics of $\xi_0$ and $\xi$ (see Fig. \ref{fig-2}).  }\label{fig-7}
\end{figure*}
\begin{figure*}[hbt]
\includegraphics[width=8cm,height=6.5cm]{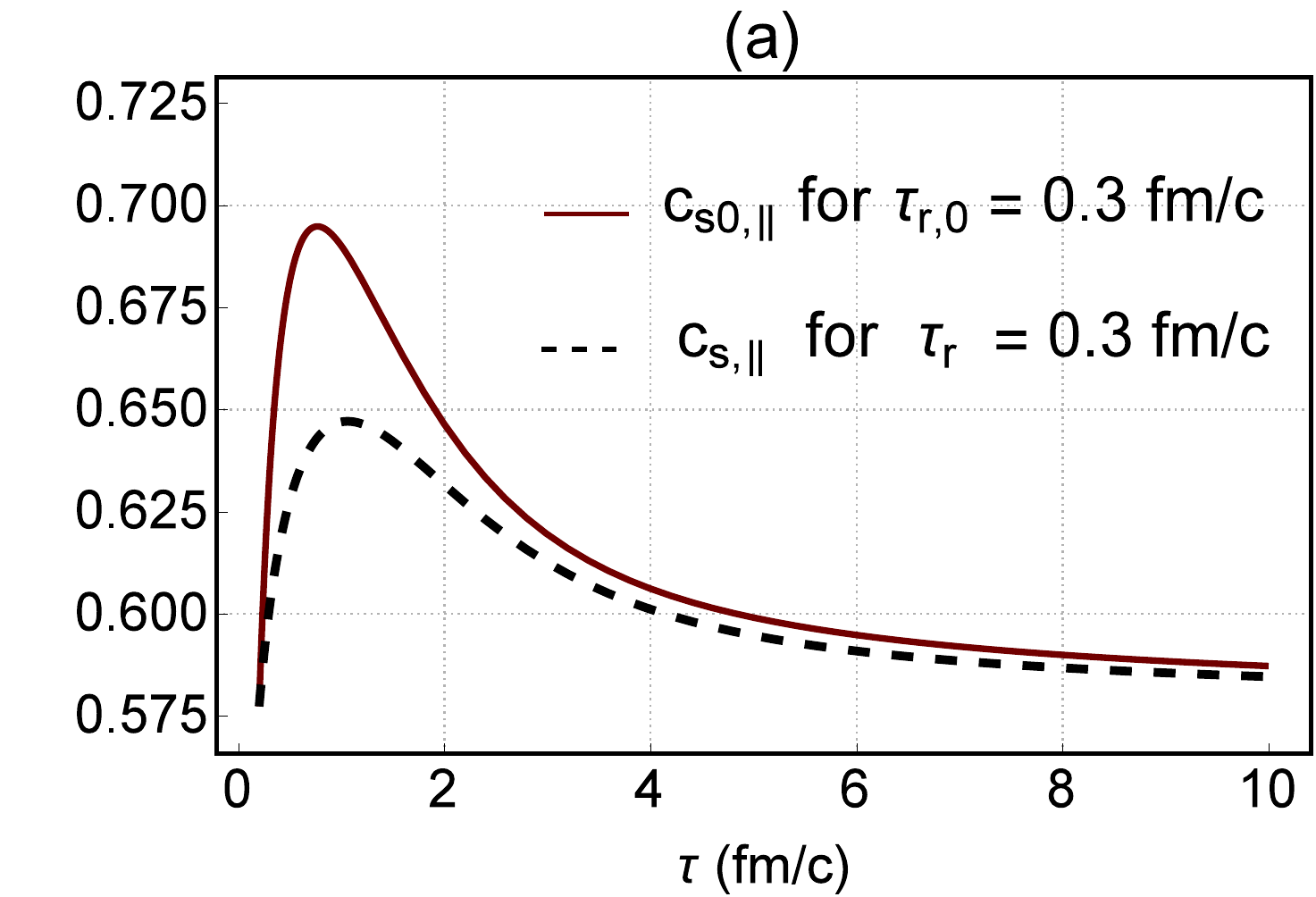}\hspace{0.8cm}
\includegraphics[width=8cm,height=6.5cm]{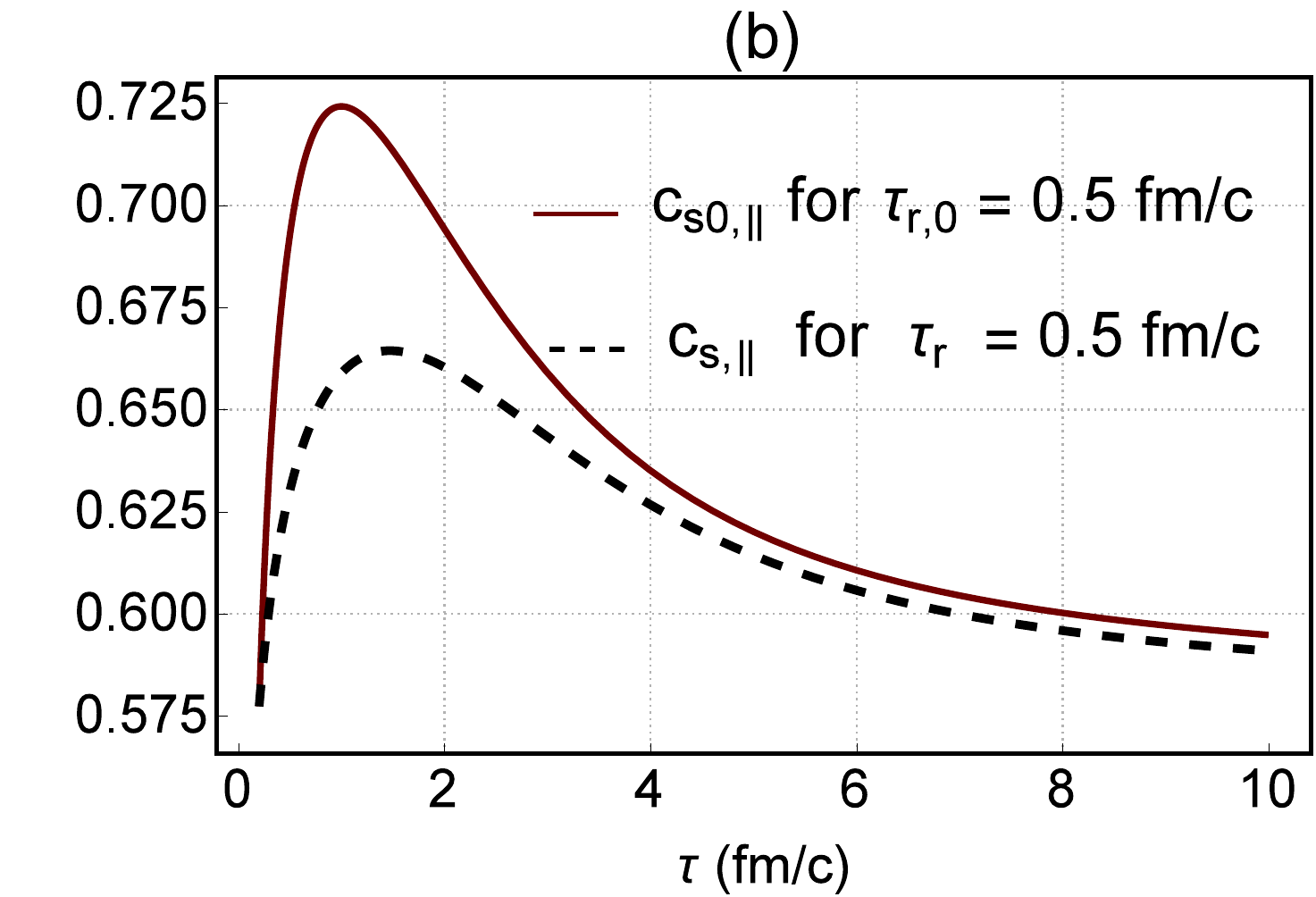}
\caption{(color online).  The $\tau$ dependence of the longitudinal speed of sound $c_{s0,\|}$ and $c_{s,\|}$ in a nondissipative (red solid curves) and dissipative (black dashed curves) fluid is plotted for the relaxation times $\tau_{r,0}$ and $\tau_{r}$ equal to $0.3$ fm/c (panel a) and $\tau_{r,0}$ and $\tau_{r}$ equal to $0.5$ fm/c (panel b). In contrast to $c_{s0,\|}$ and $c_{s,\|}$ from Fig. \ref{fig-7}, $c_{s0,\|}$ and $c_{s,\|}$ increase first to a maximum in the early stages after the collision and then decrease to a constant value at late time.}\label{fig-8}
\end{figure*}
\begin{figure*}[hbt]
\includegraphics[width=8cm,height=6.5cm]{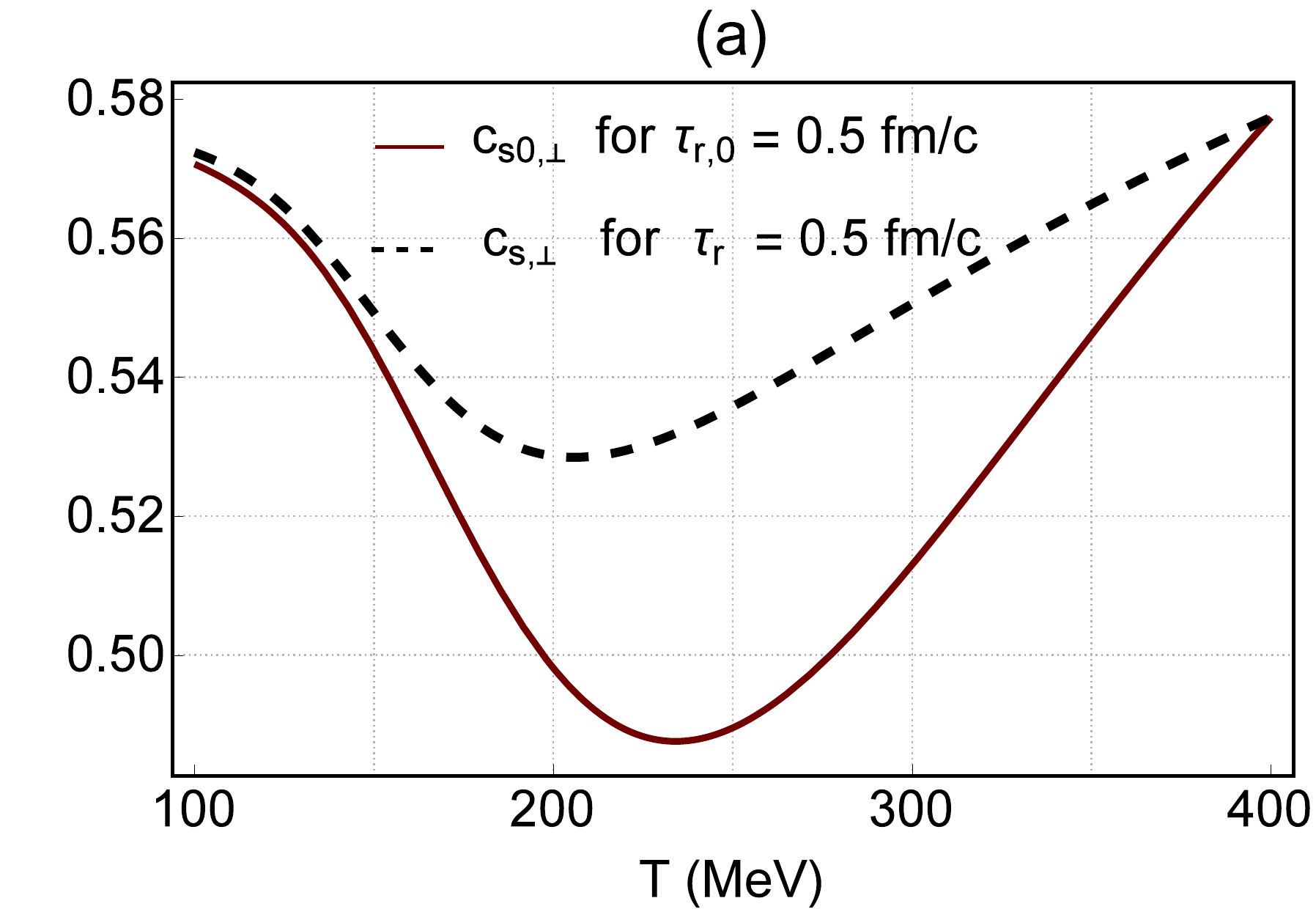}\hspace{0.8cm}
\includegraphics[width=8cm,height=6.5cm]{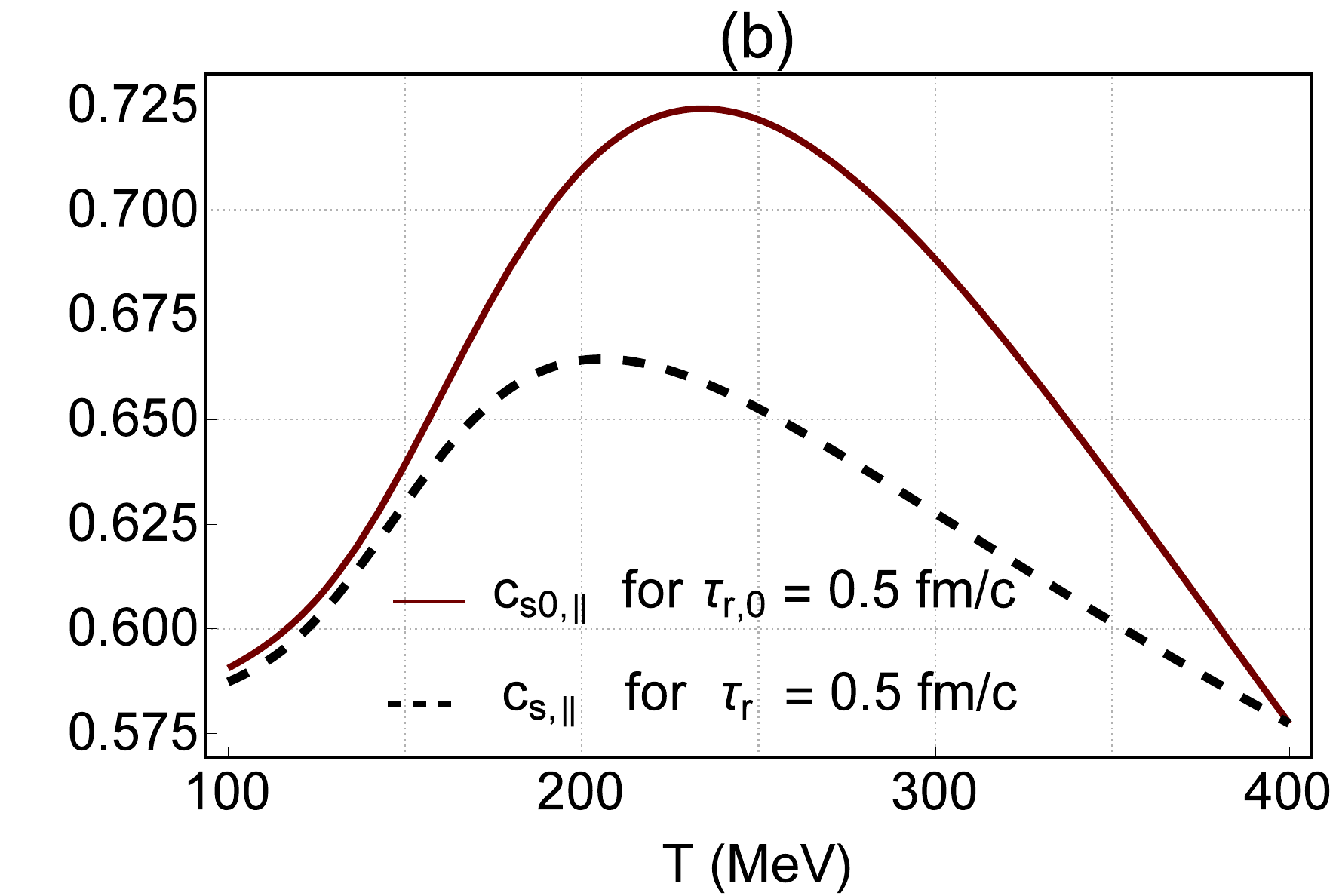}
\caption{(color online).  The $T$ dependence of $c_{s0,\perp}$ and $c_{s,\perp}$ (panel a) as well as $c_{s0,\|}$ and $c_{s,\|}$ (panel b) is plotted for the relaxation times $\tau_{r,0}$ (red solid curves) and $\tau_{r}$ (black dashed curves) equal to $0.5$ fm/c in a nondissipative and dissipative fluid.  Assuming that the QCD phase transition occurs at a critical temperature $200<T_{c}\sim 250$ MeV, the $c_{s,\perp}$ ($c_{c,\|}$) decreases (increases)  before the transition, and increases (decreases) after the transition as the fluid cools. }\label{fig-9}
\end{figure*}
\begin{figure*}[hbt]
\includegraphics[width=8cm,height=6.5cm]{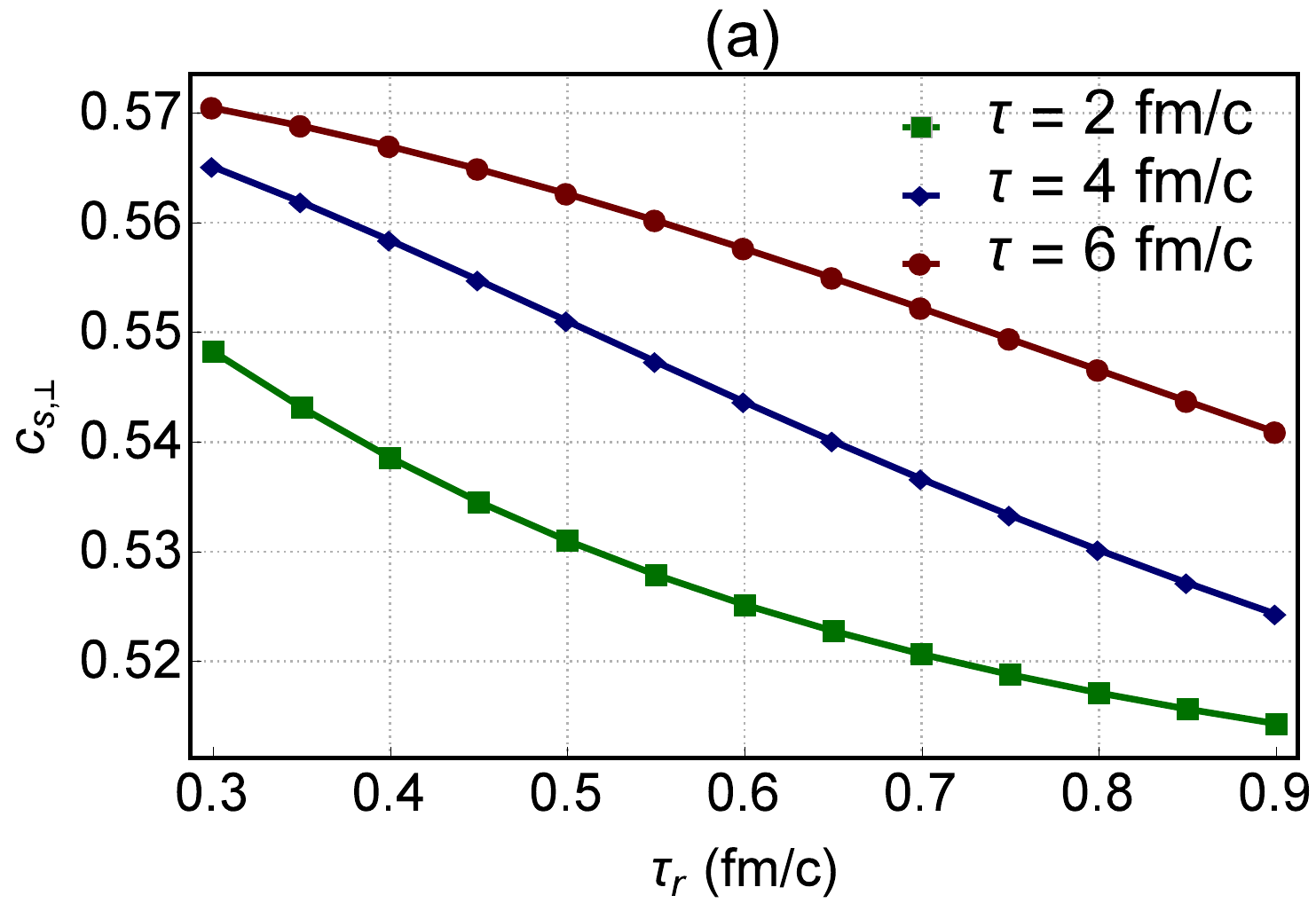}\hspace{0.8cm}
\includegraphics[width=8cm,height=6.5cm]{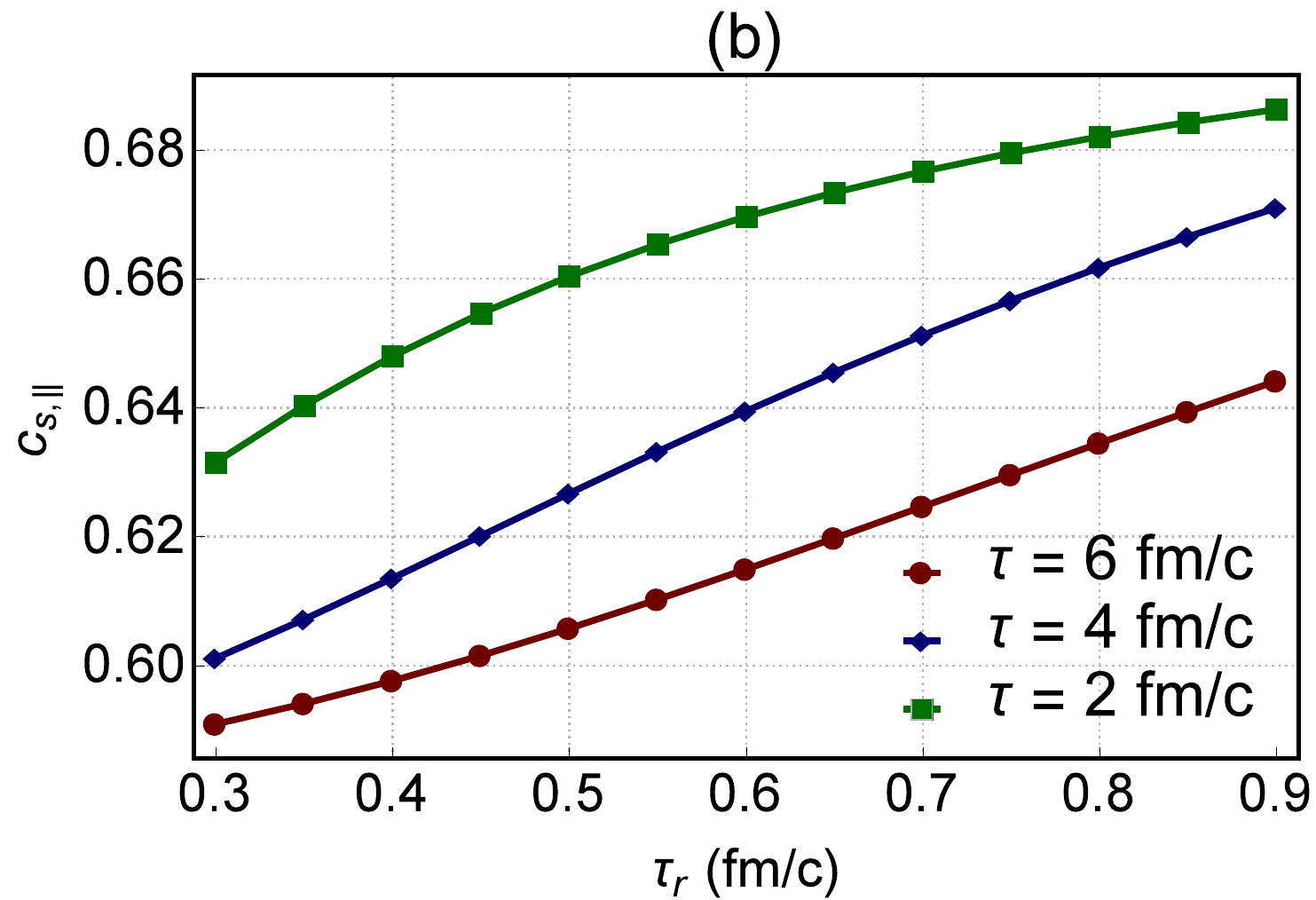}
\caption{(color online).  The $\tau_{r}$ dependence of $c_{s,\perp}$ (panel a) and $c_{s,\|}$ (panel b) for fixed proper times $\tau=2,4,6$ fm/c (red circles, blue rectangles and green squares). As expected from Figs. \ref{fig-7} and \ref{fig-8} for each fixed value of $\tau$,  $c_{s,\perp}$ ($c_{s,\|}$) decreases (increases) with increasing relaxation time $\tau_{r}$.  }\label{fig-10}
\end{figure*}
\begin{figure*}[hbt]
\includegraphics[width=8cm,height=6.5cm]{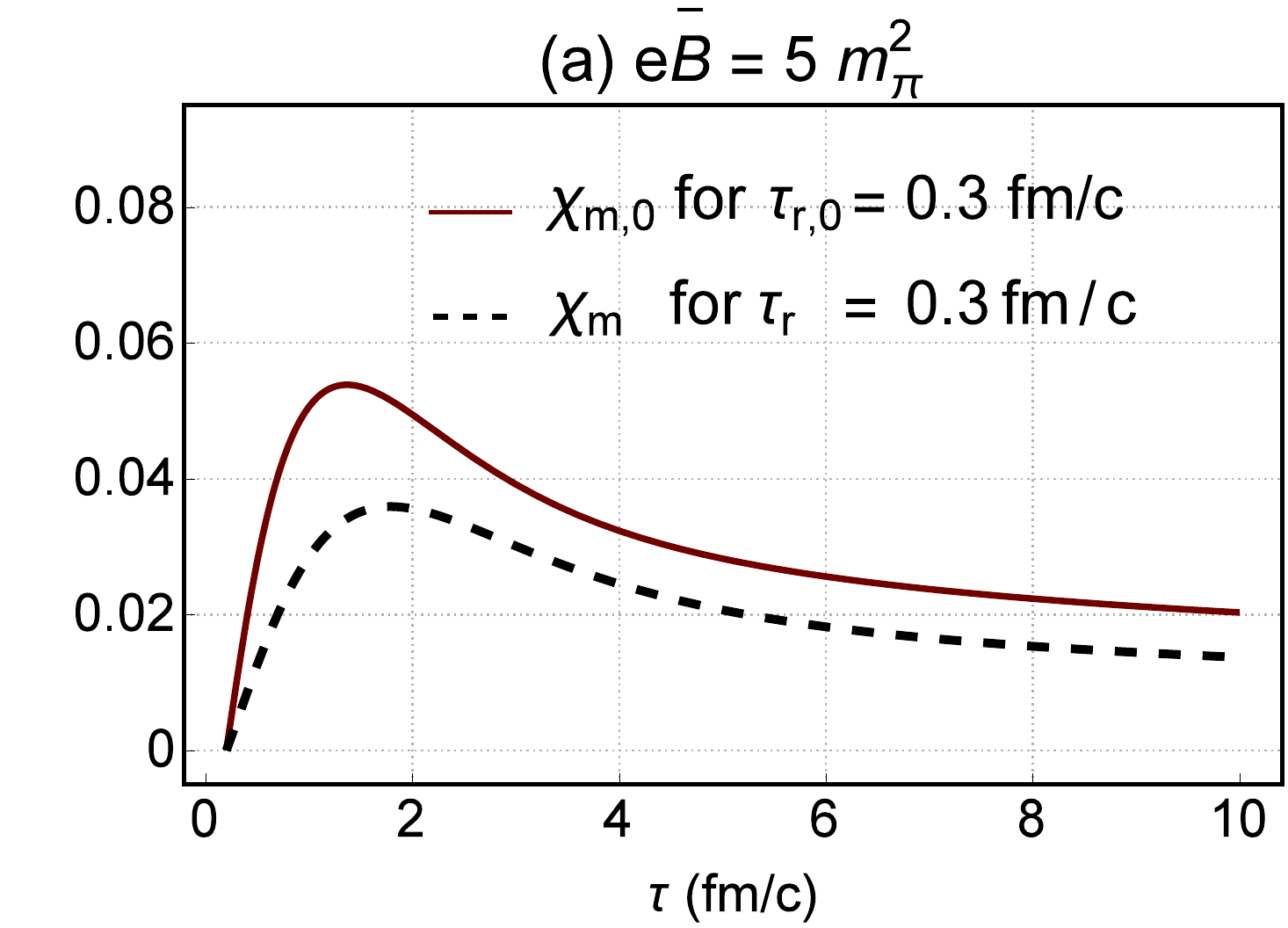}\hspace{0.8cm}
\includegraphics[width=8cm,height=6.5cm]{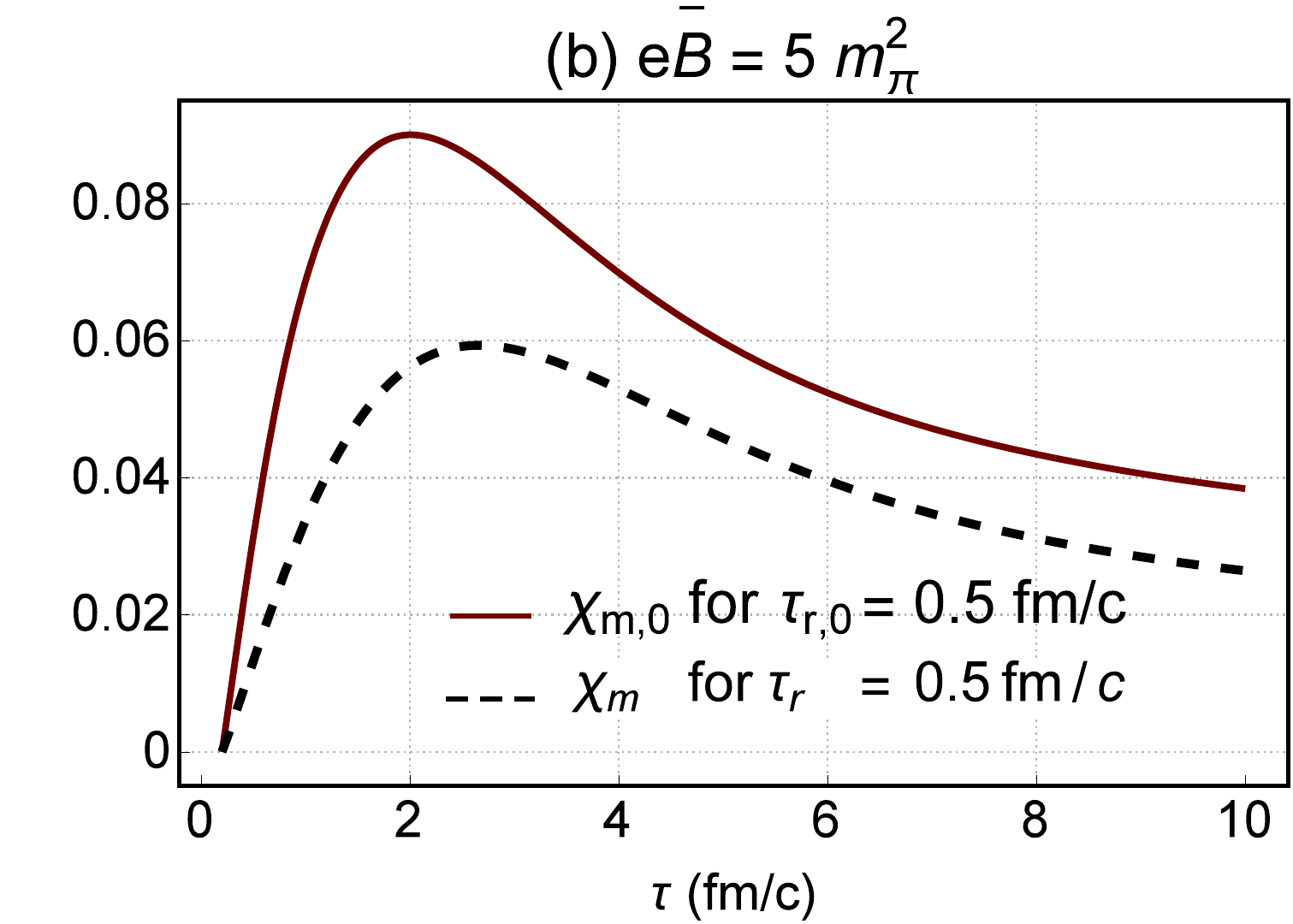}
\caption{(color online). The $\tau$ dependence of the magnetic susceptibility $\chi_{m,0}$ (nondissipative fluid) and $\chi_{m}$ (dissipative fluid) is plotted for relaxation times $\tau_{r,0}=0.3$ fm/c (panel a) and $\tau_{r,0}=0.5$ fm/c (panel b) (red solid curves) and $\tau_{r}=0.3$ fm/c  and $\tau_{r}=0.5$ fm/c (panel b) (black dashed curves) and $e\bar{B}=5 m_{\pi}^{2}$  with $m_{\pi}\sim 140$ MeV. It turns out that for larger values of relaxation time, the magnetic susceptibility is larger. Moreover, independent of $\tau_{r}$, a finite dissipation decreases the value of magnetic susceptibility, and the position of the maximum value of $\chi_{m}$ is shifted slightly to later times.  }\label{fig-11}
\end{figure*}
\begin{figure*}[hbt]
\includegraphics[width=8cm,height=6.5cm]{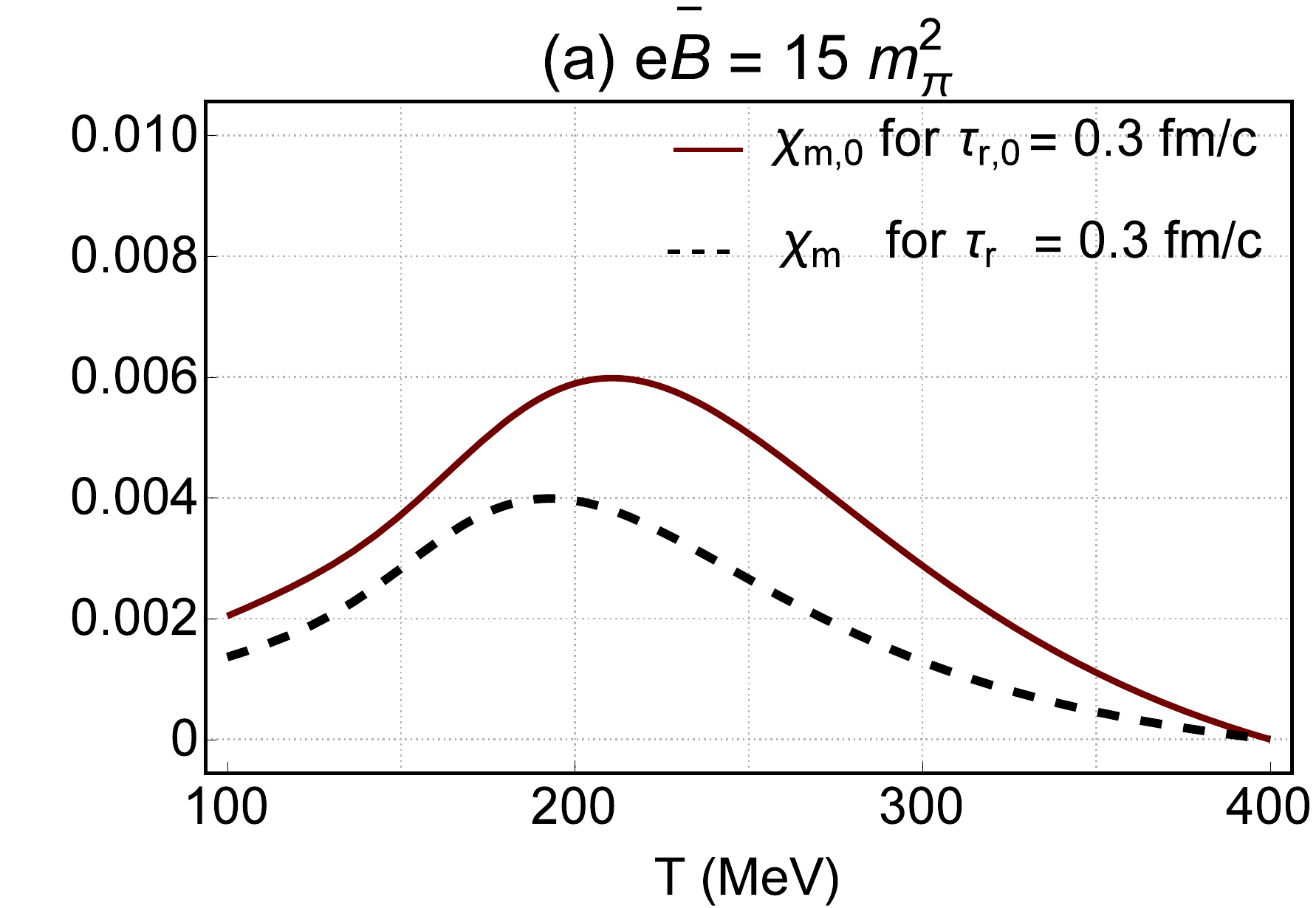}\hspace{0.8cm}
\includegraphics[width=8cm,height=6.5cm]{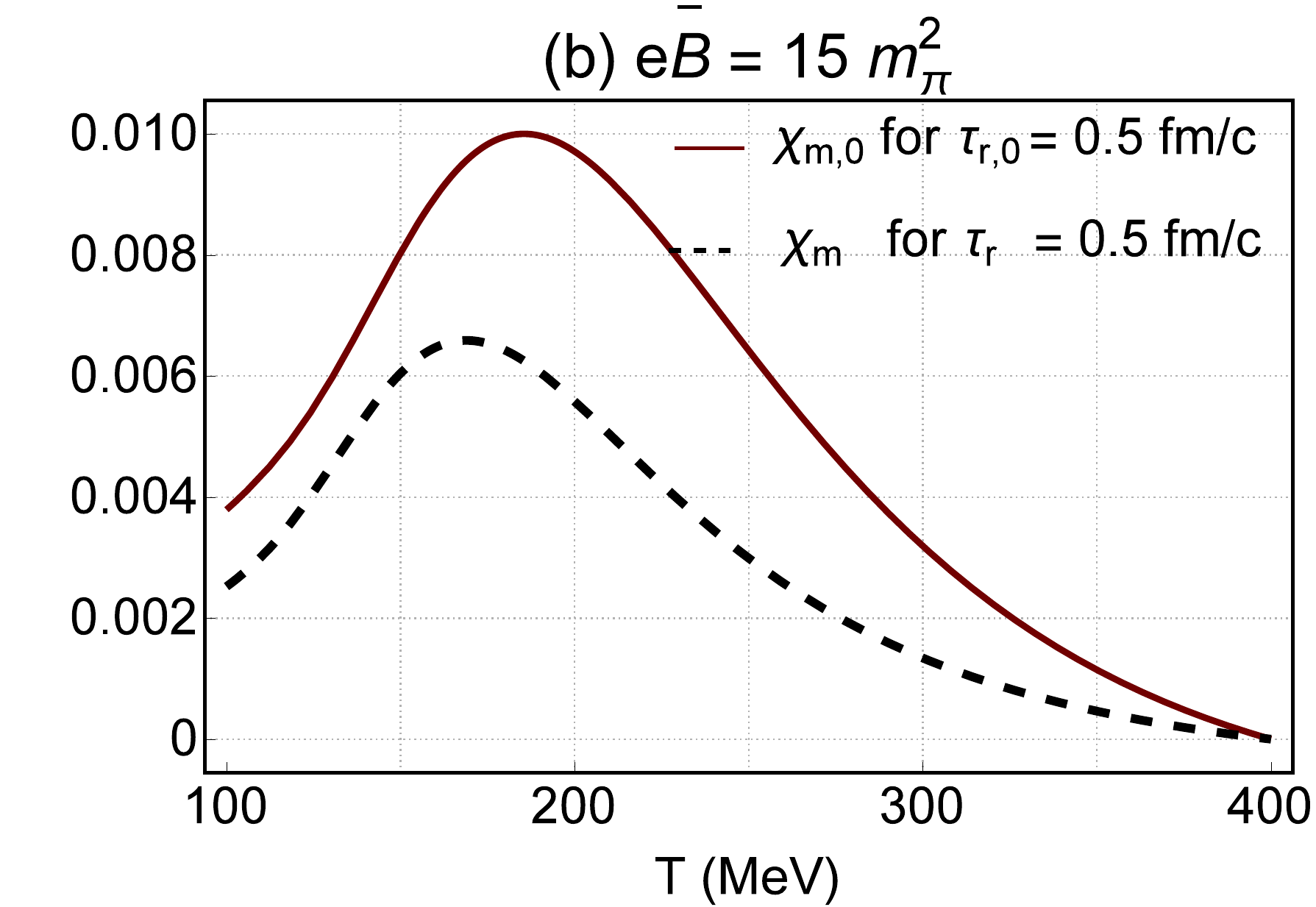}
\caption{(color online). The $T$ dependence of $\chi_{m,0}$ and $\chi_{m}$ is plotted for relaxation times $\tau_{r,0}$ (red solid curves) and $\tau_{r}$ (black dashed curves) equal to $0.5$ fm/c in a nondissipative and dissipative fluid. Here, $e\bar{B}=15 m_{\pi}^{2}$ with $m_{\pi}=140$ MeV. Assuming that the QCD phase transition occurs at a critical temperature $200<T_{c}\sim 250$ MeV, $\chi_{m,0}$ and $\chi_{m}$ increase before the transition ($T>T_{c}$), and decrease after the transition as the fluid cools to $T<T_{c}$.}\label{fig-12}
\end{figure*}
\begin{figure}[hbt]
\includegraphics[width=8cm,height=6.5cm]{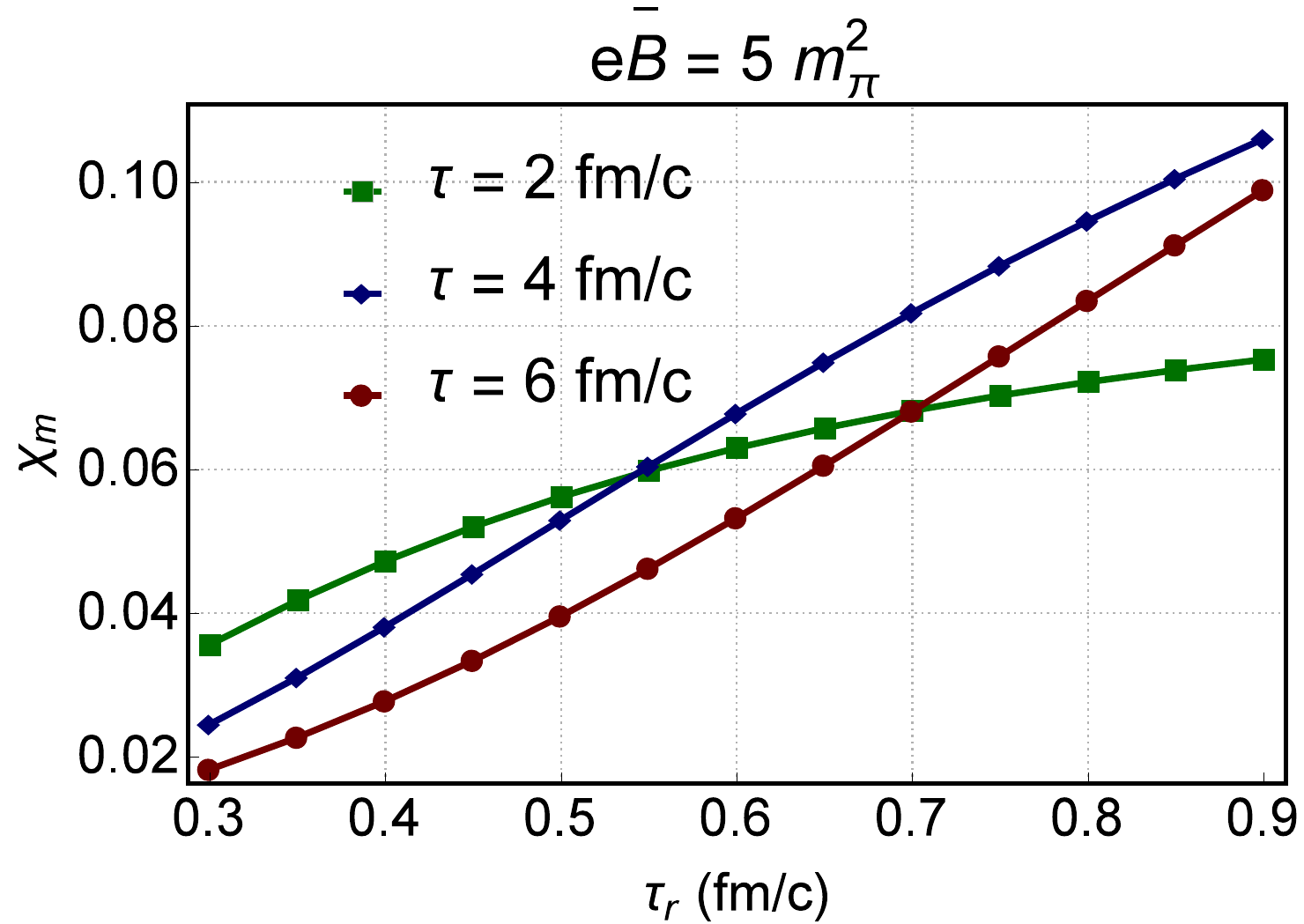}
\caption{(color online).  The $\tau_{r}$ dependence of $\chi_{m}$ is plotted for fixed proper times $\tau=2,4,6$ fm/c (red circles, blue rectangles and green squares) and $e\bar{B}=5 m_{\pi}^{2}$ with $m_{\pi}=140$ MeV. For fixed value of $\tau$, the magnetic susceptibility increases, in general, with increasing relaxation time $\tau_{r}$. }\label{fig-13}
\end{figure}
\begin{figure}[hbt]
\includegraphics[width=8cm,height=6cm]{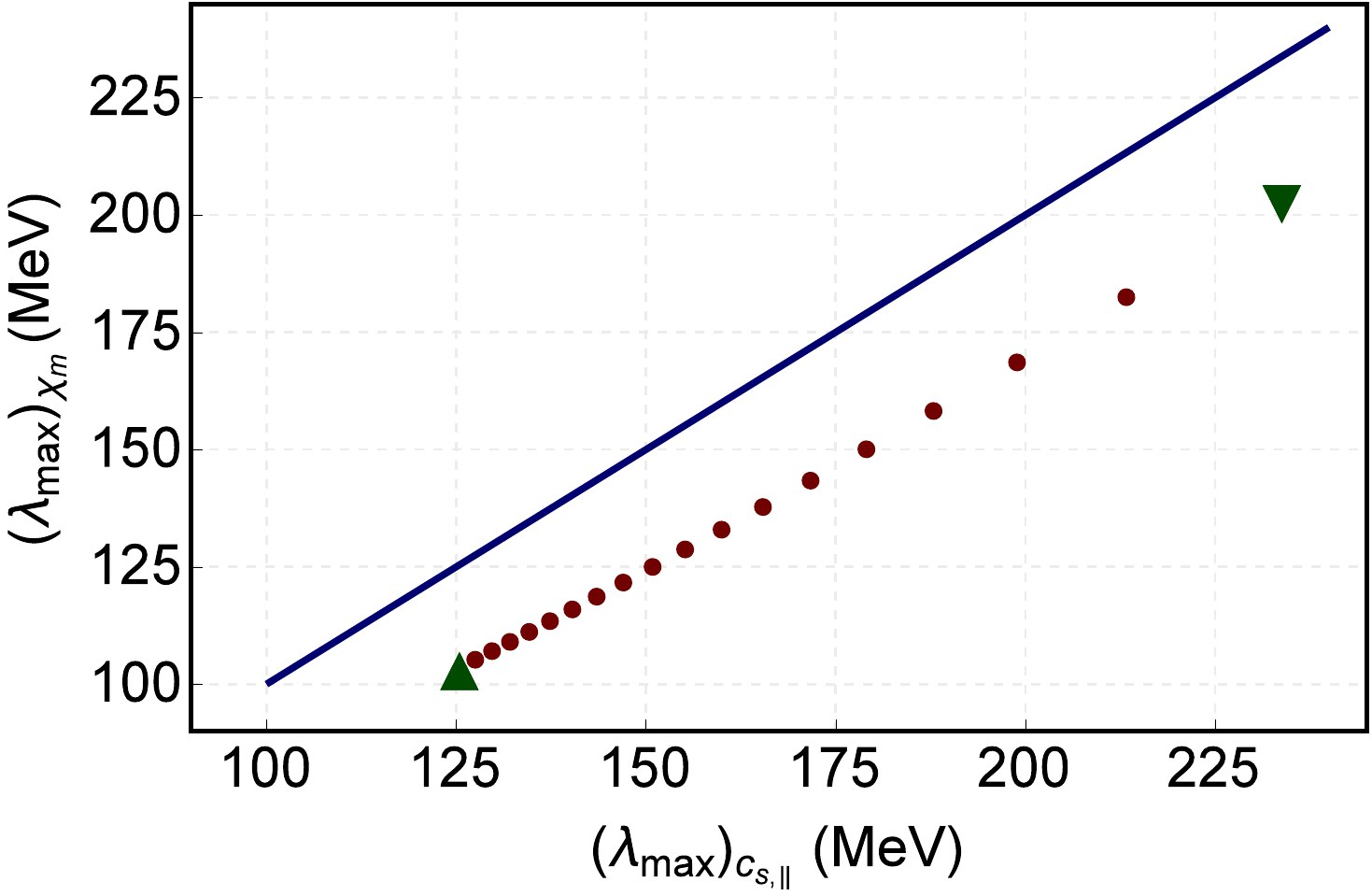}
\caption{(color online).
In this plot, $\lambda_{\mbox{\tiny{max}}}$ corresponding to $\chi_{m}$ is plotted versus $\lambda_{\mbox{\tiny{max}}}$ corresponding to $c_{s,\|}$ for a number of relaxation times $\tau_{r}=0.2,\cdots,2$ fm/c in $\Delta{\tau_r}=0.1$ fm/c steps.
The initial magnetic field $e\bar{B}=5 m_{\pi}^{2}$.
The green down-triangle at $((\lambda_{\mbox{\tiny{max}}})_{c_{s,\|}}, (\lambda_{\mbox{\tiny{max}}})_{\chi_{m}})=(234, 203)$ MeV corresponds to $\tau_{r}=0.2$ fm/c, and the green up-triangle at  $((\lambda_{\mbox{\tiny{max}}})_{c_{s,\|}}, (\lambda_{\mbox{\tiny{max}}})_{\chi_{m}})=(125,103)$ MeV corresponds to $\tau_{r}=2$ fm/c. The blue solid line is the line $(\lambda_{\mbox{\tiny{max}}})_{c_{s,\|}}= (\lambda_{\mbox{\tiny{max}}})_{\chi_{m}}$. The deviation of our result from this line indicates that $(\lambda_{\mbox{\tiny{max}}})_{c_{s,\|}}>(\lambda_{\mbox{\tiny{max}}})_{\chi_{m}}$ for each fixed value of $\tau_{r}\in [0.2,2]$ fm/c.  Moreover, it turns out that for larger relaxation times, $(\lambda_{\mbox{\tiny{max}}})_{c_{s,\|}}$ and $(\lambda_{\mbox{\tiny{max}}})_{\chi_{m}}$ are shifted to smaller temperatures.}\label{fig-14}%
\end{figure}
\begin{figure*}[hbt]
\includegraphics[width=8cm,height=6.5cm]{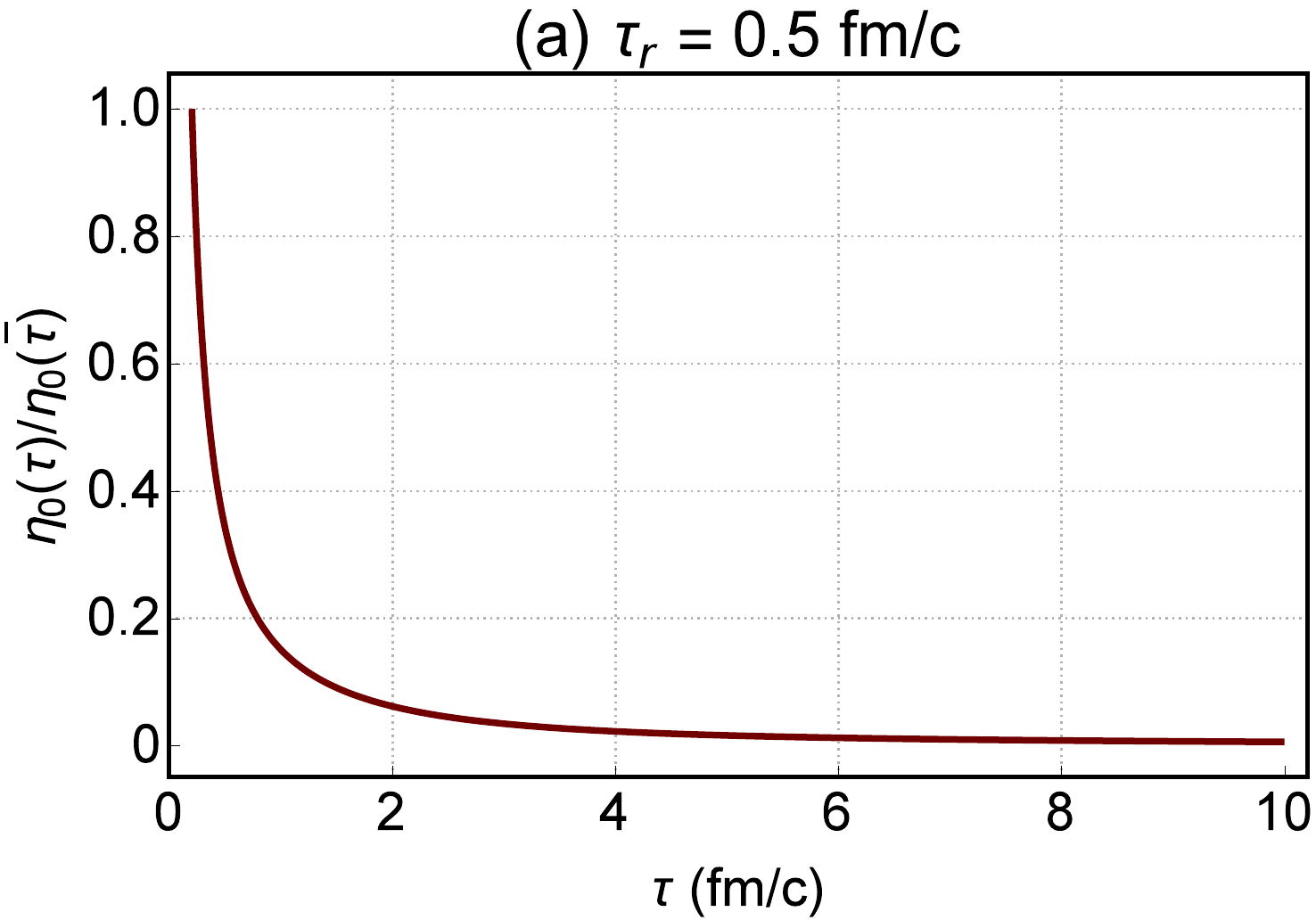}\hspace{0.8cm}
\includegraphics[width=8cm,height=6.5cm]{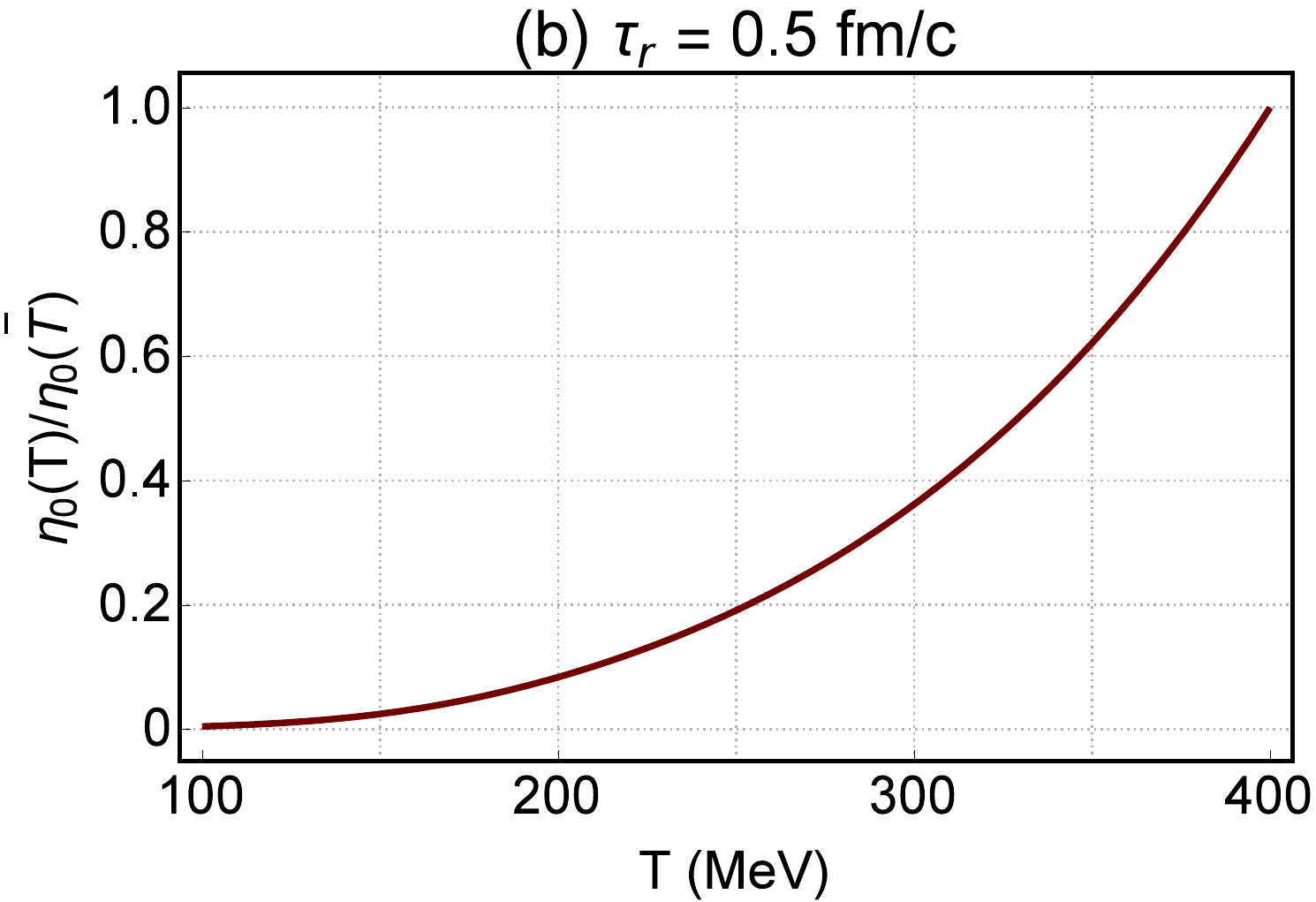}
\caption{(color online).   (a) The $\tau$ dependence of the shear viscosity $\eta_{0}(\tau)$ normalized by its value at the initial time $\bar{\tau}=0.2$ fm/c is plotted for $\tau_{r}=0.5$ fm/c. (b) The $T$ dependence of the shear viscosity $\eta_{0}(T)$ normalized by its value at the initial temperature $\bar{T}=400$ MeV is plotted for $\tau_{r}=0.5$ fm/c. }\label{fig-15}
\end{figure*}
\begin{figure*}[hbt]
\includegraphics[width=8cm,height=6.5cm]{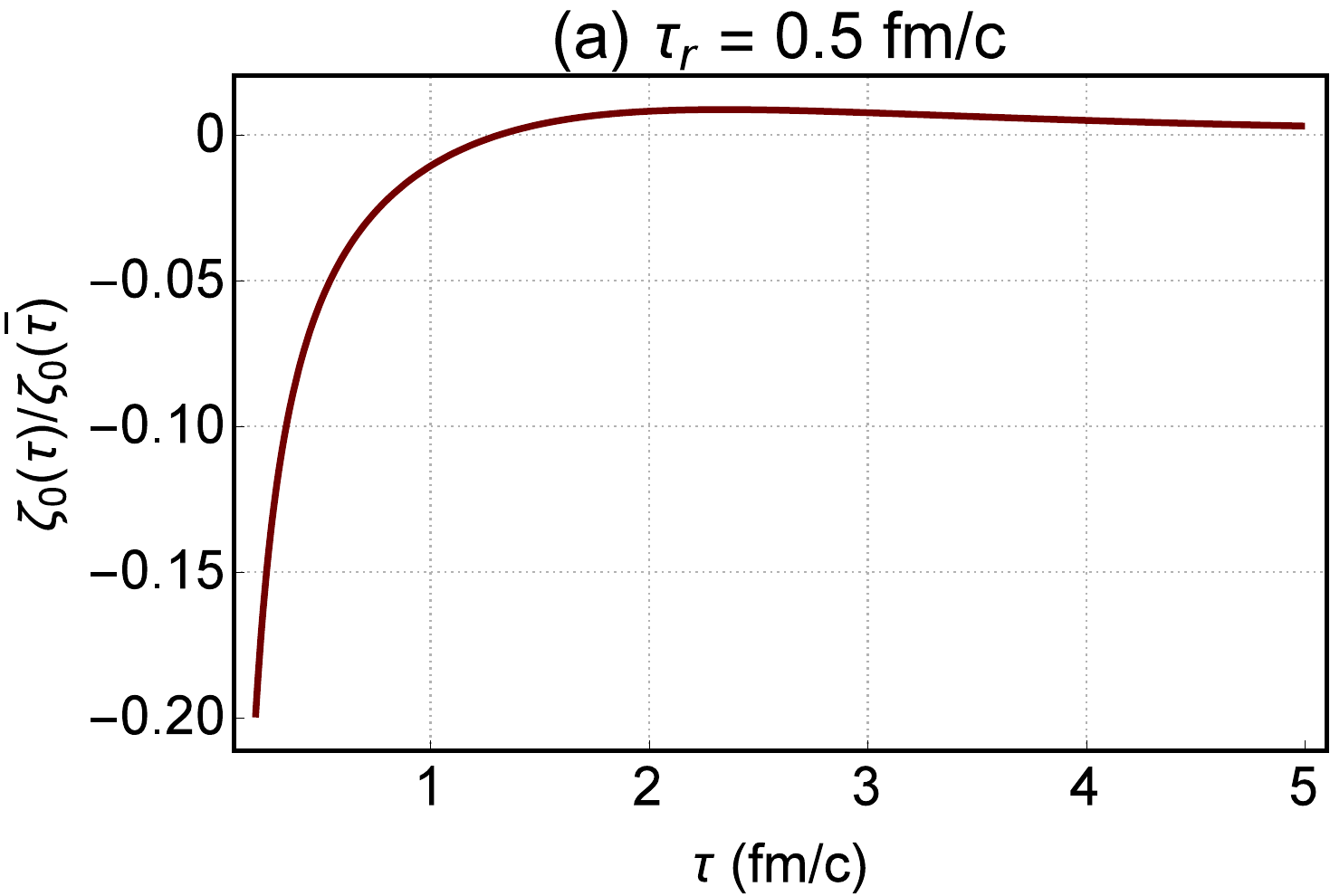}\hspace{0.8cm}
\includegraphics[width=8cm,height=6.5cm]{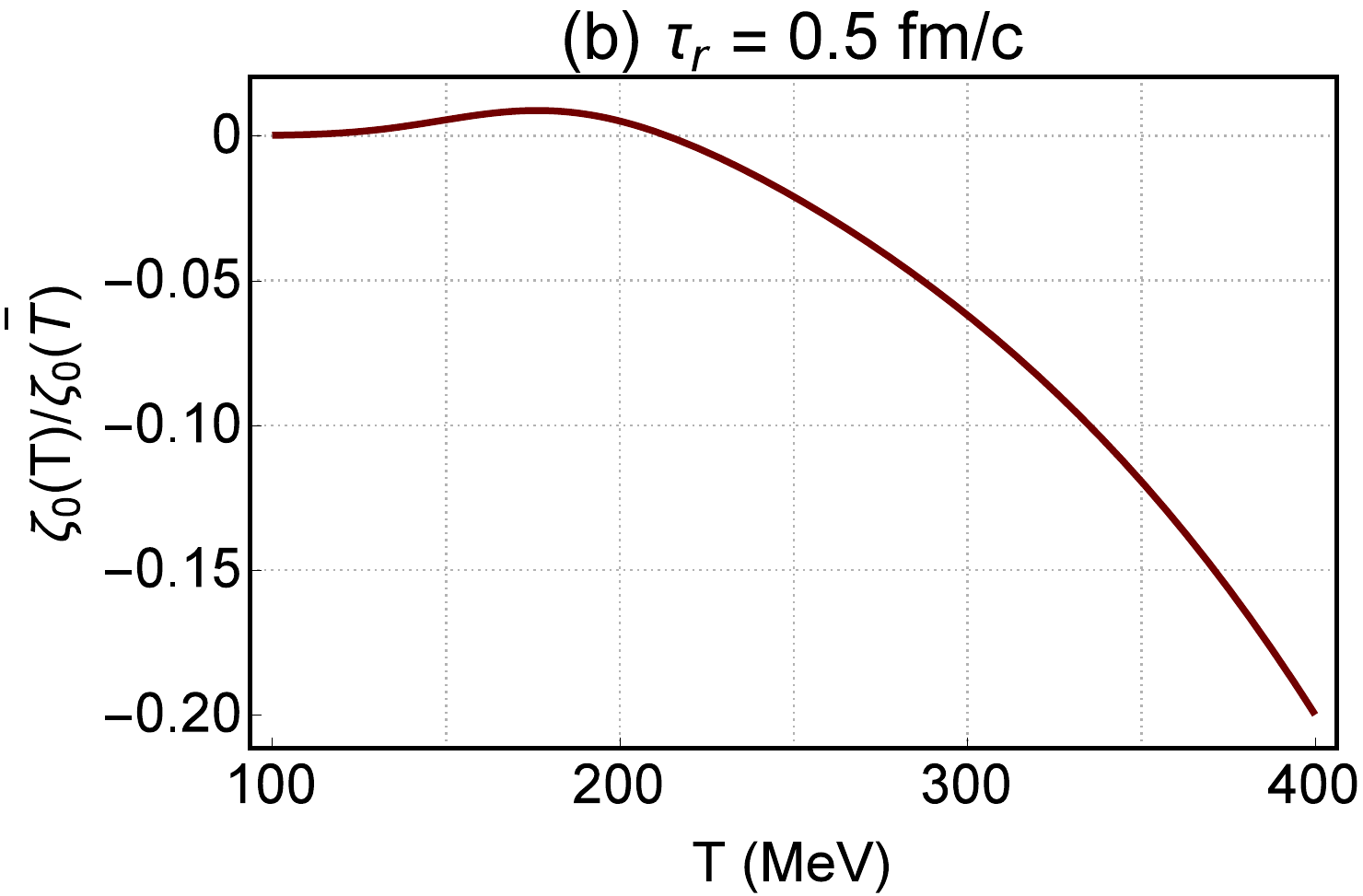}
\caption{(color online).  (a) The $\tau$ dependence of the bulk viscosity $\zeta_{0}(\tau)$ normalized by its value at the initial time $\bar{\tau}=0.2$ fm/c is plotted for $\tau_{r}=0.5$ fm/c. (b) The $T$ dependence of the shear viscosity $\zeta_{0}(T)$ normalized by its value at the initial temperature  $\bar{T}=400$ MeV is plotted for $\tau_{r}=0.5$ fm/c.  }\label{fig-16}
\end{figure*}
In this section, we present the numerical solutions of two sets of coupled differential equations \eqref{E27} and \eqref{E28} for nondissipative, as well as \eqref{D45} and \eqref{D54} for dissipative fluids. They are determined by using the analogy between the energy-momentum tensor in ideal MHD with nonzero magnetization and the energy-momentum tensor of an anisotropic fluid. The only free parameter in these differential equations is the relaxation time $\tau_{r,0}$ and $\tau_{r}$ appearing in \eqref{E27} and \eqref{D54} for nondissipative and dissipative QGP, respectively. Bearing in mind that the onset of hydrodynamical expansion occurs approximately at $\bar{\tau}\sim 0.2$ fm/c, we set, in what follows, $\tau_{r,0}$ and $\tau_{r}$ equal to $0.3,0.5$ fm/c, and compare the corresponding results for the proper time evolution of $\xi_0, \xi$ as well as $\lambda_{0},\lambda$, separately. These results are then used to determine the evolution of longitudinal and transverse pressures $p_{\|}$ and $p_{\perp}$, that, together with the expressions for the energy density $\epsilon$, lead to the transverse and longitudinal speeds of sound, $c_{s,i}=p_i/\epsilon, i=\|,\perp$ in nondissipative and dissipative cases. Moreover, combining these results with the corresponding results to the magnetization $M_{0}$ and $M$, the evolution of the magnetic susceptibilities $\chi_{m,0}$ and $\chi_{m}$ is determined in these two cases for fixed values of $e\bar{B}=5m_{\pi}^{2}$ and $e\bar{B}=15 m_{\pi}^{2}$. These are the values of magnetic fields that are believed to be created in noncentral HICs at the RHIC and LHC, respectively. In order to compare our results with the results arising from lattice QCD, we use the proper time evolution of the temperature \eqref{N29}, and plot $\chi_{m,0}$ and $\chi_{m}$ as a function of $T$. The corresponding results are in good agreement with lattice QCD results from \cite{delia2013} in the regime $T<T_{c}$, where $T_{c}\simeq 200$ MeV is the critical temperature of the QCD phase transition. We discuss the apparent discrepancy for $T>T_{c}$, and argue that it may lie on the effect of dynamical magnetic field created in HICs. We perform the same analysis for transverse and longitudinal speeds of sound, and present the temperature dependence of these velocities in the interval $T\in \{100,400\}$ MeV.
To study the effect of the relaxation time $\tau_{r}$ on the evolution of thermodynamic quantities, we plot the $\tau_{r}$ dependence of $\chi_{m}$ for fixed $e\bar{B}=5 m_{\pi}^{2}$, and $c_{s,i}=p_i/\epsilon, i=\|,\perp$ at fixed proper times $\tau=2,4,6$ fm/c. We also focus on the position of the maxima appearing in $\chi_{m}(T)$ and $c_{s,\|}(T)$, and study their dependence on the relaxation time $\tau_r$.
We finally present the $\tau$ as well as $T$ dependence of shear and bulk viscosities $\eta_{0}$ and $\zeta_{0}$. As aforementioned, the latter is given as a linear combination of $\tilde{\zeta}_{0}$ and $\alpha_{0}$, whose analytical expressions are presented in the previous section.
\par
To start, let us consider the differential equations \eqref{E27} [\eqref{D45}] and \eqref{E28} [\eqref{D54}], whose solution yields the anisotropy parameter and the effective temperature $\xi_{0}$ and $\lambda_{0}$ ($\xi$ and $\lambda$) for nondissipative (dissipative) magnetized fluid.  In Fig. \ref{fig-2}, the proper time evolution of $\xi_{0}$ and $\xi$ is plotted for $\tau_{r,0}$ (blue solid curves) and $\tau_{r}$ (black dashed curves) equal to $0.3$ fm/c [Fig. \ref{fig-2}(a)] and $0.5$ fm/c [Fig. \ref{fig-2}(b)]. To solve the corresponding differential equations, the initial values of $\xi_{0},\xi$ and $\lambda_{0},\lambda$ in the initial time $\bar{\tau}=0.2$ fm/c are chosen to be $\xi_{0}(\bar{\tau})=\xi(\bar{\tau})=10^{-7}$ and $\lambda_{0}(\bar{\tau})=\lambda(\bar{\tau})=400$ MeV. The comparison between $\xi_{0}$ and $\xi$ for each fixed $\tau_{r,0}$ and $\tau_{r}$ shows that, qualitatively, nonvanishing dissipation does not change the $\tau$ dependence of the anisotropy function $\xi_{0}$ and $\xi$. They sharply decrease in the early stages of the evolution, and then increase, and approach asymptotically a fixed value. The corresponding minimum of $\xi_{0}$ is however deeper. This specific feature, which does not obviously depend on the choice of the relaxation time, may show that the effect of pressure anisotropies arising from the magnetization of the fluid is diminished by the dissipation.
The two plots demonstrated in Fig. \ref{fig-2} have the same scale. It is thus possible to compare $\xi_{0}$ (blue solid curves) for $\tau_{r,0}=0.3$ fm/c in Fig. \ref{fig-2}(a) and $\xi_{0}$ for $\tau_{r,0}=0.5$ fm/c in Fig. \ref{fig-2}(b). As it turns out, the minima of $\xi_{0}$ and $\xi$ become deeper with increasing $\tau_{r,0}$ and $\tau_{r}$. Moreover, $\xi_{0}$ and $\xi$ need longer to reach their asymptotic value $\approx -0.1$ at $\tau\simeq 10$ fm/c.
\par
Using the same initial values for $\xi_0,\xi$ and $\lambda_0,\lambda$ at $\bar{\tau}=0.2$ fm/c, the $\tau$ dependence of the effective temperature $\lambda_0$ (nondissipative case) and $\lambda$ (dissipative case) is plotted in Figs. \ref{fig-3}(a) and \ref{fig-3}(b). It turns out that neither for small value of $\tau_{r,0}$ and $\tau_{r}$ equal to $0.3$ fm/c (blue solid curves) nor for larger value of $\tau_{r,0}$ and $\tau_{r}$ equal to $0.5$ fm/c (black dashed curves), the dissipation does affect the effective temperature significantly. In Fig. \ref{fig-4}, we compare the temperature $T=\bar{T}\left(\bar{\tau}/\tau\right)^{1/3}$ from \eqref{N29} (green dashed curve),  with the effective temperatures, $\lambda_{0}$ and $\lambda$ for $\tau_{r,0}=\tau_{r}=0.5$ fm/c. Except in the interval $\tau\in [\sim 0.5,\sim 4]$ fm/c, the dynamics of $T,\lambda_{0}$, and $\lambda$ coincides. Bearing in mind that $T$ is the temperature for an isotropic ideal fluid, the result presented in Fig. \ref{fig-4} indicates that neither the magnetization nor the dissipation affect the late time dynamics of the temperature in a magnetized fluid.
\par
In Fig. \ref{fig-5}, the evolution of the ratio $p_{0,\perp}/p_{0,\|}$  and $p_{b,\perp}/p_{b,\|}$ for a nondissipative and a dissipative magnetized fluid is plotted for relaxation times  $\tau_{r,0}$ (red solid curves) and $\tau_{r}$ (black dashed curves) equal to  $0.3$ fm/c [Fig. \ref{fig-5}(a)] and $0.5$ fm/c [Fig. \ref{fig-5}(b)]. Here, $p_{0}$ and $M_{0}$ in $p_{0,\perp}=p_{0,\|}-BM_{0}$ with $p_{0,\|}=p_{0}$ are given in \eqref{E12} in terms of $\xi_{0}$ and $\lambda_{0}$, and $p_{b}$ as well as $M_{b}$ in $p_{b,\perp}=p_{b,\|}-BM_{b}$ with $p_{b,\|}=p_{b}$ are given in \eqref{D47} in terms of $\xi$ and $\lambda$. As it turns out, independent of the choice of the relaxation time, $p_{0,\perp}/p_{0,\|}$  and $p_{b,\perp}/p_{b,\|}$ decrease abruptly at the beginning of the evolution. After reaching a minimum at $\tau_{\mbox{\tiny{min}}}$, they increase with increasing $\tau$, and become almost constant at a certain proper time $\tau_{c}$. The values of $\tau_{\mbox{\tiny{min}}}$ and $\tau_{c}$, as well as the values of $p_{0,\perp}/p_{0,\|}$  and $p_{b,\perp}/p_{b,\|}$ at these times, depend on the value of the relaxation time $\tau_{r,0}$ and $\tau_{r}$. For larger values of $\tau_{r,0}$ and $\tau_{r}$, the minima are deeper, and $\tau_{c}$ is larger. A comparison between the evolution of $p_{0,\perp}/p_{0,\|}$  and $p_{b,\perp}/p_{b,\|}$ with the evolution of $\xi_{0}$ and $\xi$ shows that the dynamics of these anisotropy parameters is strongly reflected in the dynamics of the ratio of transverse over longitudinal pressures in a magnetized fluid with and without dissipation.
Let us notice, at this stage, that the evolution of $p_{0,\perp}/p_{0,\|}$  and $p_{b,\perp}/p_{b,\|}$ in Fig. \ref{fig-5} is similar to the evolution of $p_{L}/p_{T}$, e.g. from Fig. 3 in \cite{strickland2017}, where the difference between the longitudinal and transverse pressure, $p_{L}$ and $p_{T}$, arises from the nonvanishing ratio of the shear viscosity to the entropy density of the fluid, $\eta/s$. In the case discussed in the present paper, however, the difference between $p_{0,\|}$ and $p_{0,\perp}$ arises because of the finite magnetization of the QGP, which has to be considered as an additional source, apart from dissipative effects, for the pressure anisotropy of the QGP in the early stages of HICs. Here, it is believed that large magnetic fields are created in noncentral collisions. The magnetization of the medium and its evolution thus plays an important role in the isotropization process which occurs in the early stages of the collision. A comparison with $p_{b,\|}$ and $p_{b,\perp}$ shows that, similar to $p_{0,\perp}$ and  $p_{0,\|}$, $p_{\perp}\lesssim p_{\|}$.  This result coincides with the results presented in \cite{strickland2010, strickland2017}.\footnote{Let us remind that, according to our descriptions in Sec. \ref{sec3}, the subscripts $\{\|,\perp\}$ in the present paper correspond to $\{T,L\}$ used in \cite{strickland2010, strickland2017}.}
\par
In Fig. \ref{fig-6}, the $\tau$ dependence of the energy density $\epsilon$ (green squares) of an isotropic ideal fluid from \eqref{N28}, $\epsilon_{0}$ (black solid curve)  of a magnetized nondissipative fluid from \eqref{E12}, and $\epsilon_{b}$ (red circles)  of a magnetized dissipative fluid from \eqref{D43} is plotted for relaxation times $\tau_{r,0}$ and $\tau_{r}$ equal to $0.3$ fm/c. The initial value of these energy densities at the initial proper time $\bar{\tau}=0.2$ fm/c is taken to be $\bar{\epsilon}=\epsilon(\bar{\tau})=1$ GeVfm$^{-3}$. As  expected from \eqref{E24} and \eqref{D42}, we have $\epsilon=\epsilon_0=\epsilon_b$.  In all these cases, the energy density decreases very fast with increasing $\tau$.
\par
Combining the results for $p_{0,\perp}, p_{b/\perp}$ and  $p_{0,\|},p_{b,\|}$ as well as $\epsilon_{0}$ and $\epsilon_{b}$, the speed of sound in the transverse and longitudinal directions with respect to the magnetic field,
\begin{eqnarray}\label{A1}
c_{s0,i}\equiv \frac{p_{0,i}}{\epsilon_{0}},\quad\mbox{and}\quad c_{s,i}\equiv \frac{p_{b,i}}{\epsilon_{b}},\quad i=\perp,\|,
\end{eqnarray}
for a nondissipative (subscript $0$) and a dissipative fluid (subscript $b$) is determined. In Figs. \ref{fig-7} and \ref{fig-8}, the $\tau$ dependence of $c_{s0,i}$ and $c_{s,i}$ with $i=\perp,\|$ in a nondissipative (red solid curves) and dissipative (black dashed curves) fluid is plotted for the relaxation times $\tau_{r,0}$ and $\tau_{r}$ equal to $0.3$ fm/c [Figs. \ref{fig-7}(a) and \ref{fig-8}(a)] and $0.5$ fm/c [Figs. \ref{fig-7}(b) and \ref{fig-8}(b)]. In contrast to transverse speed of sound from Fig. \ref{fig-7}, whose $\tau$ dependence is similar to the $\tau$ dependence of $\xi_{0}$ and $\xi$ from Fig.  \ref{fig-2} as well as $p_{0,\perp}/p_{0,\|}$ and $p_{b,\perp}/p_{b,\|}$ from Fig. \ref{fig-5}, nondissipative and dissipative longitudinal speed of sound $c_{s0,\|}$ and $c_{s,\|}$ increase very fast at the beginning of the expansion to a maximum at $\tau_{\mbox{\tiny{max}}}$, and then decrease slowly to certain constant values at $\tau_{c}$ (see Fig. \ref{fig-8}). The values of $\tau_{\mbox{\tiny{max}}}$ and $\tau_{c}$ depend on the relaxation times $\tau_{r,0}$ and $\tau_{r}$. The same is also true for the proper time at which $c_{s0,\perp}$ and $c_{s,\perp}$ reach their minima, and become approximately constant. Moreover, as it turns out, for larger values of $\tau_{r,0}$ and $\tau_{r}$ the minima (maxima) of $c_{s0,\perp}$ and $c_{s,\perp}$ ($c_{s0,\|}$ and $c_{s,\|}$) are deeper (higher).  A comparison between the results presented in Figs. \ref{fig-7} and \ref{fig-8} shows that the transverse speed of sound is in general smaller than the longitudinal speed of sound.
Replacing, at this stage, the proper time $\tau$ arising in the corresponding expressions to $c_{s0,i}$ and $c_{s,i}$, $i=\perp,\|$, with $\tau=\bar{\tau}\left(\bar{T}/T\right)^{3}$ from \eqref{N29} with the initial time $\bar{\tau}=0.2$ fm/c and the initial temperature $\bar{T}=400$ MeV,\footnote{In \eqref{N29}, $\kappa=3$ is chosen.} we arrive at the $T$ dependence of $c_{s0,i}$ and $c_{s,i}$, $i=\perp,\|$. This is demonstrated in Fig. \ref{fig-9}. In Fig. \ref{fig-9}(a) [Fig. \ref{fig-9}(b)] the transverse (longitudinal) speed of sound is plotted for the relaxation times $\tau_{r,0}$ (red solid curves) and $\tau_{r}$ (black dashed curves) equal to $0.5$ fm/c. Assuming that the QCD phase transition occurs at a critical temperature $T_{c}\sim 200$ MeV, and focusing on the results for $c_{s,i}$, $i=\perp,\|$ of a dissipative fluid (black dashed curves) in Figs. \ref{fig-9}(a) and \ref{fig-9}(b), it turns out that the transverse (longitudinal) speed of sound decreases (increases) before the transition, and increases (decreases) after the transition as the fluid slowly cools.  We have also plotted the $\lambda_{0}$ ($\lambda$) dependence of $c_{s0,\perp}$ and $c_{s0,\|}$ ($c_{s,\perp}$ and $c_{s,\|}$), with $\lambda_{0}$ and $\lambda$ being the effective temperatures. The resulting plots do not differ qualitatively from the plots demonstrated in Fig. \ref{fig-9}. The only difference is the position of the minimum (maximum) appearing for $c_{s0,\perp}$ and $c_{s,\perp}$ ($c_{s0,\|}$ and $c_{s,\|}$) in Fig. \ref{fig-9}(a) [Fig. \ref{fig-9}(b)], which is shifted to smaller value of effective temperatures in nondissipative and dissipative cases. This is because of the difference between $T$ and $\lambda_{0}$ and $\lambda$ at early $\tau\in[\sim 0.5,\sim 4]$ fm/c, where $T\sim 150$-$200$ MeV  [see Fig. \ref{fig-4}]. The question whether the position of the maximum appearing in $c_{s,\|}$ for a fixed relaxation time is related to the temperature of the QCD phase transition cannot be answered at this stage (see in the description of Fig. \ref{fig-14}, for more details).
\par
To study the effect of increasing relaxation time on the qualitative behavior of the speed of sound, the $\tau_{r}$ dependence of $c_{s,\perp}$ and $c_{s,\|}$ is plotted in Figs. \ref{fig-10}(a) and \ref{fig-10}(b) for fixed proper times $\tau=2,4,6$ fm/c (red circles, blue rectangles and green squares). As expected from Figs. \ref{fig-7} and \ref{fig-8}, for each fixed value of $\tau$,  $c_{s,\perp}$ ($c_{s,\|}$) decreases (increases) with increasing relaxation time $\tau_{r}$.
\par
Using, at this stage, the combinations $BM_{0}$ from \eqref{E11} for the nondissipative case and $BM_{b}$ from \eqref{D47} for the dissipative case, together with relations $M_0=\chi_{m,0}B$ and $M_b=\chi_m B$ for these two cases, the $\tau$ dependence of magnetic susceptibilities $\chi_{m,0}$ and $\chi_{m}$ are determined. In Fig. \ref{fig-11}, the $\tau$ dependence of $\chi_{m,0}$ and $\chi_{m}$ is plotted for relaxation times $\tau_{r,0}$ and $\tau_{r}$ equal to $0.3$ fm/c (red solid curves) and $0.5$ fm/c (black dashed curves) and an initial magnetic field $e\bar{B}=5 m_{\pi}^{2}$  with $m_{\pi}\sim 140$ MeV. It turns out that for larger values of relaxation time, the magnetic susceptibility is larger.  Moreover, independent of $\tau_{r}$, a finite dissipation diminishes the value of the magnetic susceptibility. According to the results demonstrated in Fig. \ref{fig-11}, the magnetic susceptibility increases with a relatively large slope at early stages of the expansion. It reaches a maximum at a certain $\tau_{\mbox{\tiny{max}}}$, and decreases slowly to a certain constant value at $\tau\sim 0$ fm/c. For larger values of relaxation times $\tau_{r}$, the position of $\tau_{\mbox{\tiny{max}}}$ is slightly shifted to larger $\tau$. The fact that the initial value of $\chi_m$ is very small is related to vanishing initial value of the anisotropy parameter $\xi\sim 0$ at the initial time $\bar{\tau}=0.2$ fm/c.
\par
Using the same method which is used to determine the $T$ dependence of the speed of sound in Fig. \ref{fig-9}, the $T$ dependence of $\chi_{m,0}$ and $\chi_{m}$ is determined, and the result is plotted in Fig. \ref{fig-12} for relaxation times $\tau_{r,0}$ (red solid curves) and $\tau_{r}$ (black dashed curves) equal to $0.3$ fm/c [Fig. \ref{fig-12}(a)] and $0.5$ fm/c [Fig. \ref{fig-12}(b)], and an initial magnetic field $e\bar{B}=15 m_{\pi}^{2}$ with $m_{\pi}\sim 140$ MeV. In comparison to the results arising for $\chi_{m,0}$  and $\chi_{m}$, appearing in Fig. \ref{fig-11}, magnetic susceptibilities arising for $e\bar{B}=15 m_{\pi}^{2}$ are $1$ order of magnitude smaller than those for $e\bar{B}=5 m_{\pi}^{2}$.
According to results presented in Fig. \ref{fig-12}, $\chi_{m,0}$ and $\chi_{m}$ increase after the collision at $T=400$ MeV. After reaching a maximum at $T\sim 220$ MeV for a nondissipative fluid and $T_{\mbox{\tiny{max}}}\sim 180-200$ MeV for a dissipative fluid, they decrease as the fluid cools. For larger value of $\tau_{r}$, $T_{\mbox{\tiny{max}}}$ is slightly shifted to smaller values of $T$. As in the case of $c_{s}$, we have also plotted $\chi_{m,0}$ and $\chi_{m}$ in terms of $\lambda_0$ and $\lambda$, respectively. Because of the slight difference between $T, \lambda_{0}$ and $\lambda$ in the regime $\tau\in [0.5, 4]$ fm/c, demonstrated in Fig. \ref{fig-4}, a shift of $\lambda_{0,max}$ and $\lambda_{\mbox{\tiny{max}}}$ to even smaller values of temperature occurs once $\chi_{m,0}$ and $\chi_{m}$ are plotted as functions of $\lambda_{0}$ and $\lambda$. Assuming the critical temperature of the QCD phase transition to be at $T_{c}\sim 180-200$ MeV, it is possible to identify $T_{\mbox{\tiny{max}}}$ with $T_{c}$. We show, however, that this interpretation depends strongly on the relaxation time $\tau_r$ (see Fig. \ref{fig-14}).
\par
In Fig. \ref{fig-13}, the $\tau_{r}$ dependence of $\chi_{m}$ is plotted for fixed $\tau=2,4,6$ fm/c (red circles, blue rectangles, and green squares). The initial value of the magnetic field is chosen to be $e\bar{B}=5 m_{\pi}^{2}$ with the pion mass given by  $m_{\pi}=140$ MeV. As expected from Fig. \ref{fig-12}, for each fixed value of $\tau$, the magnetic susceptibility increases with increasing relaxation time $\tau_{r}$. Similar results arise for $e\bar{B}=15 m_{\pi}^{2}$.
\par
In Fig. \ref{fig-14}, the correlation of the position of the maxima appearing in $c_{s,\|}(\lambda)$ with those appearing in $\chi_{m}(\lambda)$ for various relaxation times $\tau_r$ is studied. To do this, we consider the $\lambda$ dependence of $c_{s,\|}$ and $\chi_{m}$, and determine the position of their maxima for a number of fixed relaxation times, $\tau_r$. Let us denote these positions by $(\lambda_{\mbox{\tiny{max}}})_{c_{s,\|}}$ and $(\lambda_{\mbox{\tiny{max}}})_{\chi_{m}}$, respectively. In Fig. \ref{fig-14}, $(\lambda_{\mbox{\tiny{max}}})_{\chi_{m}}$ is then plotted versus $(\lambda_{\mbox{\tiny{max}}})_{c_{s,\|}}$ for $\tau_{r}\in [0.2,2]$ fm/c in $\Delta \tau_{r}=0.1$ fm/c steps (see the red points in Fig. \ref{fig-14}). For the latter, we choose the initial magnetic field $e\bar{B}=5 m_{\pi}^{2}$ with $m_{\pi}=140$ MeV.  The green down-triangle  $(\blacktriangledown)$ at $((\lambda_{\mbox{\tiny{max}}})_{c_{s,\|}}, (\lambda_{\mbox{\tiny{max}}})_{\chi_{m}})=(234, 203)$ MeV corresponds to $\tau_{r}=0.2$ fm/c, and the green up-triangle $(\blacktriangle$ at  $((\lambda_{\mbox{\tiny{max}}})_{c_{s,\|}}, (\lambda_{\mbox{\tiny{max}}})_{\chi_{m}})=(125,103)$ MeV corresponds to $\tau_{r}=2$ fm/c. The blue solid line is characterized by $(\lambda_{\mbox{\tiny{max}}})_{c_{s,\|}}= (\lambda_{\mbox{\tiny{max}}})_{\chi_{m}}$. The deviation of our result from this line indicates that $(\lambda_{\mbox{\tiny{max}}})_{c_{s,\|}}>(\lambda_{\mbox{\tiny{max}}})_{\chi_{m}}$ for all values of $\tau_{r}\in [0.2,2]$ fm/c.
According to these results, for larger relaxation times $(\lambda_{\mbox{\tiny{max}}})_{c_{s,\|}}$ and $(\lambda_{\mbox{\tiny{max}}})_{\chi_{m}}$ are shifted to smaller effective temperatures.
Moreover, as it turns out, the relation between the position of the maxima appearing in $\chi_{m}$ and $c_{s_{\|}}$ and the temperature of the QCD phase transition at $T_{c}\sim 180-200$ MeV strongly depend on the relaxation time $\tau_{r}$.
\par
We finally focus on the proper time evolution of the transport coefficients $\eta_{0}$ and $\xi_{0}$. In Figs. \ref{fig-15}(a) and \ref{fig-16}(a), the $\tau$ dependence of the viscosities $\eta_{0}(\tau)/\eta_{0}(\bar{\tau})$ and $\zeta_{0}(\tau)/\zeta_{0}(\bar{\tau})$ is plotted for a fixed relaxation time $\tau_{r}=0.5$ fm/c. Here, the initial time is $\bar{\tau}_{r}=0.2$ fm/c. The corresponding expression for the shear viscosity $\eta_{0}$ is given in \eqref{D52}. The bulk viscosity $\zeta_{0}$ arises by combining $\alpha_{0}$ and $\tilde{\zeta}_{0}$ from \eqref{D51} and \eqref{D53} as $\zeta_{0}=\alpha_{0}\tau+\tilde{\zeta}_{0}$.  According to these results $\eta_{0}$ ($\zeta_{0}$) decreases (increases) with increasing $\tau$.
To determine the temperature dependence of $\eta_{0}$ and $\zeta_{0}$, we use, as in previous cases, $\tau=\bar{\tau}\left(\bar{T}/T\right)^{3}$ from \eqref{N29}. The resulting $T$ dependence of $\eta_{0}(\tau)/\eta_{0}(\bar{\tau})$ and $\zeta_{0}(\tau)/\zeta_{0}(\bar{\tau})$ is plotted in Figs. \ref{fig-15}(b) and \ref{fig-16}(b). As it turns out, $\eta_{0}$ ($\zeta_{0}$) decreases (increases) with decreasing $T$. Bearing in mind that shear viscosity is proportional to the mean free path of quarks in the fluid, $\lambda_{\mbox{\tiny{mfp}}}$ \cite{weise2012,sadooghi2014}, the fact that $\eta_{0}$ increases with increasing temperature indicates that  $\lambda_{\mbox{\tiny{mfp}}}$ also increases with increasing temperature. We also notice that the result arising in Fig. \ref{fig-15}(b) for the temperature dependence of $\eta_0(T)/\eta_{0}(\bar{T})$ is in good agreement with the expected $\eta_0(T)/\eta_{0}(\bar{T})\sim (T/\bar{T})^{3}$ from \cite{arnold2000} with the initial temperature $\bar{T}=400$ MeV. As concerns the temperature dependence of $\zeta_0(T)/\zeta_{0}(\bar{T})$ from Fig. \ref{fig-16}(b), however, it does not coincide with $\zeta(T)\propto \eta(T)\left(\frac{1}{3}-c_{s}^{2}\right)$ from \cite{buchel2005}, arising from gauge/gravity duality. Whereas, plugging $c_s=c_{s,\|}$ from Fig. \ref{fig-9}(b) into this expression $\zeta_0(T)/\zeta_{0}(\bar{T})$ turns out to be always positive, for $c_s=c_{s,\perp}$ the resulting negative values for $\zeta_0(T)/\zeta_{0}(\bar{T})$ were several orders of magnitude larger than the result presented in Fig. \ref{fig-16}(b). Despite this discrepancy in the anisotropic case, in the isotropic limit, the expression $\zeta_{0}=\alpha_{0}\tau+\tilde{\zeta}_{0}$ includes the expected factor $\left(\frac{1}{3}-c_{s}^{2}\right)$, as expected. To see this, let us combine, $\alpha_{0}$ and $\tilde{\zeta}_{0}$ from (\ref{D31}) with $\ell_{0}=\nu_{H}\frac{D\lambda}{\lambda}$ from \eqref{D13} and  $\ell_{0}=\frac{\nu_{H}}{3}$ from \eqref{D26} to arrive at
\begin{eqnarray}\label{A2}
\zeta_{0}=\frac{1}{3}\int d\tilde{k}~\nu_{H}\left(\frac{D\lambda}{\lambda}\tau+\frac{1}{3}\right)|\boldsymbol{k}|^{4}.
\end{eqnarray}
Setting, in the isotropic limit, the ratio $\frac{D\lambda}{\lambda}=\frac{DT}{T}$, and bearing in mind that in this case $\frac{DT}{T}=-c_{s}^{2}\partial_{\mu}u^{\mu}=-\frac{c_{s}^{2}}{\tau}$ with $c_{s}^{2}=\frac{\partial p}{\partial\epsilon}$ \cite{kajantie1985},\footnote{For the Bjorken flow the four-divergence of $u^{\mu}$ is given by $1/\tau$.} $\zeta_{0}$ from \eqref{A2} becomes proportional to $\left(-c_{s}^{2}+\frac{1}{3}\right)$, as claimed. Let us also notice that the transport coefficients of hot and magnetized quark matter are recently computed in \cite{chandra2018}. In contrast to our computation, where the evolution of the magnetic field is implemented in the computation, the magnetic field in \cite{chandra2018} is assumed to be aligned in a fixed direction, and remains constant during the evolution of the QGP.
\section{Concluding remarks}\label{sec6}
\setcounter{equation}{0}
The main purpose of the present paper was to study the role played by the finite magnetization of the paramagnetic QGP in the production of pressure anisotropies, and to quantify the possible interplay between the effects caused by this magnetization and those arising from nonvanishing dissipations in the isotropization of this medium. We were motivated by the wide belief that very large magnetic fields are produced in the early stages of noncentral HICs \cite{warringa2007}, and that the QCD matter produced in these collisions is paramagnetic \cite{endrodi2013}. Paramagnetic squeezing was previously studied in \cite{bali2013} in the framework of lattice QCD for a \textit{static} QCD matter in the presence of a \textit{constant} magnetic field. We generalized the same proposal to a uniformly expanding QGP in the presence of a dynamical magnetic field, by making use of standard methods from aHydro \cite{strickland2010}. In particular, we used the similarity between \eqref{E1} and \eqref{E2}, the energy-momentum tensor of an ideal paramagnetic fluid and of a longitudinally expanding fluid in the framework of aHydro, and introduced the magneto-anisotropic one-particle distribution function $f_{b}$ in terms of the unit vector in the direction of the magnetic field $b^{\mu}$, an anisotropy parameter $\xi_{0}$, and an effective temperature $\lambda_{0}$. In this way, the effect of the anisotropies caused, in particular, by the magnetization of the QGP is phenomenologically taken into account.
\par
Using $f_{b}$, we determined, similar to the standard aHydro method, described, e.g., in  \cite{strickland2010,strickland2017}, the first two moments of the Boltzmann equation satisfied by $f_{b}$ in the RTA, and derived a set of coupled differential equations for $\xi_{0}$ and $\lambda_{0}$ in terms of the relaxation time $\tau_{r,0}$. The latter is taken to be a free parameter, apart from the initial proper time $\bar{\tau}$ and magnetic field $\bar{B}$. The uniform expansion of the fluid was described by the $1+1$ dimensional Bjorken flow \cite{bjorken}, which is only valid when (i) the fluid expands only in the longitudinal direction with respect to the beam direction and (ii) the system is boost invariant along this direction. Moreover, by making the assumption that the magnetic field is transverse to the direction of the fluid velocity (see Fig. \ref{fig1}), it was possible to use the solution $B(\tau)\sim \tau^{-1}$, which arises in the framework of ideal transverse MHD \cite{rischke2015}.
\par
We used appropriate initial values for $\xi_{0}$ and $\lambda_{0}$, and solved numerically the aforementioned differential equations. In this way, we first determined the proper time dependence of $\xi_{0}$ and $\lambda_{0}$ for various fixed $\tau_{r,0}$. Using the dependence of various thermodynamical quantities on $f_{b}(x,p;\xi_0,\lambda_{0})$, it was then possible to determine the evolution of transverse and longitudinal pressures $p_{0\perp}$ and $p_{0\|}$, the energy density $\epsilon_{0}$, transverse and longitudinal speeds of sound $c_{s0,\perp}$ and $c_{s0,\|}$, as well as the magnetic susceptibility $\chi_{m,0}$ for an ideal nondissipative, and longitudinally expanding QGP.\footnote{As described in previous sections, the symbols $\perp$ and $\|$ describe the ``transverse'' and ``longitudinal'' directions with respect to the magnetic field. The latter is assumed to be perpendicular to the beam direction.} The results are presented in Sec. \ref{sec5}.
\par
To take the viscous effects, apart from the magnetization of the fluid, into account, we extended our method to a dissipative QGP. To do this, we first derived the dissipative correction to $f_{b}$ in a first-order derivative expansion by making use of a number of results from \cite{tuchin2013,tabatabaee2016,rischke2009,roy2018}. Because of the presence of an additional four-vector $b^{\mu}$, apart from the velocity four-vector $u^{\mu}$, a large number of transport coefficients were defined in a magnetized fluid, as expected. Performing then the same steps that led to the aforementioned differential equations for $\xi_{0}$ and $\lambda_{0}$, we arrived at the corresponding coupled differential equations for $\xi$ and $\lambda$ in the dissipative case. We numerically solved these equations for different choices of the relaxation time $\tau_{r}$ in the dissipative case.
\par
According to the plots demonstrated in Figs. \ref{fig-2}, \ref{fig-5}, and \ref{fig-7}, for fixed values of $\tau_{r,0}$ and $\tau_{r}$, the anisotropy induced by the magnetization in the early stages of the evolution of a nondissipative fluid is quite large, and, as it turns out, it is compensated by dissipative effects. As concerns the longitudinal and transverse pressures, for instance, the longitudinal pressure is in the absence of dissipation larger than the transverse pressure. Here, the terms longitudinal and transverse are with respect to the direction of the magnetic field. Using the same terminology as in the aHydro literature, i.e. using these two terms with respect to the beam line, our results indicate that the dissipation diminishes the effect of magnetization in making $p_{T}$ larger than $p_{L}$.
We used the proper time evolution of the energy density and transverse as well as longitudinal pressures to determine the transverse and longitudinal speeds of sound. The completely different proper time dependence of these two velocities is demonstrated in Figs. \ref{fig-7} and \ref{fig-8}.
\par
Parallel to the above results, we were interested in the temperature dependence of the transverse and longitudinal speeds of sound, $c_{s,\perp}$ and $c_{s,\|}$, magnetic susceptibility $\chi_m$, shear and bulk viscosities $\eta$ and $\zeta$. We used the simple proper time dependence of the temperature $T$, $T\sim \tau^{-1/3}$ from \eqref{N29}, and converted the proper time dependence of these quantities into their $T$ dependence. This was possible because according to our results from Fig. \ref{fig-4}, there are almost no differences between $T,\lambda_0$ and $\lambda$. In other words, the finite magnetization of the fluid does not practically affect the evolution of its temperature. The same is also true for dissipative effects.
According to the results for the $T$ dependence of $\chi_{m}$ from Fig. \ref{fig-12}, $\chi_{m}$ increases with increasing $T$ up to a maximum value, and then decreases with increasing $T$. This result may not be expected from lattice QCD results in \cite{delia2013}, for instance, but this may lie on the fact that in the previous computations of $\chi_{m}$, the
quark matter and background magnetic fields are assumed to be static. In the present paper, however, we considered the $\tau^{-1}$ decay of the magnetic field, and the results demonstrated in Fig. \ref{fig-12} for the $T$ dependence of $\chi_{m}$ comprise this crucial difference. To describe the backreaction arising from the dynamical evolution of the magnetic field, let us consider Fig. \ref{fig-11}. Multiplying the curve plotted in this figure with $B=B_{0}(\tau_0/\tau)$, we arrive at a $\tau$ dependent magnetization $M$, whose $\tau$ dependence is qualitatively similar to the $\tau$ dependence of $\chi_m$ plotted in this figure. The result shows that $M$ increases with increasing $\tau$, reaches a maximum at an early stage ($\tau\sim 2$ fm/c), and decreases then with increasing $\tau$. The same kind of backreaction also occurs in the $T$ dependence of the shear and bulk viscosities from Figs. \ref{fig-15} and \ref{fig-16}, which are qualitatively in agreement with similar results in the literature, as is described in the previous section.
\par
Despite these promising results, there is one remaining point to be noticed. As aforementioned, there are a small number of free parameters in our numerical computations. The value of the initial time $\bar{\tau}$ and initial magnetic field $\bar{B}$, which we have used, comply with the existing numbers in the literature related to HICs. The values of the relaxation times, $\tau_{r,0}$ in the nondissipative case and $\tau_{r}$ in the dissipative case, are, however, arbitrarily chosen to be $0.3$ and $0.5$ fm/c. It is not clear how close these numbers are to the real relaxation times in the expanding QGP. In particular, their dependence on the magnitude of a dynamical magnetic field is yet unknown.\footnote{In \cite{chandra2018}, the thermal relaxation time in a static QGP in the presence of a constant magnetic field is computed in the lowest Landau level approximation. } It is thus necessary to separately determine these parameters in an expanding magnetized QGP. Notwithstanding this caveat, the magneto-anisotropic one-particle distribution function $f_{b}$ proposed in the present paper can, in principal, be used to determine a large number of observables in HIC experiments.
Recently, using an appropriate anisotropic distribution function, the dilepton production rate is computed within the aHydro framework \cite{kasmaei2018}. It would be interesting to generalize this computation for a magnetized QGP to take, in particular, the effect of anisotropies caused by its magnetization and the evolution of the QGP as well as the dynamics of the background magnetic field into account. The result may be then compared with those presented in \cite{taghinavaz2016,ghosh2018}, where the QGP and the background magnetic field are assumed to be static. Another possibility to extend the results presented in this work is to allow the fluid to possess, apart from the longitudinal expansion assumed in the framework of transverse MHD, an expansion in the transverse direction with respect to the beam line. To do this, one should replace the $\tau^{-1}$ solution of $B(\tau)$ with the recently found $3+1$ dimensional solution to the conformal (Gubser) MHD, presented in \cite{shokri2018-2}. We postpone all these computations to our future publications.
\section{acknowledgments}
The authors thank M. Shokri for useful discussions. S. M. A. Tabatabaee thanks R. Ryblewski for discussions during the XIIIth conference on Quark Confinement and the Hadron Spectrum, Ireland, July 2018, where he presented preliminary results from the present work in a poster. This work is supported by Sharif University of Technology's Office of Vice President for Research under Grant No. G960212/Sadooghi.

\end{document}